\newif\ifanon\anonfalse\newif\ifextended\extendedtrue\newif\ifarxivbuild\arxivbuildtrue\newif\ifarxiv\arxivtrue

\pdfoutput=1  %
\PassOptionsToPackage{svgnames}{xcolor}

\newif\ifarxivsinglecolumn
\arxivsinglecolumntrue

\ifanon
  \documentclass[sigconf, screen, review,anonymous]{acmart}
\else
  \ifarxiv
    \ifarxivsinglecolumn
      \documentclass[acmsmall,screen,timestamp,nonacm]{acmart}
    \else
      \documentclass[sigconf, screen, review, nonacm]{acmart}
    \fi
  \else
    \documentclass[sigconf, screen, review]{acmart}
  \fi
\fi

\usepackage{natbib}

\newif\ifclean
\cleanfalse

\newif\ifoldbits
\oldbitsfalse

\setcopyright{none}
\copyrightyear{2024}
\acmYear{2024}
\ifanon
\acmDOI{XXXXXXX.XXXXXXX}

\acmConference[]{}
\acmISBN{978-1-4503-XXXX-X/18/06}
\else
\ifarxiv
\acmDOI{}
\acmConference{}
\acmISBN{}
\else
\acmConference{}
\acmISBN{}
\fi%
\fi%

\usepackage{multicol}
\usepackage{cleveref}
\usepackage{tikz}
\usetikzlibrary{decorations.pathreplacing, calligraphy, arrows.meta}
\usetikzlibrary{patterns}

\usepackage[listings]{tcolorbox}
\usepackage{enumitem}
\usepackage{bbding}
\usepackage{etoolbox}
\usepackage{stmaryrd}
\usepackage{array}
\usepackage{longtable}
\usepackage{caption}
\usepackage{subcaption}
\usepackage{placeins}
\usepackage{etoc}

\usepackage[T1]{fontenc}
\usepackage[scaled=0.82]{beramono}

\definecolor{ibmcolourblind1}{RGB}{100, 143, 255} %
\definecolor{ibmcolourblind2}{RGB}{120, 94, 240} %
\definecolor{ibmcolourblind3}{RGB}{220, 38, 127} %
\definecolor{ibmcolourblind4}{RGB}{254, 97, 0} %
\definecolor{ibmcolourblind5}{RGB}{255, 176, 0} %

\definecolor{commentcolour}{RGB}{0, 176, 0} %
\colorlet{grey}{gray} %

\ifclean
\else

\fi

\ifanon
\newcommand\nonanonhref[2]{{\color{DarkGreen}#2}}
\else
\let\nonanonhref=\href
\fi

\newcommand{\allow}{A}
\newcommand{\forbid}{F}
\newcommand{\tableheader}[1]{\rotatebox{90}{#1}}
\expandafter\newcommand \csname RPi 3B+-MP+dmb+ctrl-rfisvceret-addr result\endcsname {0/22M}
\expandafter\newcommand \csname RPi 4B-MP+dmb+ctrl-rfisvceret-addr result\endcsname {0/108M}
\expandafter\newcommand \csname RPi 5-MP+dmb+ctrl-rfisvceret-addr result\endcsname {0/39M}
\expandafter\newcommand \csname ODROID N2+ (big)-MP+dmb+ctrl-rfisvceret-addr result\endcsname {0/18M}
\expandafter\newcommand \csname RPi 3B+-MP+dmb+ctrlelr result\endcsname {0/22M}
\expandafter\newcommand \csname RPi 4B-MP+dmb+ctrlelr result\endcsname {0/108M}
\expandafter\newcommand \csname RPi 5-MP+dmb+ctrlelr result\endcsname {0/39M}
\expandafter\newcommand \csname ODROID N2+ (big)-MP+dmb+ctrlelr result\endcsname {0/18M}
\expandafter\newcommand \csname RPi 3B+-MP.EL1+dmb+ctrlvbarsvc result\endcsname {0/21M}
\expandafter\newcommand \csname RPi 4B-MP.EL1+dmb+ctrlvbarsvc result\endcsname {0/108M}
\expandafter\newcommand \csname RPi 5-MP.EL1+dmb+ctrlvbarsvc result\endcsname {0/39M}
\expandafter\newcommand \csname ODROID N2+ (big)-MP.EL1+dmb+ctrlvbarsvc result\endcsname {0/17M}
\expandafter\newcommand \csname RPi 3B+-SB+dmb+rfieret-addr result\endcsname {591K/21M}
\expandafter\newcommand \csname RPi 4B-SB+dmb+rfieret-addr result\endcsname {8.0K/107M}
\expandafter\newcommand \csname RPi 5-SB+dmb+rfieret-addr result\endcsname {54/39M}
\expandafter\newcommand \csname ODROID N2+ (big)-SB+dmb+rfieret-addr result\endcsname {16K/16M}
\expandafter\newcommand \csname RPi 3B+-SB+dmb+rfisvc-addr result\endcsname {839K/21M}
\expandafter\newcommand \csname RPi 4B-SB+dmb+rfisvc-addr result\endcsname {2.9K/106M}
\expandafter\newcommand \csname RPi 5-SB+dmb+rfisvc-addr result\endcsname {135K/39M}
\expandafter\newcommand \csname ODROID N2+ (big)-SB+dmb+rfisvc-addr result\endcsname {18K/16M}
\expandafter\newcommand \csname RPi 3B+-SB+svcs result\endcsname {534K/21M}
\expandafter\newcommand \csname RPi 4B-SB+svcs result\endcsname {0/106M}
\expandafter\newcommand \csname RPi 5-SB+svcs result\endcsname {646K/39M}
\expandafter\newcommand \csname ODROID N2+ (big)-SB+svcs result\endcsname {2.0K/16M}
\expandafter\newcommand \csname RPi 3B+-MP+svc+eret result\endcsname {1.4K/21M}
\expandafter\newcommand \csname RPi 4B-MP+svc+eret result\endcsname {33/107M}
\expandafter\newcommand \csname RPi 5-MP+svc+eret result\endcsname {77/39M}
\expandafter\newcommand \csname ODROID N2+ (big)-MP+svc+eret result\endcsname {22K/17M}
\expandafter\newcommand \csname RPi 3B+-MP+eret+svc result\endcsname {244/21M}
\expandafter\newcommand \csname RPi 4B-MP+eret+svc result\endcsname {256K/107M}
\expandafter\newcommand \csname RPi 5-MP+eret+svc result\endcsname {20/39M}
\expandafter\newcommand \csname ODROID N2+ (big)-MP+eret+svc result\endcsname {9.2K/18M}
\expandafter\newcommand \csname RPi 3B+-SB+dmb+svc result\endcsname {351K/33M}
\expandafter\newcommand \csname RPi 4B-SB+dmb+svc result\endcsname {458/11M}
\expandafter\newcommand \csname RPi 5-SB+dmb+svc result\endcsname {63K/11M}
\expandafter\newcommand \csname ODROID N2+ (big)-SB+dmb+svc result\endcsname {195/16M}
\expandafter\newcommand \csname RPi 3B+-S+dmb+svc result\endcsname {0/33M}
\expandafter\newcommand \csname RPi 4B-S+dmb+svc result\endcsname {0/12M}
\expandafter\newcommand \csname RPi 5-S+dmb+svc result\endcsname {0/11M}
\expandafter\newcommand \csname ODROID N2+ (big)-S+dmb+svc result\endcsname {0/17M}
\expandafter\newcommand \csname RPi 3B+-SB+dmb+eret result\endcsname {162K/33M}
\expandafter\newcommand \csname RPi 4B-SB+dmb+eret result\endcsname {85K/12M}
\expandafter\newcommand \csname RPi 5-SB+dmb+eret result\endcsname {2.0K/11M}
\expandafter\newcommand \csname ODROID N2+ (big)-SB+dmb+eret result\endcsname {38/17M}
\expandafter\newcommand \csname RPi 3B+-S+dmb+eret result\endcsname {0/33M}
\expandafter\newcommand \csname RPi 4B-S+dmb+eret result\endcsname {0/12M}
\expandafter\newcommand \csname RPi 5-S+dmb+eret result\endcsname {0/11M}
\expandafter\newcommand \csname ODROID N2+ (big)-S+dmb+eret result\endcsname {0/17M}
\expandafter\newcommand \csname RPi 3B+-MP+eret+dmb result\endcsname {1.3K/33M}
\expandafter\newcommand \csname RPi 4B-MP+eret+dmb result\endcsname {262/19M}
\expandafter\newcommand \csname RPi 5-MP+eret+dmb result\endcsname {20/11M}
\expandafter\newcommand \csname ODROID N2+ (big)-MP+eret+dmb result\endcsname {18K/18M}
\expandafter\newcommand \csname RPi 3B+-MP+dmb+svc-eret result\endcsname {0/33M}
\expandafter\newcommand \csname RPi 4B-MP+dmb+svc-eret result\endcsname {0/19M}
\expandafter\newcommand \csname RPi 5-MP+dmb+svc-eret result\endcsname {0/11M}
\expandafter\newcommand \csname ODROID N2+ (big)-MP+dmb+svc-eret result\endcsname {0/18M}
\expandafter\newcommand \csname RPi 3B+-MP.EL1+dmb+svc result\endcsname {0/33M}
\expandafter\newcommand \csname RPi 4B-MP.EL1+dmb+svc result\endcsname {0/12M}
\expandafter\newcommand \csname RPi 5-MP.EL1+dmb+svc result\endcsname {0/11M}
\expandafter\newcommand \csname ODROID N2+ (big)-MP.EL1+dmb+svc result\endcsname {29/17M}
\expandafter\newcommand \csname RPi 3B+-LB+svc-dmb-erets result\endcsname {0/1.5M}
\expandafter\newcommand \csname RPi 4B-LB+svc-dmb-erets result\endcsname {0/19M}
\expandafter\newcommand \csname RPi 5-LB+svc-dmb-erets result\endcsname {0/11M}
\expandafter\newcommand \csname ODROID N2+ (big)-LB+svc-dmb-erets result\endcsname {0/18M}
\expandafter\newcommand \csname RPi 3B+-LB+svc-erets result\endcsname {0/1.5M}
\expandafter\newcommand \csname RPi 4B-LB+svc-erets result\endcsname {0/19M}
\expandafter\newcommand \csname RPi 5-LB+svc-erets result\endcsname {0/11M}
\expandafter\newcommand \csname ODROID N2+ (big)-LB+svc-erets result\endcsname {0/18M}
\expandafter\newcommand \csname RPi 3B+-LB+svcs result\endcsname {0/1M}
\expandafter\newcommand \csname RPi 4B-LB+svcs result\endcsname {0/19M}
\expandafter\newcommand \csname RPi 5-LB+svcs result\endcsname {0/11M}
\expandafter\newcommand \csname ODROID N2+ (big)-LB+svcs result\endcsname {388/18M}
\expandafter\newcommand \csname RPi 3B+-MP+dmb+ctrl-eret result\endcsname {0/1M}
\expandafter\newcommand \csname RPi 4B-MP+dmb+ctrl-eret result\endcsname {0/19M}
\expandafter\newcommand \csname RPi 5-MP+dmb+ctrl-eret result\endcsname {0/11M}
\expandafter\newcommand \csname ODROID N2+ (big)-MP+dmb+ctrl-eret result\endcsname {0/18M}
\expandafter\newcommand \csname RPi 3B+-MP+dmb+ctrl-svc result\endcsname {0/1M}
\expandafter\newcommand \csname RPi 4B-MP+dmb+ctrl-svc result\endcsname {0/19M}
\expandafter\newcommand \csname RPi 5-MP+dmb+ctrl-svc result\endcsname {0/11M}
\expandafter\newcommand \csname ODROID N2+ (big)-MP+dmb+ctrl-svc result\endcsname {0/18M}
\expandafter\newcommand \csname RPi 3B+-MP+dmb+data-svc result\endcsname {0/1M}
\expandafter\newcommand \csname RPi 4B-MP+dmb+data-svc result\endcsname {0/19M}
\expandafter\newcommand \csname RPi 5-MP+dmb+data-svc result\endcsname {0/11M}
\expandafter\newcommand \csname ODROID N2+ (big)-MP+dmb+data-svc result\endcsname {0/18M}
\expandafter\newcommand \csname RPi 3B+-MP+dmb+dmb-eret result\endcsname {0/1M}
\expandafter\newcommand \csname RPi 4B-MP+dmb+dmb-eret result\endcsname {0/19M}
\expandafter\newcommand \csname RPi 5-MP+dmb+dmb-eret result\endcsname {0/11M}
\expandafter\newcommand \csname ODROID N2+ (big)-MP+dmb+dmb-eret result\endcsname {0/18M}
\expandafter\newcommand \csname RPi 3B+-MP+dmb+eret result\endcsname {0/1M}
\expandafter\newcommand \csname RPi 4B-MP+dmb+eret result\endcsname {0/19M}
\expandafter\newcommand \csname RPi 5-MP+dmb+eret result\endcsname {0/11M}
\expandafter\newcommand \csname ODROID N2+ (big)-MP+dmb+eret result\endcsname {0/18M}
\expandafter\newcommand \csname RPi 3B+-MP+dmb+eret-dmb result\endcsname {0/1M}
\expandafter\newcommand \csname RPi 4B-MP+dmb+eret-dmb result\endcsname {0/19M}
\expandafter\newcommand \csname RPi 5-MP+dmb+eret-dmb result\endcsname {0/11M}
\expandafter\newcommand \csname ODROID N2+ (big)-MP+dmb+eret-dmb result\endcsname {0/18M}
\expandafter\newcommand \csname RPi 3B+-MP+dmb+eret-svc result\endcsname {0/1M}
\expandafter\newcommand \csname RPi 4B-MP+dmb+eret-svc result\endcsname {0/3M}
\expandafter\newcommand \csname RPi 5-MP+dmb+eret-svc result\endcsname {0/5.5M}
\expandafter\newcommand \csname ODROID N2+ (big)-MP+dmb+eret-svc result\endcsname {0/4M}
\expandafter\newcommand \csname RPi 3B+-MP+dmb+eret=addr result\endcsname {0/1M}
\expandafter\newcommand \csname RPi 4B-MP+dmb+eret=addr result\endcsname {0/19M}
\expandafter\newcommand \csname RPi 5-MP+dmb+eret=addr result\endcsname {0/11M}
\expandafter\newcommand \csname ODROID N2+ (big)-MP+dmb+eret=addr result\endcsname {0/18M}
\expandafter\newcommand \csname RPi 3B+-MP+dmb+svc result\endcsname {0/1M}
\expandafter\newcommand \csname RPi 4B-MP+dmb+svc result\endcsname {0/19M}
\expandafter\newcommand \csname RPi 5-MP+dmb+svc result\endcsname {0/11M}
\expandafter\newcommand \csname ODROID N2+ (big)-MP+dmb+svc result\endcsname {84/18M}
\expandafter\newcommand \csname RPi 3B+-MP+dmb+svc-addreret result\endcsname {0/1M}
\expandafter\newcommand \csname RPi 4B-MP+dmb+svc-addreret result\endcsname {0/19M}
\expandafter\newcommand \csname RPi 5-MP+dmb+svc-addreret result\endcsname {0/11M}
\expandafter\newcommand \csname ODROID N2+ (big)-MP+dmb+svc-addreret result\endcsname {0/18M}
\expandafter\newcommand \csname RPi 3B+-MP+dmb+svc-dmb result\endcsname {0/1M}
\expandafter\newcommand \csname RPi 4B-MP+dmb+svc-dmb result\endcsname {0/19M}
\expandafter\newcommand \csname RPi 5-MP+dmb+svc-dmb result\endcsname {0/11M}
\expandafter\newcommand \csname ODROID N2+ (big)-MP+dmb+svc-dmb result\endcsname {0/18M}
\expandafter\newcommand \csname RPi 3B+-MP+dmb+svc-dmb-eret result\endcsname {0/1M}
\expandafter\newcommand \csname RPi 4B-MP+dmb+svc-dmb-eret result\endcsname {0/19M}
\expandafter\newcommand \csname RPi 5-MP+dmb+svc-dmb-eret result\endcsname {0/11M}
\expandafter\newcommand \csname ODROID N2+ (big)-MP+dmb+svc-dmb-eret result\endcsname {0/18M}
\expandafter\newcommand \csname RPi 3B+-MP+dmb+svcnoeis result\endcsname {0/1M}
\expandafter\newcommand \csname RPi 4B-MP+dmb+svcnoeis result\endcsname {0/19M}
\expandafter\newcommand \csname RPi 5-MP+dmb+svcnoeis result\endcsname {0/11M}
\expandafter\newcommand \csname ODROID N2+ (big)-MP+dmb+svcnoeis result\endcsname {23/18M}
\expandafter\newcommand \csname RPi 3B+-MP+eret+addr result\endcsname {63/1M}
\expandafter\newcommand \csname RPi 4B-MP+eret+addr result\endcsname {0/19M}
\expandafter\newcommand \csname RPi 5-MP+eret+addr result\endcsname {2/11M}
\expandafter\newcommand \csname ODROID N2+ (big)-MP+eret+addr result\endcsname {22K/18M}
\expandafter\newcommand \csname RPi 3B+-MP+erets result\endcsname {30/1M}
\expandafter\newcommand \csname RPi 4B-MP+erets result\endcsname {59/19M}
\expandafter\newcommand \csname RPi 5-MP+erets result\endcsname {16/11M}
\expandafter\newcommand \csname ODROID N2+ (big)-MP+erets result\endcsname {29K/18M}
\expandafter\newcommand \csname RPi 3B+-MP+svc+addr result\endcsname {59/1M}
\expandafter\newcommand \csname RPi 4B-MP+svc+addr result\endcsname {0/19M}
\expandafter\newcommand \csname RPi 5-MP+svc+addr result\endcsname {8/11M}
\expandafter\newcommand \csname ODROID N2+ (big)-MP+svc+addr result\endcsname {17K/18M}
\expandafter\newcommand \csname RPi 3B+-MP+svc+dmb result\endcsname {80/1M}
\expandafter\newcommand \csname RPi 4B-MP+svc+dmb result\endcsname {3/19M}
\expandafter\newcommand \csname RPi 5-MP+svc+dmb result\endcsname {876/11M}
\expandafter\newcommand \csname ODROID N2+ (big)-MP+svc+dmb result\endcsname {18K/17M}
\expandafter\newcommand \csname RPi 3B+-MP+svc-dmb+addr result\endcsname {0/1M}
\expandafter\newcommand \csname RPi 4B-MP+svc-dmb+addr result\endcsname {0/19M}
\expandafter\newcommand \csname RPi 5-MP+svc-dmb+addr result\endcsname {0/11M}
\expandafter\newcommand \csname ODROID N2+ (big)-MP+svc-dmb+addr result\endcsname {0/17M}
\expandafter\newcommand \csname RPi 3B+-MP+svc-dmb-eret+addr result\endcsname {0/1M}
\expandafter\newcommand \csname RPi 4B-MP+svc-dmb-eret+addr result\endcsname {0/19M}
\expandafter\newcommand \csname RPi 5-MP+svc-dmb-eret+addr result\endcsname {0/11M}
\expandafter\newcommand \csname ODROID N2+ (big)-MP+svc-dmb-eret+addr result\endcsname {0/17M}
\expandafter\newcommand \csname RPi 3B+-MP+svc-eret+addr result\endcsname {52/1M}
\expandafter\newcommand \csname RPi 4B-MP+svc-eret+addr result\endcsname {0/19M}
\expandafter\newcommand \csname RPi 5-MP+svc-eret+addr result\endcsname {2/11M}
\expandafter\newcommand \csname ODROID N2+ (big)-MP+svc-eret+addr result\endcsname {13K/17M}
\expandafter\newcommand \csname RPi 3B+-MP+svc-erets result\endcsname {42/1M}
\expandafter\newcommand \csname RPi 4B-MP+svc-erets result\endcsname {2/19M}
\expandafter\newcommand \csname RPi 5-MP+svc-erets result\endcsname {8/11M}
\expandafter\newcommand \csname ODROID N2+ (big)-MP+svc-erets result\endcsname {3.9K/17M}
\expandafter\newcommand \csname RPi 3B+-MP+svcs result\endcsname {31/1M}
\expandafter\newcommand \csname RPi 4B-MP+svcs result\endcsname {0/19M}
\expandafter\newcommand \csname RPi 5-MP+svcs result\endcsname {20/11M}
\expandafter\newcommand \csname ODROID N2+ (big)-MP+svcs result\endcsname {8.5K/17M}
\expandafter\newcommand \csname RPi 3B+-MP.EL1+dmb+eret result\endcsname {0/1M}
\expandafter\newcommand \csname RPi 4B-MP.EL1+dmb+eret result\endcsname {0/3M}
\expandafter\newcommand \csname RPi 5-MP.EL1+dmb+eret result\endcsname {0/5.5M}
\expandafter\newcommand \csname ODROID N2+ (big)-MP.EL1+dmb+eret result\endcsname {0/4M}
\expandafter\newcommand \csname RPi 3B+-MP.EL1+dmb+eret-svc result\endcsname {0/1M}
\expandafter\newcommand \csname RPi 4B-MP.EL1+dmb+eret-svc result\endcsname {0/3M}
\expandafter\newcommand \csname RPi 5-MP.EL1+dmb+eret-svc result\endcsname {0/5.5M}
\expandafter\newcommand \csname ODROID N2+ (big)-MP.EL1+dmb+eret-svc result\endcsname {0/4M}
\expandafter\newcommand \csname RPi 3B+-MP.EL1+dmb+svc-eret result\endcsname {0/1M}
\expandafter\newcommand \csname RPi 4B-MP.EL1+dmb+svc-eret result\endcsname {0/3M}
\expandafter\newcommand \csname RPi 5-MP.EL1+dmb+svc-eret result\endcsname {0/5.5M}
\expandafter\newcommand \csname ODROID N2+ (big)-MP.EL1+dmb+svc-eret result\endcsname {0/4M}
\expandafter\newcommand \csname RPi 3B+-S+erets result\endcsname {0/1M}
\expandafter\newcommand \csname RPi 4B-S+erets result\endcsname {0/19M}
\expandafter\newcommand \csname RPi 5-S+erets result\endcsname {0/11M}
\expandafter\newcommand \csname ODROID N2+ (big)-S+erets result\endcsname {0/17M}
\expandafter\newcommand \csname RPi 3B+-S+svc-dmb-erets result\endcsname {0/1M}
\expandafter\newcommand \csname RPi 4B-S+svc-dmb-erets result\endcsname {0/19M}
\expandafter\newcommand \csname RPi 5-S+svc-dmb-erets result\endcsname {0/11M}
\expandafter\newcommand \csname ODROID N2+ (big)-S+svc-dmb-erets result\endcsname {0/17M}
\expandafter\newcommand \csname RPi 3B+-S+svc-erets result\endcsname {0/1M}
\expandafter\newcommand \csname RPi 4B-S+svc-erets result\endcsname {0/19M}
\expandafter\newcommand \csname RPi 5-S+svc-erets result\endcsname {0/11M}
\expandafter\newcommand \csname ODROID N2+ (big)-S+svc-erets result\endcsname {0/17M}
\expandafter\newcommand \csname RPi 3B+-S+svcs result\endcsname {0/1M}
\expandafter\newcommand \csname RPi 4B-S+svcs result\endcsname {0/19M}
\expandafter\newcommand \csname RPi 5-S+svcs result\endcsname {0/11M}
\expandafter\newcommand \csname ODROID N2+ (big)-S+svcs result\endcsname {0/17M}
\expandafter\newcommand \csname RPi 3B+-SB+dmb+rfi-ctrl-eret result\endcsname {10K/1M}
\expandafter\newcommand \csname RPi 4B-SB+dmb+rfi-ctrl-eret result\endcsname {42K/19M}
\expandafter\newcommand \csname RPi 5-SB+dmb+rfi-ctrl-eret result\endcsname {46/11M}
\expandafter\newcommand \csname ODROID N2+ (big)-SB+dmb+rfi-ctrl-eret result\endcsname {17K/17M}
\expandafter\newcommand \csname RPi 3B+-SB+dmb+rfi-ctrl-svc result\endcsname {15K/1M}
\expandafter\newcommand \csname RPi 4B-SB+dmb+rfi-ctrl-svc result\endcsname {2.1K/18M}
\expandafter\newcommand \csname RPi 5-SB+dmb+rfi-ctrl-svc result\endcsname {58K/11M}
\expandafter\newcommand \csname ODROID N2+ (big)-SB+dmb+rfi-ctrl-svc result\endcsname {13K/17M}
\expandafter\newcommand \csname RPi 3B+-SB+svc-dmb-erets result\endcsname {0/1M}
\expandafter\newcommand \csname RPi 4B-SB+svc-dmb-erets result\endcsname {0/18M}
\expandafter\newcommand \csname RPi 5-SB+svc-dmb-erets result\endcsname {0/11M}
\expandafter\newcommand \csname ODROID N2+ (big)-SB+svc-dmb-erets result\endcsname {0/16M}
\expandafter\newcommand \csname RPi 3B+-SB+svc-erets result\endcsname {22K/1M}
\expandafter\newcommand \csname RPi 4B-SB+svc-erets result\endcsname {7.8K/18M}
\expandafter\newcommand \csname RPi 5-SB+svc-erets result\endcsname {38/11M}
\expandafter\newcommand \csname ODROID N2+ (big)-SB+svc-erets result\endcsname {9.9K/16M}
\expandafter\newcommand \csname RPi 3B+-SB.EL1+erets result\endcsname {25K/1M}
\expandafter\newcommand \csname RPi 4B-SB.EL1+erets result\endcsname {148/2.5M}
\expandafter\newcommand \csname RPi 5-SB.EL1+erets result\endcsname {43/5M}
\expandafter\newcommand \csname ODROID N2+ (big)-SB.EL1+erets result\endcsname {15K/3.5M}
\expandafter\newcommand \csname RPi 3B+-SB.EL1+svc-erets result\endcsname {19K/1M}
\expandafter\newcommand \csname RPi 4B-SB.EL1+svc-erets result\endcsname {1.7K/2.5M}
\expandafter\newcommand \csname RPi 5-SB.EL1+svc-erets result\endcsname {17/5M}
\expandafter\newcommand \csname ODROID N2+ (big)-SB.EL1+svc-erets result\endcsname {3.6K/3.5M}
\expandafter\newcommand \csname RPi 3B+-SEA_R_detect result\endcsname {0/1M}
\expandafter\newcommand \csname RPi 4B-SEA_R_detect result\endcsname {0/18M}
\expandafter\newcommand \csname RPi 5-SEA_R_detect result\endcsname {0/10M}
\expandafter\newcommand \csname ODROID N2+ (big)-SEA_R_detect result\endcsname {0/16M}
\expandafter\newcommand \csname RPi 3B+-SEA_W_detect result\endcsname {0/1M}
\expandafter\newcommand \csname RPi 4B-SEA_W_detect result\endcsname {0/18M}
\expandafter\newcommand \csname RPi 5-SEA_W_detect result\endcsname {0/10M}
\expandafter\newcommand \csname ODROID N2+ (big)-SEA_W_detect result\endcsname {0/16M}

\expandafter\newcommand \csname ref_none-LB+svc-dmb-erets ref\endcsname {\forbid{}}
\expandafter\newcommand \csname ref_exs-LB+svc-dmb-erets ref\endcsname {\forbid{}}
\expandafter\newcommand \csname ref_sea_r-LB+svc-dmb-erets ref\endcsname {\forbid{}}
\expandafter\newcommand \csname ref_sea_w-LB+svc-dmb-erets ref\endcsname {\forbid{}}
\expandafter\newcommand \csname ref_sea_rw-LB+svc-dmb-erets ref\endcsname {\forbid{}}
\expandafter\newcommand \csname ref_none-LB+svc-erets ref\endcsname {\allow{}}
\expandafter\newcommand \csname ref_exs-LB+svc-erets ref\endcsname {\allow{}}
\expandafter\newcommand \csname ref_sea_r-LB+svc-erets ref\endcsname {\forbid{}}
\expandafter\newcommand \csname ref_sea_w-LB+svc-erets ref\endcsname {\allow{}}
\expandafter\newcommand \csname ref_sea_rw-LB+svc-erets ref\endcsname {\forbid{}}
\expandafter\newcommand \csname ref_none-LB+svcs ref\endcsname {\allow{}}
\expandafter\newcommand \csname ref_exs-LB+svcs ref\endcsname {\allow{}}
\expandafter\newcommand \csname ref_sea_r-LB+svcs ref\endcsname {\forbid{}}
\expandafter\newcommand \csname ref_sea_w-LB+svcs ref\endcsname {\allow{}}
\expandafter\newcommand \csname ref_sea_rw-LB+svcs ref\endcsname {\forbid{}}
\expandafter\newcommand \csname ref_none-MP+daifset+dmb ref\endcsname {\allow{}}
\expandafter\newcommand \csname ref_exs-MP+daifset+dmb ref\endcsname {\allow{}}
\expandafter\newcommand \csname ref_sea_r-MP+daifset+dmb ref\endcsname {\allow{}}
\expandafter\newcommand \csname ref_sea_w-MP+daifset+dmb ref\endcsname {\forbid{}}
\expandafter\newcommand \csname ref_sea_rw-MP+daifset+dmb ref\endcsname {\forbid{}}
\expandafter\newcommand \csname ref_none-MP+dmb+ctrl-eret ref\endcsname {\forbid{}}
\expandafter\newcommand \csname ref_exs-MP+dmb+ctrl-eret ref\endcsname {\allow{}}
\expandafter\newcommand \csname ref_sea_r-MP+dmb+ctrl-eret ref\endcsname {\forbid{}}
\expandafter\newcommand \csname ref_sea_w-MP+dmb+ctrl-eret ref\endcsname {\forbid{}}
\expandafter\newcommand \csname ref_sea_rw-MP+dmb+ctrl-eret ref\endcsname {\forbid{}}
\expandafter\newcommand \csname ref_none-MP+dmb+ctrl-rfisvceret-addr ref\endcsname {\forbid{}}
\expandafter\newcommand \csname ref_exs-MP+dmb+ctrl-rfisvceret-addr ref\endcsname {\allow{}}
\expandafter\newcommand \csname ref_sea_r-MP+dmb+ctrl-rfisvceret-addr ref\endcsname {\forbid{}}
\expandafter\newcommand \csname ref_sea_w-MP+dmb+ctrl-rfisvceret-addr ref\endcsname {\forbid{}}
\expandafter\newcommand \csname ref_sea_rw-MP+dmb+ctrl-rfisvceret-addr ref\endcsname {\forbid{}}
\expandafter\newcommand \csname ref_none-MP+dmb+ctrl-svc ref\endcsname {\forbid{}}
\expandafter\newcommand \csname ref_exs-MP+dmb+ctrl-svc ref\endcsname {\allow{}}
\expandafter\newcommand \csname ref_sea_r-MP+dmb+ctrl-svc ref\endcsname {\forbid{}}
\expandafter\newcommand \csname ref_sea_w-MP+dmb+ctrl-svc ref\endcsname {\forbid{}}
\expandafter\newcommand \csname ref_sea_rw-MP+dmb+ctrl-svc ref\endcsname {\forbid{}}
\expandafter\newcommand \csname ref_none-MP+dmb+ctrlelr ref\endcsname {\forbid{}}
\expandafter\newcommand \csname ref_exs-MP+dmb+ctrlelr ref\endcsname {\allow{}}
\expandafter\newcommand \csname ref_sea_r-MP+dmb+ctrlelr ref\endcsname {\forbid{}}
\expandafter\newcommand \csname ref_sea_w-MP+dmb+ctrlelr ref\endcsname {\forbid{}}
\expandafter\newcommand \csname ref_sea_rw-MP+dmb+ctrlelr ref\endcsname {\forbid{}}
\expandafter\newcommand \csname ref_none-MP+dmb+daifset ref\endcsname {\allow{}}
\expandafter\newcommand \csname ref_exs-MP+dmb+daifset ref\endcsname {\allow{}}
\expandafter\newcommand \csname ref_sea_r-MP+dmb+daifset ref\endcsname {\allow{}}
\expandafter\newcommand \csname ref_sea_w-MP+dmb+daifset ref\endcsname {\allow{}}
\expandafter\newcommand \csname ref_sea_rw-MP+dmb+daifset ref\endcsname {\allow{}}
\expandafter\newcommand \csname ref_none-MP+dmb+dmb-eret ref\endcsname {\forbid{}}
\expandafter\newcommand \csname ref_exs-MP+dmb+dmb-eret ref\endcsname {\forbid{}}
\expandafter\newcommand \csname ref_sea_r-MP+dmb+dmb-eret ref\endcsname {\forbid{}}
\expandafter\newcommand \csname ref_sea_w-MP+dmb+dmb-eret ref\endcsname {\forbid{}}
\expandafter\newcommand \csname ref_sea_rw-MP+dmb+dmb-eret ref\endcsname {\forbid{}}
\expandafter\newcommand \csname ref_none-MP+dmb+eret-dmb ref\endcsname {\forbid{}}
\expandafter\newcommand \csname ref_exs-MP+dmb+eret-dmb ref\endcsname {\forbid{}}
\expandafter\newcommand \csname ref_sea_r-MP+dmb+eret-dmb ref\endcsname {\forbid{}}
\expandafter\newcommand \csname ref_sea_w-MP+dmb+eret-dmb ref\endcsname {\forbid{}}
\expandafter\newcommand \csname ref_sea_rw-MP+dmb+eret-dmb ref\endcsname {\forbid{}}
\expandafter\newcommand \csname ref_none-MP+dmb+eret-svc ref\endcsname {\allow{}}
\expandafter\newcommand \csname ref_exs-MP+dmb+eret-svc ref\endcsname {\allow{}}
\expandafter\newcommand \csname ref_sea_r-MP+dmb+eret-svc ref\endcsname {\forbid{}}
\expandafter\newcommand \csname ref_sea_w-MP+dmb+eret-svc ref\endcsname {\allow{}}
\expandafter\newcommand \csname ref_sea_rw-MP+dmb+eret-svc ref\endcsname {\forbid{}}
\expandafter\newcommand \csname ref_none-MP+dmb+eret ref\endcsname {\allow{}}
\expandafter\newcommand \csname ref_exs-MP+dmb+eret ref\endcsname {\allow{}}
\expandafter\newcommand \csname ref_sea_r-MP+dmb+eret ref\endcsname {\forbid{}}
\expandafter\newcommand \csname ref_sea_w-MP+dmb+eret ref\endcsname {\allow{}}
\expandafter\newcommand \csname ref_sea_rw-MP+dmb+eret ref\endcsname {\forbid{}}
\expandafter\newcommand \csname ref_none-MP+dmb+eret=addr ref\endcsname {\forbid{}}
\expandafter\newcommand \csname ref_exs-MP+dmb+eret=addr ref\endcsname {\forbid{}}
\expandafter\newcommand \csname ref_sea_r-MP+dmb+eret=addr ref\endcsname {\forbid{}}
\expandafter\newcommand \csname ref_sea_w-MP+dmb+eret=addr ref\endcsname {\forbid{}}
\expandafter\newcommand \csname ref_sea_rw-MP+dmb+eret=addr ref\endcsname {\forbid{}}
\expandafter\newcommand \csname ref_none-MP+dmb+svc-addreret ref\endcsname {\allow{}}
\expandafter\newcommand \csname ref_exs-MP+dmb+svc-addreret ref\endcsname {\allow{}}
\expandafter\newcommand \csname ref_sea_r-MP+dmb+svc-addreret ref\endcsname {\forbid{}}
\expandafter\newcommand \csname ref_sea_w-MP+dmb+svc-addreret ref\endcsname {\allow{}}
\expandafter\newcommand \csname ref_sea_rw-MP+dmb+svc-addreret ref\endcsname {\forbid{}}
\expandafter\newcommand \csname ref_none-MP+dmb+svc-dmb-eret ref\endcsname {\forbid{}}
\expandafter\newcommand \csname ref_exs-MP+dmb+svc-dmb-eret ref\endcsname {\forbid{}}
\expandafter\newcommand \csname ref_sea_r-MP+dmb+svc-dmb-eret ref\endcsname {\forbid{}}
\expandafter\newcommand \csname ref_sea_w-MP+dmb+svc-dmb-eret ref\endcsname {\forbid{}}
\expandafter\newcommand \csname ref_sea_rw-MP+dmb+svc-dmb-eret ref\endcsname {\forbid{}}
\expandafter\newcommand \csname ref_none-MP+dmb+svc-dmb ref\endcsname {\forbid{}}
\expandafter\newcommand \csname ref_exs-MP+dmb+svc-dmb ref\endcsname {\forbid{}}
\expandafter\newcommand \csname ref_sea_r-MP+dmb+svc-dmb ref\endcsname {\forbid{}}
\expandafter\newcommand \csname ref_sea_w-MP+dmb+svc-dmb ref\endcsname {\forbid{}}
\expandafter\newcommand \csname ref_sea_rw-MP+dmb+svc-dmb ref\endcsname {\forbid{}}
\expandafter\newcommand \csname ref_none-MP+dmb+svc-eret ref\endcsname {\allow{}}
\expandafter\newcommand \csname ref_exs-MP+dmb+svc-eret ref\endcsname {\allow{}}
\expandafter\newcommand \csname ref_sea_r-MP+dmb+svc-eret ref\endcsname {\forbid{}}
\expandafter\newcommand \csname ref_sea_w-MP+dmb+svc-eret ref\endcsname {\allow{}}
\expandafter\newcommand \csname ref_sea_rw-MP+dmb+svc-eret ref\endcsname {\forbid{}}
\expandafter\newcommand \csname ref_none-MP+dmb+svc ref\endcsname {\allow{}}
\expandafter\newcommand \csname ref_exs-MP+dmb+svc ref\endcsname {\allow{}}
\expandafter\newcommand \csname ref_sea_r-MP+dmb+svc ref\endcsname {\forbid{}}
\expandafter\newcommand \csname ref_sea_w-MP+dmb+svc ref\endcsname {\allow{}}
\expandafter\newcommand \csname ref_sea_rw-MP+dmb+svc ref\endcsname {\forbid{}}
\expandafter\newcommand \csname ref_none-MP+dmb+svcnoeis ref\endcsname {\allow{}}
\expandafter\newcommand \csname ref_exs-MP+dmb+svcnoeis ref\endcsname {\allow{}}
\expandafter\newcommand \csname ref_sea_r-MP+dmb+svcnoeis ref\endcsname {\forbid{}}
\expandafter\newcommand \csname ref_sea_w-MP+dmb+svcnoeis ref\endcsname {\allow{}}
\expandafter\newcommand \csname ref_sea_rw-MP+dmb+svcnoeis ref\endcsname {\forbid{}}
\expandafter\newcommand \csname ref_none-MP+eret+addr ref\endcsname {\allow{}}
\expandafter\newcommand \csname ref_exs-MP+eret+addr ref\endcsname {\allow{}}
\expandafter\newcommand \csname ref_sea_r-MP+eret+addr ref\endcsname {\allow{}}
\expandafter\newcommand \csname ref_sea_w-MP+eret+addr ref\endcsname {\forbid{}}
\expandafter\newcommand \csname ref_sea_rw-MP+eret+addr ref\endcsname {\forbid{}}
\expandafter\newcommand \csname ref_none-MP+eret+dmb ref\endcsname {\allow{}}
\expandafter\newcommand \csname ref_exs-MP+eret+dmb ref\endcsname {\allow{}}
\expandafter\newcommand \csname ref_sea_r-MP+eret+dmb ref\endcsname {\allow{}}
\expandafter\newcommand \csname ref_sea_w-MP+eret+dmb ref\endcsname {\forbid{}}
\expandafter\newcommand \csname ref_sea_rw-MP+eret+dmb ref\endcsname {\forbid{}}
\expandafter\newcommand \csname ref_none-MP+eret+svc ref\endcsname {\allow{}}
\expandafter\newcommand \csname ref_exs-MP+eret+svc ref\endcsname {\allow{}}
\expandafter\newcommand \csname ref_sea_r-MP+eret+svc ref\endcsname {\allow{}}
\expandafter\newcommand \csname ref_sea_w-MP+eret+svc ref\endcsname {\allow{}}
\expandafter\newcommand \csname ref_sea_rw-MP+eret+svc ref\endcsname {\forbid{}}
\expandafter\newcommand \csname ref_none-MP+erets ref\endcsname {\allow{}}
\expandafter\newcommand \csname ref_exs-MP+erets ref\endcsname {\allow{}}
\expandafter\newcommand \csname ref_sea_r-MP+erets ref\endcsname {\allow{}}
\expandafter\newcommand \csname ref_sea_w-MP+erets ref\endcsname {\allow{}}
\expandafter\newcommand \csname ref_sea_rw-MP+erets ref\endcsname {\forbid{}}
\expandafter\newcommand \csname ref_none-MP+svc+addr ref\endcsname {\allow{}}
\expandafter\newcommand \csname ref_exs-MP+svc+addr ref\endcsname {\allow{}}
\expandafter\newcommand \csname ref_sea_r-MP+svc+addr ref\endcsname {\allow{}}
\expandafter\newcommand \csname ref_sea_w-MP+svc+addr ref\endcsname {\forbid{}}
\expandafter\newcommand \csname ref_sea_rw-MP+svc+addr ref\endcsname {\forbid{}}
\expandafter\newcommand \csname ref_none-MP+svc+dmb ref\endcsname {\allow{}}
\expandafter\newcommand \csname ref_exs-MP+svc+dmb ref\endcsname {\allow{}}
\expandafter\newcommand \csname ref_sea_r-MP+svc+dmb ref\endcsname {\allow{}}
\expandafter\newcommand \csname ref_sea_w-MP+svc+dmb ref\endcsname {\forbid{}}
\expandafter\newcommand \csname ref_sea_rw-MP+svc+dmb ref\endcsname {\forbid{}}
\expandafter\newcommand \csname ref_none-MP+svc+eret ref\endcsname {\allow{}}
\expandafter\newcommand \csname ref_exs-MP+svc+eret ref\endcsname {\allow{}}
\expandafter\newcommand \csname ref_sea_r-MP+svc+eret ref\endcsname {\allow{}}
\expandafter\newcommand \csname ref_sea_w-MP+svc+eret ref\endcsname {\allow{}}
\expandafter\newcommand \csname ref_sea_rw-MP+svc+eret ref\endcsname {\forbid{}}
\expandafter\newcommand \csname ref_none-MP+svc-dmb+addr ref\endcsname {\forbid{}}
\expandafter\newcommand \csname ref_exs-MP+svc-dmb+addr ref\endcsname {\forbid{}}
\expandafter\newcommand \csname ref_sea_r-MP+svc-dmb+addr ref\endcsname {\forbid{}}
\expandafter\newcommand \csname ref_sea_w-MP+svc-dmb+addr ref\endcsname {\forbid{}}
\expandafter\newcommand \csname ref_sea_rw-MP+svc-dmb+addr ref\endcsname {\forbid{}}
\expandafter\newcommand \csname ref_none-MP+svc-dmb-eret+addr ref\endcsname {\forbid{}}
\expandafter\newcommand \csname ref_exs-MP+svc-dmb-eret+addr ref\endcsname {\forbid{}}
\expandafter\newcommand \csname ref_sea_r-MP+svc-dmb-eret+addr ref\endcsname {\forbid{}}
\expandafter\newcommand \csname ref_sea_w-MP+svc-dmb-eret+addr ref\endcsname {\forbid{}}
\expandafter\newcommand \csname ref_sea_rw-MP+svc-dmb-eret+addr ref\endcsname {\forbid{}}
\expandafter\newcommand \csname ref_none-MP+svc-eret+addr ref\endcsname {\allow{}}
\expandafter\newcommand \csname ref_exs-MP+svc-eret+addr ref\endcsname {\allow{}}
\expandafter\newcommand \csname ref_sea_r-MP+svc-eret+addr ref\endcsname {\allow{}}
\expandafter\newcommand \csname ref_sea_w-MP+svc-eret+addr ref\endcsname {\forbid{}}
\expandafter\newcommand \csname ref_sea_rw-MP+svc-eret+addr ref\endcsname {\forbid{}}
\expandafter\newcommand \csname ref_none-MP+svc-erets ref\endcsname {\allow{}}
\expandafter\newcommand \csname ref_exs-MP+svc-erets ref\endcsname {\allow{}}
\expandafter\newcommand \csname ref_sea_r-MP+svc-erets ref\endcsname {\allow{}}
\expandafter\newcommand \csname ref_sea_w-MP+svc-erets ref\endcsname {\allow{}}
\expandafter\newcommand \csname ref_sea_rw-MP+svc-erets ref\endcsname {\forbid{}}
\expandafter\newcommand \csname ref_none-MP+svcs ref\endcsname {\allow{}}
\expandafter\newcommand \csname ref_exs-MP+svcs ref\endcsname {\allow{}}
\expandafter\newcommand \csname ref_sea_r-MP+svcs ref\endcsname {\allow{}}
\expandafter\newcommand \csname ref_sea_w-MP+svcs ref\endcsname {\allow{}}
\expandafter\newcommand \csname ref_sea_rw-MP+svcs ref\endcsname {\forbid{}}
\expandafter\newcommand \csname ref_none-MP.EL1+dmb+ctrlvbarsvc ref\endcsname {\forbid{}}
\expandafter\newcommand \csname ref_exs-MP.EL1+dmb+ctrlvbarsvc ref\endcsname {\allow{}}
\expandafter\newcommand \csname ref_sea_r-MP.EL1+dmb+ctrlvbarsvc ref\endcsname {\forbid{}}
\expandafter\newcommand \csname ref_sea_w-MP.EL1+dmb+ctrlvbarsvc ref\endcsname {\forbid{}}
\expandafter\newcommand \csname ref_sea_rw-MP.EL1+dmb+ctrlvbarsvc ref\endcsname {\forbid{}}
\expandafter\newcommand \csname ref_none-MP.EL1+dmb+eret-svc ref\endcsname {\allow{}}
\expandafter\newcommand \csname ref_exs-MP.EL1+dmb+eret-svc ref\endcsname {\allow{}}
\expandafter\newcommand \csname ref_sea_r-MP.EL1+dmb+eret-svc ref\endcsname {\forbid{}}
\expandafter\newcommand \csname ref_sea_w-MP.EL1+dmb+eret-svc ref\endcsname {\allow{}}
\expandafter\newcommand \csname ref_sea_rw-MP.EL1+dmb+eret-svc ref\endcsname {\forbid{}}
\expandafter\newcommand \csname ref_none-MP.EL1+dmb+eret ref\endcsname {\allow{}}
\expandafter\newcommand \csname ref_exs-MP.EL1+dmb+eret ref\endcsname {\allow{}}
\expandafter\newcommand \csname ref_sea_r-MP.EL1+dmb+eret ref\endcsname {\forbid{}}
\expandafter\newcommand \csname ref_sea_w-MP.EL1+dmb+eret ref\endcsname {\allow{}}
\expandafter\newcommand \csname ref_sea_rw-MP.EL1+dmb+eret ref\endcsname {\forbid{}}
\expandafter\newcommand \csname ref_none-MP.EL1+dmb+svc-eret ref\endcsname {\allow{}}
\expandafter\newcommand \csname ref_exs-MP.EL1+dmb+svc-eret ref\endcsname {\allow{}}
\expandafter\newcommand \csname ref_sea_r-MP.EL1+dmb+svc-eret ref\endcsname {\forbid{}}
\expandafter\newcommand \csname ref_sea_w-MP.EL1+dmb+svc-eret ref\endcsname {\allow{}}
\expandafter\newcommand \csname ref_sea_rw-MP.EL1+dmb+svc-eret ref\endcsname {\forbid{}}
\expandafter\newcommand \csname ref_none-MP.EL1+dmb+svc ref\endcsname {\forbid{}}
\expandafter\newcommand \csname ref_exs-MP.EL1+dmb+svc ref\endcsname {\forbid{}}
\expandafter\newcommand \csname ref_sea_r-MP.EL1+dmb+svc ref\endcsname {\forbid{}}
\expandafter\newcommand \csname ref_sea_w-MP.EL1+dmb+svc ref\endcsname {\forbid{}}
\expandafter\newcommand \csname ref_sea_rw-MP.EL1+dmb+svc ref\endcsname {\forbid{}}
\expandafter\newcommand \csname ref_none-S+dmb+eret ref\endcsname {\allow{}}
\expandafter\newcommand \csname ref_exs-S+dmb+eret ref\endcsname {\allow{}}
\expandafter\newcommand \csname ref_sea_r-S+dmb+eret ref\endcsname {\forbid{}}
\expandafter\newcommand \csname ref_sea_w-S+dmb+eret ref\endcsname {\allow{}}
\expandafter\newcommand \csname ref_sea_rw-S+dmb+eret ref\endcsname {\forbid{}}
\expandafter\newcommand \csname ref_none-S+dmb+svc ref\endcsname {\allow{}}
\expandafter\newcommand \csname ref_exs-S+dmb+svc ref\endcsname {\allow{}}
\expandafter\newcommand \csname ref_sea_r-S+dmb+svc ref\endcsname {\forbid{}}
\expandafter\newcommand \csname ref_sea_w-S+dmb+svc ref\endcsname {\allow{}}
\expandafter\newcommand \csname ref_sea_rw-S+dmb+svc ref\endcsname {\forbid{}}
\expandafter\newcommand \csname ref_none-S+erets ref\endcsname {\allow{}}
\expandafter\newcommand \csname ref_exs-S+erets ref\endcsname {\allow{}}
\expandafter\newcommand \csname ref_sea_r-S+erets ref\endcsname {\allow{}}
\expandafter\newcommand \csname ref_sea_w-S+erets ref\endcsname {\allow{}}
\expandafter\newcommand \csname ref_sea_rw-S+erets ref\endcsname {\forbid{}}
\expandafter\newcommand \csname ref_none-S+svc-dmb-erets ref\endcsname {\forbid{}}
\expandafter\newcommand \csname ref_exs-S+svc-dmb-erets ref\endcsname {\forbid{}}
\expandafter\newcommand \csname ref_sea_r-S+svc-dmb-erets ref\endcsname {\forbid{}}
\expandafter\newcommand \csname ref_sea_w-S+svc-dmb-erets ref\endcsname {\forbid{}}
\expandafter\newcommand \csname ref_sea_rw-S+svc-dmb-erets ref\endcsname {\forbid{}}
\expandafter\newcommand \csname ref_none-S+svc-erets ref\endcsname {\allow{}}
\expandafter\newcommand \csname ref_exs-S+svc-erets ref\endcsname {\allow{}}
\expandafter\newcommand \csname ref_sea_r-S+svc-erets ref\endcsname {\allow{}}
\expandafter\newcommand \csname ref_sea_w-S+svc-erets ref\endcsname {\allow{}}
\expandafter\newcommand \csname ref_sea_rw-S+svc-erets ref\endcsname {\forbid{}}
\expandafter\newcommand \csname ref_none-S+svcs ref\endcsname {\allow{}}
\expandafter\newcommand \csname ref_exs-S+svcs ref\endcsname {\allow{}}
\expandafter\newcommand \csname ref_sea_r-S+svcs ref\endcsname {\allow{}}
\expandafter\newcommand \csname ref_sea_w-S+svcs ref\endcsname {\allow{}}
\expandafter\newcommand \csname ref_sea_rw-S+svcs ref\endcsname {\forbid{}}
\expandafter\newcommand \csname ref_none-SB+daifsets ref\endcsname {\allow{}}
\expandafter\newcommand \csname ref_exs-SB+daifsets ref\endcsname {\allow{}}
\expandafter\newcommand \csname ref_sea_r-SB+daifsets ref\endcsname {\allow{}}
\expandafter\newcommand \csname ref_sea_w-SB+daifsets ref\endcsname {\allow{}}
\expandafter\newcommand \csname ref_sea_rw-SB+daifsets ref\endcsname {\allow{}}
\expandafter\newcommand \csname ref_none-SB+dmb+eret ref\endcsname {\allow{}}
\expandafter\newcommand \csname ref_exs-SB+dmb+eret ref\endcsname {\allow{}}
\expandafter\newcommand \csname ref_sea_r-SB+dmb+eret ref\endcsname {\allow{}}
\expandafter\newcommand \csname ref_sea_w-SB+dmb+eret ref\endcsname {\forbid{}}
\expandafter\newcommand \csname ref_sea_rw-SB+dmb+eret ref\endcsname {\forbid{}}
\expandafter\newcommand \csname ref_none-SB+dmb+rfi-ctrl-eret ref\endcsname {\allow{}}
\expandafter\newcommand \csname ref_exs-SB+dmb+rfi-ctrl-eret ref\endcsname {\allow{}}
\expandafter\newcommand \csname ref_sea_r-SB+dmb+rfi-ctrl-eret ref\endcsname {\allow{}}
\expandafter\newcommand \csname ref_sea_w-SB+dmb+rfi-ctrl-eret ref\endcsname {\forbid{}}
\expandafter\newcommand \csname ref_sea_rw-SB+dmb+rfi-ctrl-eret ref\endcsname {\forbid{}}
\expandafter\newcommand \csname ref_none-SB+dmb+rfi-ctrl-svc ref\endcsname {\allow{}}
\expandafter\newcommand \csname ref_exs-SB+dmb+rfi-ctrl-svc ref\endcsname {\allow{}}
\expandafter\newcommand \csname ref_sea_r-SB+dmb+rfi-ctrl-svc ref\endcsname {\allow{}}
\expandafter\newcommand \csname ref_sea_w-SB+dmb+rfi-ctrl-svc ref\endcsname {\forbid{}}
\expandafter\newcommand \csname ref_sea_rw-SB+dmb+rfi-ctrl-svc ref\endcsname {\forbid{}}
\expandafter\newcommand \csname ref_none-SB+dmb+rfieret-addr ref\endcsname {\allow{}}
\expandafter\newcommand \csname ref_exs-SB+dmb+rfieret-addr ref\endcsname {\allow{}}
\expandafter\newcommand \csname ref_sea_r-SB+dmb+rfieret-addr ref\endcsname {\allow{}}
\expandafter\newcommand \csname ref_sea_w-SB+dmb+rfieret-addr ref\endcsname {\forbid{}}
\expandafter\newcommand \csname ref_sea_rw-SB+dmb+rfieret-addr ref\endcsname {\forbid{}}
\expandafter\newcommand \csname ref_none-SB+dmb+rfisvc-addr ref\endcsname {\allow{}}
\expandafter\newcommand \csname ref_exs-SB+dmb+rfisvc-addr ref\endcsname {\allow{}}
\expandafter\newcommand \csname ref_sea_r-SB+dmb+rfisvc-addr ref\endcsname {\allow{}}
\expandafter\newcommand \csname ref_sea_w-SB+dmb+rfisvc-addr ref\endcsname {\forbid{}}
\expandafter\newcommand \csname ref_sea_rw-SB+dmb+rfisvc-addr ref\endcsname {\forbid{}}
\expandafter\newcommand \csname ref_none-SB+dmb+svc ref\endcsname {\allow{}}
\expandafter\newcommand \csname ref_exs-SB+dmb+svc ref\endcsname {\allow{}}
\expandafter\newcommand \csname ref_sea_r-SB+dmb+svc ref\endcsname {\allow{}}
\expandafter\newcommand \csname ref_sea_w-SB+dmb+svc ref\endcsname {\forbid{}}
\expandafter\newcommand \csname ref_sea_rw-SB+dmb+svc ref\endcsname {\forbid{}}
\expandafter\newcommand \csname ref_none-SB+svc+dmb-erets ref\endcsname {\forbid{}}
\expandafter\newcommand \csname ref_exs-SB+svc+dmb-erets ref\endcsname {\forbid{}}
\expandafter\newcommand \csname ref_sea_r-SB+svc+dmb-erets ref\endcsname {\forbid{}}
\expandafter\newcommand \csname ref_sea_w-SB+svc+dmb-erets ref\endcsname {\forbid{}}
\expandafter\newcommand \csname ref_sea_rw-SB+svc+dmb-erets ref\endcsname {\forbid{}}
\expandafter\newcommand \csname ref_none-SB+svcs ref\endcsname {\allow{}}
\expandafter\newcommand \csname ref_exs-SB+svcs ref\endcsname {\allow{}}
\expandafter\newcommand \csname ref_sea_r-SB+svcs ref\endcsname {\allow{}}
\expandafter\newcommand \csname ref_sea_w-SB+svcs ref\endcsname {\forbid{}}
\expandafter\newcommand \csname ref_sea_rw-SB+svcs ref\endcsname {\forbid{}}
\expandafter\newcommand \csname ref_none-SB.EL1+erets ref\endcsname {\allow{}}
\expandafter\newcommand \csname ref_exs-SB.EL1+erets ref\endcsname {\allow{}}
\expandafter\newcommand \csname ref_sea_r-SB.EL1+erets ref\endcsname {\allow{}}
\expandafter\newcommand \csname ref_sea_w-SB.EL1+erets ref\endcsname {\forbid{}}
\expandafter\newcommand \csname ref_sea_rw-SB.EL1+erets ref\endcsname {\forbid{}}
\expandafter\newcommand \csname ref_none-SB.EL1+svc-erets ref\endcsname {\allow{}}
\expandafter\newcommand \csname ref_exs-SB.EL1+svc-erets ref\endcsname {\allow{}}
\expandafter\newcommand \csname ref_sea_r-SB.EL1+svc-erets ref\endcsname {\allow{}}
\expandafter\newcommand \csname ref_sea_w-SB.EL1+svc-erets ref\endcsname {\forbid{}}
\expandafter\newcommand \csname ref_sea_rw-SB.EL1+svc-erets ref\endcsname {\forbid{}}
\expandafter\newcommand \csname ref_none-MP+dmb+ctrl-int ref\endcsname {\forbid{}}
\expandafter\newcommand \csname ref_exs-MP+dmb+ctrl-int ref\endcsname {\forbid{}}
\expandafter\newcommand \csname ref_sea_r-MP+dmb+ctrl-int ref\endcsname {\forbid{}}
\expandafter\newcommand \csname ref_sea_w-MP+dmb+ctrl-int ref\endcsname {\forbid{}}
\expandafter\newcommand \csname ref_sea_rw-MP+dmb+ctrl-int ref\endcsname {\forbid{}}
\expandafter\newcommand \csname ref_none-MP+dmb+int ref\endcsname {\allow{}}
\expandafter\newcommand \csname ref_exs-MP+dmb+int ref\endcsname {\allow{}}
\expandafter\newcommand \csname ref_sea_r-MP+dmb+int ref\endcsname {\allow{}}
\expandafter\newcommand \csname ref_sea_w-MP+dmb+int ref\endcsname {\allow{}}
\expandafter\newcommand \csname ref_sea_rw-MP+dmb+int ref\endcsname {\allow{}}
\expandafter\newcommand \csname ref_none-MP+int+dmb ref\endcsname {\allow{}}
\expandafter\newcommand \csname ref_exs-MP+int+dmb ref\endcsname {\allow{}}
\expandafter\newcommand \csname ref_sea_r-MP+int+dmb ref\endcsname {\allow{}}
\expandafter\newcommand \csname ref_sea_w-MP+int+dmb ref\endcsname {\allow{}}
\expandafter\newcommand \csname ref_sea_rw-MP+int+dmb ref\endcsname {\allow{}}

\newcommand\hwrefsName{hw-refs}
\newcommand\modelsName{param-refs}
\newcommand\hwrefs[1]{%
\small%
\begin{tabular}{r l r l}
  \multicolumn{2}{c}{\hwrefsName} & \multicolumn{2}{c}{\modelsName} \\
  RPi 3B+           &  \csname RPi 3B+-#1 result\endcsname
  & ExS & \csname ref_exs-#1 ref\endcsname \\
  RPi 4B            &  \csname RPi 4B-#1 result\endcsname
  & \SEAR & \csname ref_sea_r-#1 ref\endcsname \\
  RPi 5             &  \csname RPi 5-#1 result\endcsname
  & \SEAW & \csname ref_sea_w-#1 ref\endcsname \\
  ODROID N2+ (big)  &  \csname ODROID N2+ (big)-#1 result\endcsname
  & \SEARW & \csname ref_sea_rw-#1 ref\endcsname
\end{tabular}
}
\newcommand{\tableTestName}[2]{%
  \hyperref[test:#1]{\texttt{#2}}
}
\newcommand{\testref}[1]{%
  \hyperref[test:#1]{\texttt{#1}}%
}

\newcommand{\asm}[1]{\texttt{#1}}

\newcommand{\herd}[1]{\texttt{#1}}

\newcommand{\FEAT}[1]{{\small\texttt{FEAT\_#1}}}
\makeatletter
\newcommand{\sail}{\begingroup\makeunderletter\@sail}
\newcommand{\@sail}[1]{\texttt{#1}\endgroup}
\makeatother

\newcommand{\noemph}[1]{#1}%

\newcommand\SEA[1]{\ensuremath{\mathbf{SEA}_{#1}}}
\newcommand\SEAW{\SEA{W}}
\newcommand\SEAR{\SEA{R}}
\newcommand\SEARW{\SEA{R+W}}

\usepackage[listings]{tcolorbox}
\newtcbox{\salmonbox}{%
  on line, arc=1pt, outer arc=1pt,colframe=black,
  colback=red!30,
  boxsep=0pt,
  left=1pt,right=1pt,top=1pt,bottom=1pt,%
  boxrule=.5pt,bottomrule=.5pt,toprule=.5pt}

\newtcbox{\greybox}{%
  on line, arc=1pt, outer arc=1pt,colframe=black,
  colback=black!50,
  coltext=white,
  boxsep=0pt,
  left=1pt,right=1pt,top=1pt,bottom=1pt,%
  boxrule=.5pt,bottomrule=.5pt,toprule=.5pt}

\usetikzlibrary{litmus, litmus.listings}
\tikzset{
  node font=\sffamily,
}
\litmusset{
  use listing={style=litmus},
  AArch64,
  every initial state/.style={fill=blue!10},
  every final state allowed/.style={fill=green!15},
  every final state forbidden/.style={fill=red!15},
  table construction/.style={
    initial state box={/tikz/above=of threads box},
    table header={/tikz/above=1.5pt of initial state box, /tikz/anchor=base},
    final state box={/tikz/below=of threads box},
    hw observations box,
  },
  hw observations box/.code={},
  every simple events/.append style={
    /litmus/event label/.code={\ifstrempty{##1}{}{##1:}},
    /litmus/every event box/.append style={outer sep=2pt},
    /tikz/column sep=2em,
    /tikz/column 2/.style=events column,
  },
  every simple full/.append style={
    /litmus/event label/.code={\ifstrempty{##1}{}{##1:}},
    /litmus/every event box/.append style={outer sep=2pt},
    /tikz/column 2/.append style={
      /litmus/every event/.append style={blue},
      /litmus/every event label/.append style={blue},
    },
    /tikz/column sep=2em,
    /tikz/column 3/.style=events column,
    /tikz/row sep={5ex,between origins},
  },
    every state/.append style={
      inner xsep=1pt,
      state prefix font=\small\sffamily,
    },
    every thread header/.append style={node font=\small\sffamily},
}

\litmusset{
    every instructions/.append style={
      every assem/.append style={
        inner xsep=0.3em, %
        inner ysep=0.5pt, %
      },
    },
}

\definecolor{IslaInitialState}{rgb}{0.92,1,1}
\definecolor{Forbid}{rgb}{1,0.92,0.92}
\definecolor{Allow}{rgb}{0.92,1,0.92}
\definecolor{darkblue}{rgb}{0,0.0,0.5}
\definecolor{lightgray}{gray}{0.87}
\definecolor{aslannotatecolor}{rgb}{0.5,0,0}

\lstdefinelanguage{none}{
    identifier  style=
}

\lstdefinelanguage{cat}{
    basicstyle=\linespread{0.8}\tt\small,
    escapechar=@,
    morekeywords=[1]{acyclic, let, irreflexive, as, empty},
    keywordstyle=[1]{\bfseries},
    morekeywords=[2]{accessor, match, define, is},
    keywordstyle=[2]{\bfseries},
    morecomment=[f][\color{DarkGreen}]{//},
    morecomment=[n][\color{DarkGreen}]{(*}{*)},
    moredelim=**[is][\color{gray}]{[*}{*]},
    showstringspaces=false,
    keepspaces=true,
    breaklines=true,
    mathescape,
    literate =
        {~}{$\sim$}1,
}

\lstdefinelanguage{sail}{
    basicstyle=\ttfamily\small,
    morekeywords={
and,
alias,
as,
bitzero,
bitone,
bits,
by,
case,
clause,
configuration,
const,
dec,
default,
deinfix,
effect,
Effect,
end,
enumerate,
else,
escape,
exit,
extern,
false,
forall,
foreach,
function,
if,
in,
inc,
IN,
let,
mapping,
match,
member,
Nat,
Order,
pure,
rec,
register,
scattered,
struct,
switch,
then,
true,
Type,
typedef,
undefined,
union,
with,
val,
div,
mod,
quot,
rem,
barr,
rreg,
wreg,
rmem,
wmem,
undef,
unspec,
nondet,
bool,unit,vector,range,list,bit,nat, int, uint8,uint16,uint32,uint64,implicit
},
    keywordstyle=\bfseries,
    morecomment=[l]{//},
    commentstyle=\color{DarkGreen},
    moredelim=**[is][\colorlet{DarkGreen}{lightgray}\color{lightgray}]{[*}{*]},
    showstringspaces=true,
    keepspaces=true,
    breaklines=true,
    mathescape,
    numbers=left,
    stepnumber=1,
    escapechar=@,
}

\lstdefinelanguage{asl}{
    basicstyle=\ttfamily\small,
    morekeywords={bits, if, then, else, when, case, of, elsif, else, assert, IN, DIV, boolean, repeat, until, integer, return, otherwise},
    keywordstyle=\bfseries,
    escapechar=@,
    morecomment=[l]{//},
    commentstyle=\color{DarkGreen},
    moredelim=**[is][\colorlet{DarkGreen}{lightgray}\color{lightgray}]{[*}{*]},
    showstringspaces=true,
    keepspaces=true,
    breaklines=true,
    mathescape,
    numbers=left,
    stepnumber=1,
}

\lstdefinelanguage{smt2}{
v    basicstyle=\linespread{0.8}\ttfamily,
    keywordstyle=\bfseries,
    escapechar=@,
    morecomment=[f][\color{DarkGreen}]{//},
    morecomment=[n][\color{DarkGreen}]{(*}{*)},
    moredelim=**[is][\color{red}]{>}{<},
    moredelim=**[is][\color{gray}]{[*}{*]},
    showstringspaces=false,
    keepspaces=true,
    breaklines=true,
    mathescape
}

\lstdefinelanguage{AArch64}{
    basicstyle=\ttfamily\small,
    keywordstyle=[1]\bfseries,
    morekeywords = [1]{str, dmb, dsb, tlbi, ldr, isb, mov, mrs, msr, eret, add, cbz, and, cmp, csel, eor, orr, svc, lsr, sub, cbnz,
                    STR, STLR, DMB, DSB, TLBI, LDR, ISB, MOV, MRS, MSR, ERET, ADD, CBZ, AND, CMP, CSEL, EOR, ORR, SVC, LSR, SUB, CBNZ, LDAR},
    keywordstyle=[2]{\color{darkblue}},
    morekeywords=[2]{X0, X1, X2, X3, X4, X5, X6, X7, X8, X9, X10, X11, X12, X13, X14, X15, X16, X17, X18, X19, X20, X21, X22, X23},
    morecomment = [l]{//},
    numbers=left,
    escapechar=@,
    commentstyle=\sffamily\itshape\footnotesize,
}

\lstdefinelanguage{pseudomath}{
    literate =
        {<|}{$\langle$}1
        {|>}{$\rangle$}1,
    morecomment = [l]{//},
    keepspaces=true,
    mathescape=true,
}

\renewenvironment{quote}{%
  \list{}{%
    \leftmargin0.25cm   %
    \rightmargin\leftmargin
  }
  \item\relax
}
{\endlist}

\definecolor{quotecolor}{RGB}{236,236,246}
\newtcolorbox{quotecolorbox}{%
    colback=quotecolor, colframe=quotecolor,%
    left=2mm, right=2mm, top=1mm, bottom=1mm, boxsep=0mm
}
\NewEnviron{quotebox}{
    \begin{quote}
    \begin{quotecolorbox}
    \begin{minipage}{1.0\textwidth}
        \BODY
    \end{minipage}
    \end{quotecolorbox}
    \end{quote}
}

\newif\ifmodelifetch
\modelifetchfalse

\newif\ifmodelexc
\modelexcfalse  %
\colorlet{exc}{ibmcolourblind1}

\setcitestyle{numbers}

\begin{abstract}
To manage exceptions, software relies on a key architectural guarantee, \emph{precision}: that exceptions appear to execute between instructions.  However, this definition, dating back over 60 years, fundamentally assumes a sequential programmers model.  Modern architectures such as Arm-A with programmer-observable relaxed behaviour make such a naive definition inadequate, and it is unclear exactly what guarantees programmers have on exception entry and exit.

In this paper, we clarify the concepts needed to discuss exceptions in the relaxed-memory setting -- a key aspect of precisely specifying the architectural interface between hardware and software.  We explore the basic relaxed behaviour across exception boundaries, and the semantics of external aborts, using Arm-A as a representative modern architecture.  We identify an important problem, present yet unexplored for decades: pinning down what it means for exceptions to be precise in a relaxed setting.  We describe key phenomena that any definition should account for.  We develop an axiomatic model for Arm-A precise exceptions, tooling for axiomatic model execution, and a library of tests. Finally we explore the relaxed semantics of software-generated interrupts, as used in sophisticated programming patterns, and sketch how they too could be modelled.
\end{abstract}

\begin{document}

\ifarxiv
\title{Relaxed exception semantics for Arm-A (extended version)}
\else
\title{Precise exceptions in relaxed architectures}
\fi

\author{Ben Simner}
\affiliation{
  \institution{University of Cambridge}
  \country{UK}
}
\email{Ben.Simner@cl.cam.ac.uk}

\author{Alasdair Armstrong}
\affiliation{
  \institution{University of Cambridge}
  \country{UK}
}
\email{Alasdair.Armstrong@cl.cam.ac.uk}

\author{Thomas Bauereiss}
\affiliation{
  \institution{University of Cambridge}
  \country{UK}
}
\email{Thomas.Bauereiss@cl.cam.ac.uk}

\author{Brian Campbell}
\affiliation{
  \institution{University of Edinburgh}
  \country{UK}
}
\email{Brian.Campbell@ed.ac.uk}

\author{Ohad Kammar}
\affiliation{
  \institution{University of Edinburgh}
  \country{UK}
}
\email{Ohad.Kammar@ed.ac.uk}

\author{Jean Pichon-Pharabod}
\affiliation{
  \institution{Aarhus University}
  \country{Denmark}
  }
\email{Jean.Pichon@cs.au.dk}

\author{Peter Sewell}
\affiliation{
  \institution{University of Cambridge}
  \country{UK}
}
\email{Peter.Sewell@cl.cam.ac.uk}

\raggedbottom

\bibliographystyle{ACM-Reference-Format}

\maketitle

\ifarxiv
\ifarxivsinglecolumn
\renewcommand{\shortauthors}{B.\ Simner, A.\ Armstrong, T.\ Bauereiss, B.\ Campbell, O.\ Kammar, J.\ Pichon-Pharabod, and P.\ Sewell}
\else
\fi
\fi

\section{Introduction}
\label{sec:intro}

Hardware exceptions (and their many variants: interrupts, traps, faults, aborts, etc.)
provide support for many exceptional situations that systems software has to manage.
This includes explicit privilege transitions via system calls,
implicit privilege transitions from trappable instructions,
inter-processor software-generated interrupts,
external interrupts from timers or devices,
recoverable faults like address translation faults,
and non-recoverable faults like memory error correction faults.

To confidently write concurrent systems code that handles exceptions,
e.g.\ mapping on demand at page faults,
programmers need
a well-defined and well-understood semantics.
The definition given
in modern architectures (e.g.\ in the current Arm-A documentation)
is basically unchanged since the IBM System/360, roughly as
\citet{HennessyPatterson12} %
state:
\emph{``An exception is imprecise if the processor state when an exception is raised does not look exactly as if the instructions were executed sequentially in strict program order''.}
However, on pipelined, out-of-order processors with observable speculative execution,
exceptions have subtle interactions with relaxed memory behaviour %
which have not previously been investigated.

\subsection{Contributions}
In this paper, we investigate the relaxed concurrency semantics of exceptions on modern high-performance architectures.
We focus on the Arm-A application-profile architecture
as a representative example,
although we expect that the challenges we describe also appear in other, similarly relaxed, architectures.
This work involved detailed discussions with Arm senior staff,
including the Arm Chief Architect and an Arm Generic Interrupt Controller (GIC) expert.
Our contributions are:
\begin{itemize}
\item
We clarify the concepts and terminology needed to discuss exceptions in relaxed-memory executions (\S\ref{sec:backandconcepts}).

\item
We explore the relaxed behaviour of exceptions: out-of-order and speculative execution, and forwarding across exception entry/exit boundaries (\S\ref{sec:behaviour}).
This is based on discussions with Arm and testing of several processor implementations, using a test harness for hardware testing of exceptions,
and a library of hand-written litmus tests.

\item
We explore the semantics of memory errors (\S\ref{subsec:SEA}). In Arm-A, these can generate \emph{external aborts}. Some implementations, including server designs, may exhibit \emph{synchronous} external aborts.
Such implementations rule out load-buffering (LB) relaxed behaviour,
which substantially curtails how relaxed observable behaviour is.

\item We develop an axiomatic model for Arm-A precise exceptions (\S\ref{sec:axiomatic}).
We extend Isla~\cite{isla-cav} to support both ISA and relaxed-memory concurrency aspects of exceptions, and we use it to evaluate the axiomatic model on tests.

\item
We identify and discuss the substantial open problem of what it means for exceptions to be precise in relaxed setting~(\S\ref{sec:prec}).
We characterise key properties that a definition should respect,
and highlight the challenge of giving
a proper definition of precision when relaxed behaviour is allowed across exception boundaries.

\item
Finally, we explore a significant use-case of exceptions
that benefits from the clarification of their interaction with relaxed memory:
the relaxed semantics of software-generated interrupts as used for sophisticated low-cost synchronisation, e.g.~in Linux's RCU~\cite{rcutxt} and Verona %
(\S\ref{sec:sgis}).  We sketch this in an axiomatic model.
\end{itemize}

This is an essential part of the necessary foundation for confidently programming systems code,
building on previous work that has clarified `user' relaxed concurrency~\cite{%
DBLP:books/daglib/0073498,%
Adir:2003,%
DBLP:conf/isca/GharachorlooLLGGH90,%
GharachorlooPhD,%
SFC91,%
ItaniumFormal,%
pldi105,%
pldi2012,%
DBLP:conf/micro/GrayKMPSS15,%
DBLP:conf/popl/FlurGPSSMDS16,%
mixed17,%
PulteFDFSS18,%
Pulte-phd,%
cacm,%
x86popl,%
cav2010,%
JadeThesis,%
AMSStacas2011,%
AlglaveDGHM21,%
DBLP:journals/toplas/AlglaveMT14,%
isla-cav}
and complementing recent work on the systems aspects of instruction fetch~\cite{SimnerFPAPMS20}
and virtual memory~\cite{relaxedVM-esop2022,alglave:hal-04567296}.
It helps put processor architecture specifications such as Arm-A on an unambiguous footing, where the allowed behaviour of systems-code idioms can be computed from a precise %
and executable-as-test-oracle definition of the architecture.

\subsection{Scope and limitations}

Our models cover
important use cases of exceptions,
but there remain several questions to be addressed by future work.
We do not give semantics to imprecise exceptions,
as it is unclear how to do so at the architectural level.

For our specific modelling of Arm:
we do not define the behaviour of `constrained unpredictable',
and merely flag when it is triggered.
Clarifying it will
require substantial extensive discussions with Arm architects,
likely affecting the wording in the architectural specifications, beyond the scope of this paper.
We do not try to precisely model the relaxed behaviour of system registers,
but merely sufficient conditions for conservative use cases in the context of exceptions
(\S\ref{subsec:ExS}).
We do not model switching between Arm FEAT\_ExS modes (\S\ref{sec:disablecontextsync}):
they are supported architecturally,
but are not commonly implemented.
We rely on a specific configuration to illustrate the use of interrupts for synchronisation (\S\ref{sec:sgis}),
without detailed modelling of the Arm Generic Interrupt Controller (GIC), %
or other system-on-chip (SoC) aspects.
The GIC
is a complex hardware component,
with a $950$-page specification~\cite[H.b]{arm-gic-v3-v4},
and modelling it in full would be a major project in itself.
This work is validated by substantial discussion and hardware testing, but more extensive testing on more devices is always desirable; we hope that our work will spur such additional testing on devices not available to us.
Finally, while we believe our models correctly capture the Arm architectural intent,
and that it gives a solid basis for programmers,
this paper is not an authoritative definition of the architecture, which is in any case subject to change.

\tikzset{%
  shorten >=1pt,
  >={Latex[length=4pt]},
  grow'=right,
  level distance=4em,
  sibling distance=3em,
  edge from parent/.style={draw,->},
  inst/.style={draw, minimum width=2.0em, minimum height=1em},
  finished/.style={fill=SeaGreen!80},
  discarded/.style={fill=grey!30},
  fill part/.style={
    path picture={\fill[SeaGreen!30] (path picture bounding box.north west) rectangle ($(path picture bounding box.south west)!#1!(path picture bounding box.south east)$);},
  }
}

\section{Arm-A architectural concepts for exceptions}\label{sec:backandconcepts}

We refine the Arm-A architectural concepts for exceptions.

\subsection{Exception taxonomy}\label{sec:taxon}
Arm-A defines multiple kinds of exception~\cite[D1.3.1, p6060%
]{Arm-K.a}: \emph{Synchronous exceptions} (supervisor/hypervisor
calls, traps, data/instruction, page faults, etc.)
and \emph{interrupts} (IRQ/FIQ from
processors/peripherals/timers and system errors).

The preferred return address of synchronous exceptions has some
architecturally defined relationship with the instruction that caused
them, and they are \emph{precise}. Their precision means roughly that
synchronous exceptions are observed at particular points in the
instruction stream, and so can use the preferred return address to
resume executing it after handling the exception.  We return to
precision below.

All interrupts are precise apart from SError (System Error)
interrupts, for which it is implementation-defined (per-kind) whether
they are precise.  SError interrupts arise from external system errors
that may or may not be recoverable. For example, an unrecoverable
imprecise SError may be generated by late detection of an uncorrectable
memory error correction error.  Exceptions stemming from such late
detection of uncorrectable memory errors are called \emph{external
aborts}.  In \S\ref{sec:behaviour}, we discuss how the mechanism an
implementation uses to report external aborts can rule out or allow
relaxed behaviour.

\subsection{Basic architectural machinery for exceptions}\label{sec:basic}

In Arm-A, when an exception is taken, execution jumps to the \noemph{exception vector}, an offset from the appropriate \noemph{vector base address register (VBAR)} value depending on the kind of exception.
The appropriate \noemph{exception syndrome register (ESR)}, \noemph{fault address register (FAR)}, and \noemph{exception link register (ELR)} are written with information about the cause and the preferred return address.
In some cases, the \noemph{exception level (EL)} register value, ranging in increasing privilege from 0 to 3, is also changed.
Exception handlers typically use ERET to return from an exception,
which restores some processor state and branches to the address in the appropriate ELR.
Most of these system registers (VBAR, ESR, etc.) are banked.

\subsection{Instructions and instruction streams}\label{sec:insts}
One often thinks of processors as executing \emph{instructions} in
some \emph{instruction sequence}, and common terminology is based on
those two concepts. For example, the Arm manual has around 60
instances of \emph{instruction stream} or \emph{execution stream}.
However, to account for relaxed behaviours and exceptions, we must
refine these concepts.
\subsubsection{From instructions to fetch-decode-execute instances}
\label{subsec:fdx_instances}

Exceptions can arise at multiple points within the fetch-decode-execute cycle,
including during the fetch and decode, when there is no `instruction'.
For Armv9.4-A, much of this is captured in an Arm top-level function
written in the Arm Architecture Specification Language (ASL).

We have then integrated this into Sail-based tooling to obtain an executable-as-test-oracle semantics of the sequential ISA aspects of Armv9.4-A with exceptions~(\S\ref{sec:isla-impl}).
A highly simplified outline of a single-instruction slice of the (400k line) instruction semantics is:
\begin{lstlisting}[language=sail, escapeinside={<@}{@>}, numbers=none, breaklines=true]
function <@\nonanonhref{https://github.com/rems-project/sail-arm/blob/19566bdb8615ae92ceea4b2a0e2bfbf59f5fbf0c/arm-v9.4-a/src/fetch.sail#L343}{\texttt{\_\_TopLevel}}@>() =
  // <@\nonanonhref{https://github.com/rems-project/sail-arm/blob/19566bdb8615ae92ceea4b2a0e2bfbf59f5fbf0c/arm-v9.4-a/src/interrupts.sail#L220}{in \texttt{TakePendingInterrupts}:}@>
  if IRQ then AArch64_TakePhysicalIRQException()
  if SE then AArch64_TakePhysicalSErrorException(...)
  // <@\nonanonhref{https://github.com/rems-project/sail-arm/blob/19566bdb8615ae92ceea4b2a0e2bfbf59f5fbf0c/arm-v9.4-a/src/v8_base.sail#L34286}{in \texttt{AArch64\_CheckPCAlignment}:}@>
  if pc[1..0] != 0b00 then AArch64_PCAlignmentFault()
  // <@\nonanonhref{https://github.com/rems-project/sail-arm/blob/19566bdb8615ae92ceea4b2a0e2bfbf59f5fbf0c/arm-v9.4-a/src/fetch.sail#L194}{in \texttt{\_\_FetchInstr}:}@>
  opcode = AArch64_MemSingle_read(pc, 4) // read memory
  // <@\nonanonhref{https://github.com/rems-project/sail-arm/blob/19566bdb8615ae92ceea4b2a0e2bfbf59f5fbf0c/arm-v9.4-a/src/decode_end.sail#L85}{in \texttt{\_\_DecodeA64}:}@>
  match opcode
    [1,_,1,1,1,0,0,1,0,1,_,_,_,_,_,_,_,_,_,_,_,_,_,_,_,_,_,_,_,_,_,_] =
      // the semantics for one family of instructions,
      // including loads LDR Xt,[Xn]
      // <@\nonanonhref{https://github.com/rems-project/sail-arm/blob/19566bdb8615ae92ceea4b2a0e2bfbf59f5fbf0c/arm-v9.4-a/src/instrs64.sail#L32819}{\texttt{execute\_aarch64\_instrs\_memory\_single\_general\_}}@>
      // <@\nonanonhref{https://github.com/rems-project/sail-arm/blob/19566bdb8615ae92ceea4b2a0e2bfbf59f5fbf0c/arm-v9.4-a/src/instrs64.sail#L32819}{\texttt{immediate\_signed\_post\_idx}(n,t,...)}@>
      let address = X_read(n, 64) // read register n
      let data : bits('datasize) = // read memory
        Mem_read(address, DIV(datasize,8))
      // write register t
      X_set(t, regsize) = ZeroExtend(data, regsize)
\end{lstlisting}
Executing this semantics may lead to one or more kinds of
exception,
calling the ASL/Sail function \texttt{AArch64\_TakeException()}.
This function writes the appropriate values to registers, e.g.\ computing the next PC, exception level, etc.~and terminates this
\\
\texttt{\_\_TopLevel()} execution.
So instead of `instruction instances',
we refer to \emph{fetch-decode-execute instances} (FDX instances),
a single execution of \texttt{\_\_TopLevel()}.

\subsubsection{Fetch-decode-execute trees and streams}
One must relate the out-of-order speculative execution of
hardware implementations and
the architectural definition of the allowed behaviours.
We will use the following concepts,
well-understood when modelling relaxed memory without
exceptions.
At any instant, each core may be processing, out-of-order and
speculatively, many instructions (really, FDX instances) from its
hardware thread.  Partially executed instances are restarted or
discarded if they would violate the intended semantics
(e.g.~mispredicted branch).

One can visualise the state of a single core abstractly as a tree of
partially and completely executed instances, as in
Fig.~\ref{fig:treehw} (top).
Abstract-microarchitectural operational
models use this abstraction~\cite{pldi105,pldi2012,DBLP:conf/micro/GrayKMPSS15,DBLP:conf/popl/FlurGPSSMDS16,mixed17,PulteFDFSS18}.
\begin{figure}
\vspace*{3mm} %
\begin{center}
 \scalebox{0.8}{
    \begin{tikzpicture}[scale=0.72]
    \node[inst, finished] (instB) {}
    child {
      node[inst, fill part=0.5] (inst0) {}
      child {
        node[inst, fill part=1] (inst1) {}
        child {
          node[inst, finished] (inst2) {}
          child {
            node[inst] (inst3) {}
            child {
              node[inst, fill part=0.2] (inst4) {}
              child {
                node[inst, fill part=0.4] (inst6) {}
                child {node[inst] (inst7) {}}
                child {node[inst] (inst8) {}}
              }
            }
            child {
              node[inst, fill part=0.8] (inst5) {}
              child {node[inst] (inst9) {}}
            }
          }
        }
      }
    }
;
    \end{tikzpicture}
}

\vspace*{5mm}

\scalebox{0.8}{
    \begin{tikzpicture}[scale=0.72]
    \node[inst, finished] (instB) {}
    child {
      node[inst, finished] (inst0) {}
      child {
        node[inst, finished] (inst1) {}
        child {
          node[inst, finished] (inst2) {}
          child {
            node[inst, finished] (inst3) {}
            child {
              node[inst, finished] (inst4) {}
              child {
                node[inst, finished] (inst6) {}
                child [draw=grey!30]{node[inst, discarded] (inst7) {}}
                child {node[inst, finished] (inst8) {}}
              }
            }
            child [draw=grey!30]{
              node[inst, discarded] (inst5) {}
              child [draw=grey!30] {node[inst, discarded] (inst9) {}}
            }
          }
        }
      }
    }
;
    \end{tikzpicture}
}
  \end{center}
  \caption{\textbf{Top.} The tree of (partially) executed FDX instances at one time, in hardware or operational model execution.
  \label{fig:treehw}%
\textbf{Bottom.} The sequence of architecturally executed FDX instances in a completed execution.\label{fig:treearch}}
  \end{figure}
We depict the retired (committed) FDX instances as solid dark green,
and partially/tentatively executed in-flight instances as light green.
The arrows depict program order.
Committed instances can be program-order after in-flight instances,
and non-committed instances may need to be restarted.
Eventually all FDX instances for this hardware thread will be either committed
or discarded, e.g.\ as in Fig.~\ref{fig:treearch} (bottom).
These are the \emph{architecturally executed} FDX instances.  The
architecture definition, and any formal semantics thereof, have to
define which such sequences are allowed for each thread.  This
definition includes the register content; memory read values; and
their relationships with other threads, as determined by the relaxed
concurrency model.
Axiomatic concurrency models, e.g.~\cite{%
DBLP:books/daglib/0073498,%
Adir:2003,%
DBLP:conf/isca/GharachorlooLLGGH90,%
GharachorlooPhD,%
SFC91,%
ItaniumFormal,%
cacm,%
x86popl,%
cav2010,%
JadeThesis,%
AMSStacas2011,%
AlglaveDGHM21,%
DBLP:journals/toplas/AlglaveMT14,%
isla-cav},
use candidate executions containing the events just
from these architecturally executed instances.

The Arm prose specification
in Fig.~\ref{fig:arm simple seq. exec.} (top)
previously attempted to capture the
relationship between implementation execution (out of order and
speculative) and the architectural definition of allowed behaviour in
terms of a notion of ``simple sequential execution''.
\begin{figure}
\begin{quotebox}
\textbf{Architecturally executed}
An instruction is architecturally executed only if it would be executed in a simple sequential execution of the
program. [...]

\textbf{Simple sequential execution}
The behavior of an implementation that fetches, decodes and completely executes each instruction before
proceeding to the next instruction. Such an implementation performs no speculative accesses to memory, including
to instruction memory. The implementation does not pipeline any phase of execution. In practice, this is the
theoretical execution model that the architecture is based on, and Arm does not expect this model to correspond to
a realistic implementation of the architecture.
\end{quotebox}
\begin{quotebox}
\textbf{Architecturally executed}
A candidate execution can be architecturally executed if it is composed of a sequence of FDX instances for each thread that together satisfy the Arm concurrency model [extended to cover exceptions, as described here, and other systems phenomena], starting from the machine initial state.
\end{quotebox}
\caption{Arm prose specification~\cite[Glossary, p14749]{Arm-K.a} (\textbf{top})
and our suggested rephrasing (\textbf{bottom}).}
\label{fig:arm simple seq. exec.}
\end{figure}
As the prose says, simple sequential execution does not hold for the
intended relaxed-memory architecture.
We propose a more correct rephrasing that allows for
exceptions and other systems phenomena in Fig.~\ref{fig:arm simple seq. exec.} (bottom).

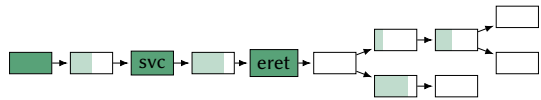
\begin{figure}
 \begin{center}
 \scalebox{0.8}{
    \begin{tikzpicture}[scale=0.72]
    \node[inst, finished] (instB) {}
    child {
      node[inst, fill part=0.5] (inst0) {}
      child {
        node[inst,finished] (exc-1)  {svc}
        child {
          node[inst,fill part=0.75] (exc0)  {}
          child {
            node[inst,finished] (exc1)  {eret}
              child {
                node[inst] (inst3) {}
                child {
                  node[inst, fill part=0.2] (inst4) {}
                  child {
                    node[inst, fill part=0.4] (inst6) {}
                    child {node[inst] (inst7) {}}
                    child {node[inst] (inst8) {}}
                  }
                }
                child {
                  node[inst, fill part=0.8] (inst5) {}
                  child {node[inst] (inst9) {}}
                }
              }
            }
          }
        }
      }
;

    \end{tikzpicture}
}
\end{center}
  \caption{The tree of partially and completely executed FDX instances with exceptions,
  in hardware or operational model execution\label{fig:treehwexec}.
  Instructions may execute out-of-order across exception boundaries,
  requiring a modern definition for precision.}
  \end{figure}

Fig.~\ref{fig:treehwexec} depicts a tree of instances involving
exception entry~(\asm{svc}) and return~(\asm{eret}).  Arm-A allows
implementations to observe the exception handling instances as
executing before program-order previous instances have been retired,
and similarly exception return.  Exception entry and return may never
be observed as starting to execute speculatively, however, and so the
three speculative branches may not observe exception entry or return
instances. Precision must account for these allowed and prohibited
relaxed behaviours. %

\section{Relaxed behaviour of precise exceptions}
\label{sec:behaviour}
\label{sec:tests}

\newcommand{\testUp}[1]{%
\scalebox{0.75}{
\begin{tabular}{c}
    \input{tests/#1.listing.tikz} \\
    \input{tests/#1.diagram.tikz}
  \end{tabular}}
  \label{test:#1}
}

\newcommand{\testRCUMPUp}[1]{%
\begin{center}
\scalebox{0.75}{\input{tests/#1.listing.tikz}}

\vspace*{-2.0\baselineskip}

\scalebox{0.75}{\input{tests/#1.diagram.tikz}}
\end{center}
  \label{test:#1}
\vspace*{-4.0\baselineskip}
}

\newcommand{\testTwoUp}[1]{%
\scalebox{0.66}{
\begin{tabular}{c}
    \input{tests/#1.listing.tikz} \\
    \input{tests/#1.diagram.tikz}
  \end{tabular}}
  \label{test:#1}
}

\newcommand{\testSide}[1]{%
\scalebox{0.75}{
\begin{tabular}{l@{}l}
    \input{tests/#1.listing.tikz}
    &
    \input{tests/#1.diagram.tikz}
  \end{tabular}}
  \label{test:#1}
}

\newcommand{\testOne}[1]{%
\scalebox{0.75}{
\begin{tabular}{c}
  \input{tests/#1} \\
\end{tabular}}
\label{test:#1}
}

\newcommand{\testTwoOne}[1]{%
\scalebox{0.66}{
\begin{tabular}{c}
  \input{tests/#1} \\
\end{tabular}}
\label{test:#1}
}

Exceptions change the control flow and processor context,
that is, the collection of system and special registers which control
the execution of the machine,
such as the current exception level (\asm{PSTATE.EL}),
masking of interrupts (\asm{PSTATE.\{D,A,I,F\}}),
processor flags, etc.
However, changes to the context may not take effect immediately,
and so, to ensure that program-order-later instructions see such changes,
exceptions usually come with context synchronisation.
It is this context synchronisation which imposes ordering,
and we show how, without such context synchronisation,
we observe reordering across exception boundaries.
For this reason,
exceptions are usually context-synchronising on Arm.

There are many things that can trigger exceptions.
The simplest way is to use an `exception-generating instruction'
such as a system call (on Arm, the \asm{SVC} instruction).
While exceptions like interrupts and page faults are more common,
they may come with extra synchronisation.
Therefore, \asm{SVC}s provide a baseline for precision,
and we use them in our exploration of the behaviour of exceptions
in the remainder of this section;
we return to discuss other exceptions later on.

In this section,
we explain relaxed behaviour of precise exceptions
through litmus tests,
the usual standard for succinctly cataloguing the relaxed behaviours
allowed by an architecture~\cite{AMSStacas2011,DBLP:journals/toplas/AlglaveMT14,isla-cav}.
Litmus tests are small programs capturing
specific software patterns or hardware mechanisms,
whose outcome depends on some kind of out-of-order execution.

Precise exceptions do not change the memory model between exception boundaries,
and so the interesting questions concern out-of-order execution across exception boundaries.

We will talk about context synchronisation in detail~(\S\ref{subsec:ExS}),
explore the baseline out-of-order execution across exception boundaries~(\S\ref{subsec:behaviours}),
then the stronger behaviour of specific types of exceptions~(\S\ref{sec:ExEts}),
touch on how the instruction semantics needs to be adapted~(\S\ref{subsec:exceptions-in-ASL}),
and finally discuss a corner case disabling context synchronisation~(\S\ref{sec:disablecontextsync}).

\subsection{Context-synchronisation}%
\label{subsec:ExS}

Updates to the context, such as writes to system registers,
need synchronisation to be guaranteed to have an effect.
We do not model the behaviour of such context-changing operations
when such synchronisation is not performed.
Instead, we merely identify when and how exceptions are context-synchronising,
and note that this has a knock-on effect on memory accesses.

Architecturally, a context synchronisation event guarantees that no
instruction program-order-after the event is observably fetched, decoded, or
executed until the context-synchronising event has happened.
A simple microarchitectural implementation for context synchronisation
is to flush the pipeline: restarting all program-order-later instances
once the context-synchronising effect occurs.
More complex implementations
may be more clever,
as long as they preserve the semantics.

Software can explicitly generate context synchronisation events
by issuing an Instruction Synchronisation Barrier (\asm{ISB}).
Context synchronisation can also happen implicitly,
for example on exception entry and exit.
This is the case in Arm,
except in a rare use case we return to in \S\ref{sec:disablecontextsync}.

The effect of context synchronisation events in exception boundaries
is that any instance after the boundary has an \asm{ISB}-equivalent
dependency on the instances before the boundary.
This mechanism implies the following fundamental invariant:
\emph{context synchronising exceptions are never taken speculatively},
and it limits speculation to the same well-understood extent as \asm{ISB}
limits speculation.
This invariant has interesting interactions with external aborts,
which we discuss in~\S\ref{subsec:SEA}.

\subsection{Relaxed behaviours}
\label{subsec:behaviours}

In this section,
we explore the relaxed behaviour of exceptions,
with a selection of litmus tests from our larger suite of 61 hand-written tests.
For each test, we include
whether the behaviour is allowed in our understanding of the architectural intent;
the relevant experimental results when available (labelled \hwrefsName);
and a candidate execution graph.  When available, the
experimental hardware results (obtained by extending the testing harness of \citet{relaxedVM-esop2022}) report the frequency of observation on
the following implementations, respectively
a Raspberry Pi 3B+ (Arm Cortex-A53 r0p4),
a Raspberry Pi 4B (Arm Cortex-A72 r0p3),
a Raspberry Pi 5 (Arm Cortex-A76 r1p4),
and an ODROID N2+ (Arm Cortex-A73 r0p2).
The latter is a big.LITTLE architecture;
our results are from the `big' A73 cores.
We mark %
behaviours as allowed/disallowed
based on discussions with Arm architects.

\subsubsection{Out-of-order execution across exception boundaries}
\label{beh:ooo-across-boundaries}
\begin{figure}
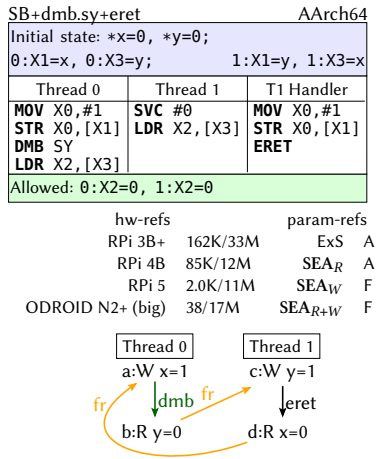

\begin{center}
\scalebox{0.75}{
\begin{tabular}{c}
    \input{tests/SB+dmb+eret.listing.tikz} \\
    \input{tests/SB+dmb+eret.diagram.tikz}
  \end{tabular}}
  \label{test:SB+dmb+eret}
\end{center}
\vspace*{-1.0\baselineskip}
\caption{%
Reads and writes may be executed out-of-order across exception entry, exit, or even both.
This shows executing a read out-of-order across exception entry+exit.
}
\label{fig:OoO-shapes}
\end{figure}

Exception boundaries do not act as memory barriers, so loads and stores may be
executed out-of-order
over
an exception entry
or an exception exit
or %
the composition of both (Figure~\ref{fig:OoO-shapes}).

\subsubsection{Speculative exception entry or return}\label{sec:specentryreturn}
\begin{figure}
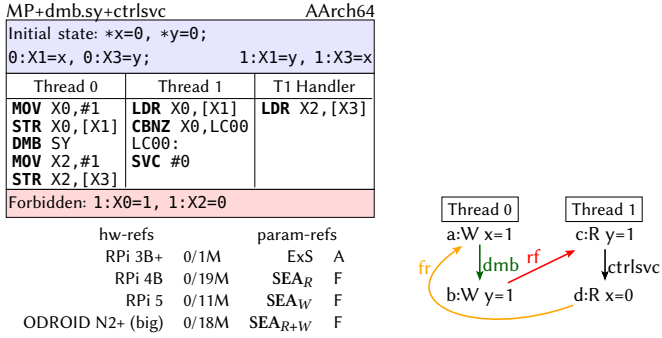
%
  \centering
\scalebox{0.75}{
\begin{tabular}{l@{}l}
    \input{tests/MP+dmb+ctrlsvc.listing.tikz}
    &
    \input{tests/MP+dmb+ctrlsvc.diagram.tikz}
  \end{tabular}}
  \label{test:MP+dmb+ctrlsvc}
  \label{test:MP+dmb+ctrl-svc}
  \caption{Context synchronising exception entry (and returns) are not executed speculatively.}
  \label{fig:mp-ctrl-exc}
\end{figure}

The invariant `context synchronising exceptions cannot be taken
speculatively' imposes the same kind of barrier as a \texttt{ctrlisb}
dependency would impose between program-order-previous instances and
the instances in the handler. The control dependency is due to the
branching to the handling code, and the \asm{ISB} dependency is due to
context synchronisation. As a consequence, the two behaviours in
Figure~\ref{fig:mp-ctrl-exc} are forbidden. On architectures that
allow the \FEAT{ExS} extension, they would be allowed when the exception entry/exit
is not context synchronising, i.e., when the corresponding
\asm{EIS}/\asm{EOS} bit is cleared. This mechanism also explains
why we do not observe load-load reordering on the Raspberry Pi
devices, but we do observe them on the ODROID-N2+ (exhibited by the
test \testref{MP+dmb+svc} which can be found in the supplementary
material\ifextended, \S\ref{app:litmus-glossary}\else\fi).  These machines
exhibit the same behaviour as they would for the
corresponding \texttt{MP+dmb+isb} behaviour from previous work.

\subsubsection{Privilege level}
The privilege level (\asm{PSTATE.EL}) has little to no additional
effect on the behaviours we present: their allowed/forbidden status
remains the same whether the privilege goes up/down in entry/exit or
remains the same.
The one exception to this principle is the effect a
privilege change has on non-faulting translation table walks. A
non-faulting translation walk for an instance program-order-before a
privilege-changing exception entry from \asm{ELn} may be reordered
with the entry, but would then also be reordered with every subsequent
exception boundary until the privilege level returns to \asm{ELn}.
Explaining this case in full detail would require substantial details
of Arm's virtual memory
architecture~\cite{relaxedVM-esop2022},
and we leave it to future work.

\subsubsection{Forwarding writes}
It is permitted for writes to be forwarded from a store to a read
across exception entry and return
\\
(\testref{SB+dmb+rfisvc-addr} in
Figure~\ref{test:SB+dmb+rfisvc-addr}).

\begin{figure}
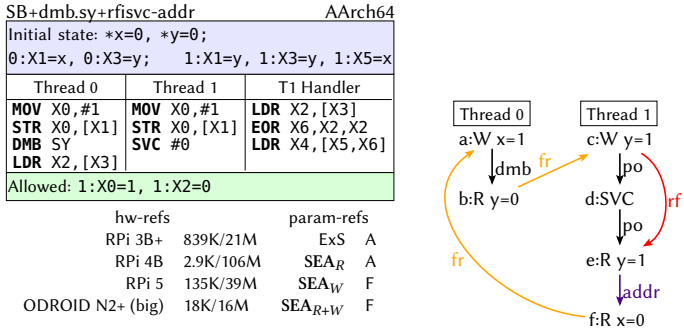
%
\scalebox{0.75}{
\begin{tabular}{l@{}l}
    \input{tests/SB+dmb+rfisvc-addr.listing.tikz}
    &
    \input{tests/SB+dmb+rfisvc-addr.diagram.tikz}
  \end{tabular}}
  \label{test:SB+dmb+rfisvc-addr}

\caption{Forwarding into a non-speculative handler.}
\label{test:SB+dmb+rfisvc-addr}
\end{figure}

\subsubsection{Dependency through system registers}
\label{subsec:exception-register-dependencies}
\begin{figure}
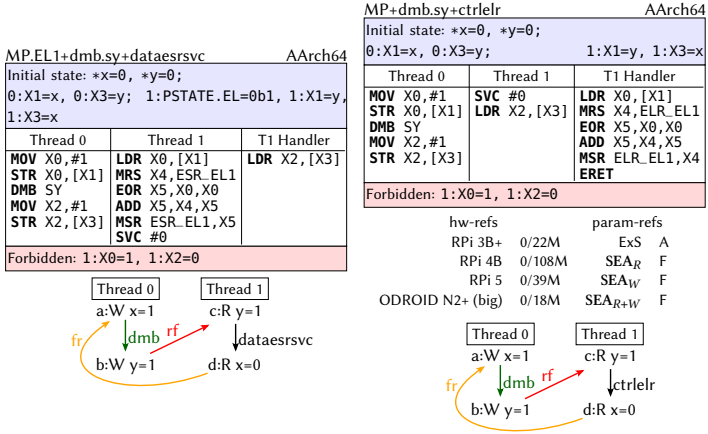

  \begin{tabular}{@{}c@{\hskip-0.3cm}c@{}}
\scalebox{0.66}{
\begin{tabular}{c}
    \input{tests/MP.EL1+dmb+dataesrsvc.listing.tikz} \\
    \input{tests/MP.EL1+dmb+dataesrsvc.diagram.tikz}
  \end{tabular}}
  \label{test:MP.EL1+dmb+dataesrsvc}

  &
\scalebox{0.66}{
\begin{tabular}{c}
    \input{tests/MP+dmb+ctrlelr.listing.tikz} \\
    \input{tests/MP+dmb+ctrlelr.diagram.tikz}
  \end{tabular}}
  \label{test:MP+dmb+ctrlelr}

  \end{tabular}
  \caption{System registers and context synchronisation}
  \label{fig:system-register-dependency}
\end{figure}

Where exceptions are taken to and returned to
are part of the context, and must be read by exception taking and returning,
and so they can be involved in register dependency chains.
Here, we do not characterise the general effect of such dependencies,
but focus on the effect exceptions have on them.
Dependencies on system register accesses compose with
ordering from context synchronisation events to program-order-later instructions.
Test
\testref{MP.EL1+dmb+dataesrsvc} in
Fig.~\ref{fig:system-register-dependency} demonstrates that a write
to the system register \asm{ESR} that depends on a read forbids
reordering this read across the boundary, even though resolving the
dependency does not affect the exception.

The \asm{ELR} register is a special-purpose register,
and is therefore `self-synchronising'.
Therefore, writes into the \asm{ELR} do not need context synchronisation
to guarantee that they are seen by program-order-later instructions,
and this means that dependencies into the \asm{ELR} are preserved
(see Fig.~\ref{fig:system-register-dependency}).

This has two related subtleties,
and is currently under investigation by Arm.
The Software Thread ID Register
(\asm{TPIDR}) is a system register in which the operating system can
store thread identifying information, but has no relevant
indirect effects. Further testing and discussions may clarify whether it
forbids reordering. %
While dependencies through special-purpose registers are preserved,
context synchronisation does not necessarily need to wait
for those writes,
and so these dependencies do not necessarily pass to
instructions after context synchronisation
(in contrast to system register writes).

\subsubsection{Ordering from asynchronous exceptions}

Asynchronous exceptions cannot be taken speculatively.
Therefore, all instructions program-order-after an asynchronous exception
happen after that exception.

\subsection{Exception-specific mechanisms}
\label{sec:ExEts}
\begin{figure}
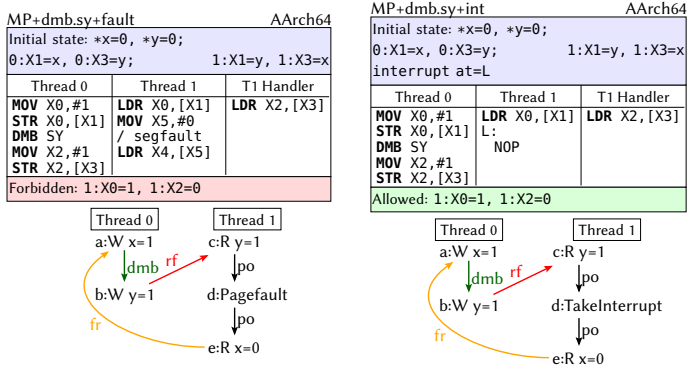
%
  \centering
  \begin{tabular}{@{}c@{}c@{}}
\scalebox{0.66}{
\begin{tabular}{c}
  \input{tests/MP+dmb.sy+fault} \\
\end{tabular}}
\label{test:MP+dmb.sy+fault}
 & %
\scalebox{0.66}{
\begin{tabular}{c}
  \input{tests/MP+dmb.sy+int} \\
\end{tabular}}
\label{test:MP+dmb.sy+int}
 \\
  \end{tabular}
  \caption{Different exception kinds can have different behaviour.}%
  \label{fig:mp_pagefault}
\end{figure}

Some exceptions on some implementations involve additional mechanisms.
For example, when an implementation supports the Enhanced Translation
Synchronisation (\FEAT{ETS2}), the
translation-table-walks which generate translation faults (pagefaults)
gain additional ordering from program-order-previous instances.
Figure~\ref{fig:mp_pagefault} compares the a Message-Passing test
involving a page-fault (\testref{MP+dmb.sy+fault}, forbidden) and the
same shape involving an \asm{SVC} exception (\testref{MP+dmb.sy+int},
allowed).

The architectural rationale for \FEAT{ETS2} is to prevent spurious
faults from old translation walks. Such faults cause difficulties for software
and require software to introduce many barriers. The \FEAT{ETS2}
extension requires hardware to always put a barrier before a translation
fault. Microarchitecturally, this can be by
restarting faulting instructions when they become non-speculative.
Implementations are required to support \FEAT{ETS2} from Armv8.8-A
onwards, and we model it.
We are aware the specification of additional mechanisms per
exception-kind is an active area for Arm, and we hope to extend the
model to match future changes in the architecture.

\subsection{Exceptions and the intra-instruction semantics}%
\label{subsec:exceptions-in-ASL}
Wherever possible, we want to interpret the intra-instruction ASL
ordering as preserved, both for conceptual simplicity, memory-model
tool execution, and reasoning.  This has previously been possible
except in a few specific cases that are inherently concurrent:
instructions that do multiple accesses,
and \asm{CSEL}, \asm{CAS}, \asm{SWAP}, etc.
Exceptions introduce a new
interesting case for instructions that do a register writeback
concurrently with a memory access.
For example, \asm{STR} (immediate)
has a ``Post-index'' and a ``Pre-index'' versions~%
\cite[C6.2.365, p2442%
]{Arm-K.a}. The post-index \asm{STR Xt, [Xn], \#8}, for example,
stores the value in \asm{Xt} to the address initially in register \asm{Xn} and adds \asm{8} to
\asm{Xn}.
The Arm ARM ASL for \asm{STR} puts that register write at the end, after
the memory access has completed.
The architectural intent is that program-order-later instances
that depend on \asm{Xn} can go ahead early, e.g.\ before the
data in register \asm{Xt} is available to be written to memory. The relevant
litmus tests have been observed (on an ODROID-N2+, with 2 Cortex A53
cores plus 4 Cortex A73 cores)~\cite{lucpc}.

Previous work captured this allowed by having
the register writeback before the memory access
in the instruction semantics.
However, exceptions require more care: when the memory
access generates an exception, the writeback register should appear
unchanged to instances after the exception boundary.

\subsection{Disabling context synchronisation}
\label{sec:disablecontextsync}

On Arm, there is an optional feature, \FEAT{ExS},
which
provides two new fields,
\asm{EIS} and \asm{EOS},
in the \asm{SCTLR\_ELx} system control register.
These allow software to disable context synchronisation on exception entry and return, respectively.
While the semantics is clear for these systems, the
programming model is unpredictable and hard to program correctly, and
so this configuration is rarely encountered in practice.

\section{Synchronous external aborts}%
\label{subsec:SEA}

The memory system may detect errors such as data corruption
independently of the MMU or Debug hardware,
e.g., using parity bits or error correcting code.
In those cases, it will signal the error by a class of exceptions called \emph{external aborts}.
The architecture does not define at which granularity implementations may report
such aborts synchronously, which we refer to as \emph{synchronous external aborts} (SEAs).
Instances program-order-after a potential cause for synchronous external aborts
are considered speculative until this external abort
can be ruled out, resulting in stronger behaviour (\S\ref{subsec:SEA-strength}).
In an implementation that always reports external aborts asynchronously,
the later instances become non-speculative earlier, allowing them to
exhibit weaker behaviours.

When external aborts are reported asynchronously, the simplest
recovery is to wind down the aborting process. To allow programmers
more reliable recovery, implementations can support the Reliability,
Availability, and Serviceability (RAS) extension.
This extension is a substantial component of the architecture,
far beyond the scope of this work.
Here, we are merely taking the first steps,
describing a baseline of behaviours in a very constrained setting,
that further work may be able to extend to account for the RAS.

\subsection{Behaviour resulting from synchronous external aborts}
\label{subsec:SEA-strength}

There is an asymmetry between reads and writes
with respect to speculation:
writes cannot be propagated speculatively,
whereas reads can be satisfied speculatively.
We will therefore consider the store and load cases separately.

If a store may generate an SEA,
then program-order-later instances are speculative until the store has (at least) propagated to memory.
In that case, write-write re-ordering (MP+po+addr) is forbidden.
Reads program-order-after writes are permitted to execute speculatively anyway,
and so the presence of these SEAs does not restrict their ability to execute early.

More interestingly, if a load may generate an SEA,
then program-order-later instances are speculative until the load has completed all its reads,
and is non-restartable.
This means that writes program-order-after that read are forbidden from executing out-of-order.
This forbids interesting tests which would otherwise be allowed,
namely load-buffering (LB+pos) and MP with a plain ISB after one load (MP+dmb.sy+isb)~\cite{acs-2022}.

\subsection{Load buffering and the out-of-thin-air problem}

This has an important and hitherto not well-understood impact %
on programming-language concurrency models.
Ruling out LB enables substantially
simpler design of programming language concurrency models:
they can execute instructions in-order and merely keep a history of the writes seen so far,
e.g.~\cite{lahav-et-al:repairing-c11},
and thereby avoid the notorious out-of-thin-air problem~\cite{BattyMNPS15}.
These simpler semantics support a line of model checkers for C/C++ and LLVM~\cite{%
kokologiannakis-et-al:effective-stateless-model-checking-for-c/c++-concurrency,%
kokologiannakis-et-al:model-checking-for-weakly-consistent-library,%
kokologiannakis-vafeiadis:genmc%
}.
In contrast, the presence of LB seems to require significant sophistication%
~\cite{Pugh99,BattyMNPS15,Pichon-PharabodS16,kang-et-al:promising-semantics,JagadeesanJR20,AlglaveDGHM21,Batty15,Chakraborty19}.

\section{An axiomatic model of exceptions}
\label{sec:axiomatic}

We now give a formal semantics
that describes the concurrent behaviour of precise exceptions on Arm-A.
We give it as an extension of the previous model of \citet{PulteFDFSS18},
a predecessor of the current Arm model~\cite{arm-cat},
in the standard `cat' format~\cite{DBLP:journals/toplas/AlglaveMT14,isla-cav},
in Figure~\ref{fig:axiomatic_exceptions}.

\begin{figure}
\centering
\begin{multicols}{2}
\begin{lstlisting}[language=cat]
"Arm-A exceptions"

include "cos.cat"
include "arm-common.cat"

(* might-be speculatively executed *)
let speculative =
    [*ctrl*]
  | [*addr; po*]
  | if "SEA_R" then [R]; po else 0
  | if "SEA_W" then [W]; po else 0

(* context-sync-events *)
let CSE =
    [*ISB*]
  | if "FEAT_ExS" & ~"EIS" then 0 else TE
  | if "FEAT_ExS" & ~"EOS" then 0 else ERET

let ASYNC =
  TakeInterrupt

(* observed by *)
[*let obs = rfe | fr | co*]

(* dependency-ordered-before *)
[*let dob =*]
    [*addr | data*]
 [* | speculative ; [W]*]
 [* | speculative ; [ISB]*]
 [* | (addr | data); rfi*]

(* atomic-ordered-before *)
[*let aob =*]
   [*rmw*]
[* | [range(rmw)]; rfi; [A|Q]*]

(* barrier-ordered-before *)
let bob =
  [*  [R] ; po ; [dmbld]*]
 [* | [W] ; po ; [dmbst]*]
 [* | [dmbst]; po; [W]*]
 [* | [dmbld]; po; [R|W]*]
 [* | [L]; po; [A]*]
 [* | [A | Q]; po; [R | W]*]
 [* | [R | W]; po; [L]*]
  | [dsb]; po

(* contextually-ordered-before *)
let ctxob =
    speculative; [MSR|CSE]
  | [MSR]; po; [CSE]
  | [CSE]; po

(* async-ordered-before *)
let asyncob =
    speculative; [ASYNC]
  | [ASYNC]; po

(* Ordered-before *)
let ob = ([*obs | dob | aob |*]
  [*bob*] | ctxob | asyncob)+

(* Internal visibility requirement *)
[*acyclic po-loc | fr | co | rf as internal*]

(* External visibility requirement *)
[*irreflexive ob as external*]

(* Atomic: Basic LDXR/STXR constraint to forbid intervening writes. *)
[*empty rmw & (fre; coe) as atomic*]
\end{lstlisting}
\end{multicols}
\caption{Arm-A exceptional model (\textcolor{grey}{greyed} out parts are unchanged from the original model).}%
\label{fig:axiomatic_exceptions}
\end{figure}

The model is parameterised along two axes:
\begin{itemize}
  \item \herd{FEAT\_ExS} corresponds to the feature of the same name being implemented;
        we do not support runtime changes of the related \asm{SCTLR\_ELx.\{EIS,EOS\}} fields,
        and so fix them as variants.
  \item \herd{SEA\_R} and \herd{SEA\_W} correspond to the \textsc{ImplementationDefined} choice
        of whether loads or stores may generate synchronous external aborts.
\end{itemize}

Most current hardware does not support \herd{FEAT\_ExS},
and moreover, we expect that most software would not use it.
However, its
semantics is relatively straight-forward as we understand it,
and so we include it in our model.

We add new events to the candidate execution:
\herd{TE} (take exception) and \herd{ERET},
which correspond to the synchronisation points
(whether they \emph{are} synchronising)
of taking or returning from an exception;
and
\herd{MRS} and \herd{MSR} events, for reading and writing system registers,
corresponding to the Arm \herd{MRS} and \herd{MSR} instructions
which change the context.

\paragraph{Exceptions and program order}
We include all the new events in program-order.
This includes the events from instructions directly before and after taking or returning from an exception.

\paragraph{Interrupts}
While this cat model does not support inter-processor interrupts and the generic interrupt controller
(see \S\ref{sec:sgis} for a draft extension to support them),
it does support other precise asynchronous exceptions (e.g. timers).

\paragraph{Ordered-before}
We expand ordered-before:
\begin{itemize}
\item
Wherever \herd{ctrl|(addr;po)} was used before,
we also include instructions program-order-after
reads or writes when in the relevant \herd{SEA} variant.
With those variants, the instructions program-order-after those events are speculative up until
the memory access has completed.
\item
The previous model's use of \herd{ISB} was purely for its context synchronisation effect.
Accordingly, wherever \herd{[ISB]} was used before,
we include exception entry (\herd{TE}) and exit (\herd{ERET}),
unless we are in the variant where context synchronisation on those events is disabled.
\item
We extend barrier-ordered-before with the \herd{DSB} barriers.
The barrier event classes are upwards-closed,
so that \herd{DSB.SY} is included in all the \herd{dmb} events.
\item
We add a context-ordered-before (\herd{ctxob}) sub-clause to the ordered-before relation,
which captures the ordering of context-changing operations and context-synchronisation:
namely, that context-changes and context-synchronisation cannot happen speculatively;
that all context-changes are ordered before any context-synchronisation;
and that no instruction program-order-after context-synchronisation can be executed until the synchronisation is complete.
\item
We add an async-ordered-before (\herd{asyncob}) clause to ordered-before,
capturing that asynchronous events (such as interrupts) cannot be done speculatively,
and instructions program-order-after them may not happen before the asynchronous event which precipitated them.
\end{itemize}

\subsection{Executable-as-a-test-oracle implementation}\label{sec:isla-impl}

We implement the model in Isla~\cite{isla-cav},
an SMT-based executable oracle for axiomatic concurrency models
(and ISA semantics).
Isla takes as input a memory model in herdtools-like cat format,
and a litmus tests.
To support tests with asynchronous exceptions,
we added a construct to
specify a label where the exception will occur,
so that
Isla then pends an interrupt at that program point.

The instruction semantics we use is a translation into the Sail language of the Armv9.4-A ASL specification, including the top-level function provided by Arm.
\ifanon
\else
\cite{sail-arm-9.4}
\fi
The translation process~\cite{sail-popl2019} is mostly automatic,
requiring select manual interventions mostly due to differences in the type systems of ASL and Sail.
We also added patches to support the integration with Isla, in particular adding hooks to expose information about exceptions being taken in a form that can be readily consumed by Isla.
In doing so, we encountered and fixed some bugs in the ASL model related to uses of uninitialised fields in data structures,
as well as missing checks for implemented processor features that led to spurious system register accesses.

For all the (non-IPI) tests presented in this paper, Isla,
the architectural intent as we understand it,
and the results of hardware testing from~\S\ref{subsec:behaviours}
are consistent.

\section{Challenges in defining precision}\label{sec:prec}

The phenomena we describe in \S\ref{sec:behaviour}
highlight that the historical, naive definition of precision
does not account for relaxed memory.
The open problem is then \emph{how to adequately define precision in a relaxed-memory setting}.
This challenge is hinted at in the way
the Arm reference manual~\cite[D1.3.1.4, p6060]{Arm-K.a} defines precision as:
\begin{quotebox}
An exception is \emph{precise} if on taking the exception, the hardware thread (aka processing element, PE) state and the memory system state is consistent with the
PE having executed all of the instructions up to but not including the point in the instruction stream where the
exception was taken from, and none afterwards.
[except that in certain specific cases some registers and memory values may be UNKNOWN]
\end{quotebox}

This definition explicitly allows various side effects of an instruction executing when an exception is taken to be visible. The details are intricate, but in outline:  registers that would be written by the instruction but which are not used by it (to compute memory access addresses) can become UNKNOWN, and for instructions that involve multiple single-copy-atomic memory writes (e.g.\ misaligned writes and store-pair instructions), where each write might generate an exception (e.g.\ a translation fault), the memory locations of the writes that do not generate exceptions become UNKNOWN.  These side effects could be observed by the exception handler, and the memory write side effects could be observed by other threads doing racy reads.
Hardware updates to page-table access flags and dirty bits, and to performance counters, could also be observable.
This means that the abstraction of a stream of instructions executed up to a given point
does not account for the relaxed-memory behaviour.

Arm \emph{classify} particular kinds of exceptions as precise or not, but all the above
makes it hard to \emph{define} in general what it means for an exception to be precise in a relaxed setting.

The ultimate architectural intent of precision is that it is
sufficient to meaningfully resume execution after the exception.
For example,
for software that does mapping on demand,
when an instruction causes a fault by accessing an address which is not currently mapped,
the exception handler will map that address and return.
This means that re-executing the original instruction will overwrite these UNKNOWNs,
and will have ordering properties much like the original instruction would have had if the mapping had already been in place.

Our models are complete enough to reason about such cases in concrete examples.
However, a general definition of precision, and the accompanying reasoning principle, would have to capture assumptions about the exception handler and its concurrent context to ensure that they do not observe the above side effects.
More straightforwardly, the above definition of what becomes UNKNOWN would have to be codified, as that is not currently in the ASL architectural pseudocode.

Exceptions may also be \emph{imprecise},
in which case the behaviour is very loosely constrained,
and the current architecture does not give well-defined guarantees
in the presence of imprecise exceptions.
All exceptions in Arm are precise except for external memory errors which are not reported synchronously~(\S\ref{subsec:SEA}),
which we do not cover.

\section{Software-generated interrupts}\label{sec:sgis}

Inter-processor interrupts (IPIs),
known as software-generated interrupts (SGIs) on Arm,
are an important synchronisation mechanism available to software.
They are
used throughout systems software to signal other threads,
including within the Linux kernel (in its RCU synchronisation mechanism),
in software (via Linux's \asm{sys\_membarrier}),
e.g.\ in JITs~\cite{SimnerFPAPMS20},
and in programming language runtimes
(e.g.\ in Microsoft's%
Verona~\cite{CheesemanEtAl2023}).
Such use of SGIs critically depends
on a detailed understanding of the interaction of exceptions
with relaxed-memory behaviour.

To manage the sending, routing, prioritisation, and delivery of interrupts,
Arm define an optional \emph{generic interrupt controller} (GIC).
The GIC provides a uniform API for sending and routing interrupts from peripherals to threads,
and comes in several versions.
We focus on GICv3 and its CPU interface,
but expect
the behaviour we describe should apply to %
GICv4.

There are many interesting questions about SGIs.
We cover just a simple baseline:
enough to reason about the synchronisation used by software,
but ignoring much of the complexity of the GIC.
We fix a relatively simple configuration,
and focus on the relaxed-memory aspects of the interaction between SGIs and the rest of the memory and processor state.

\subsection{The Generic Interrupt Controller -- basic machinery}

We begin by introducing the context of the basic Arm GIC machinery, before addressing its relaxed ordering in later subsections.
An interrupt is \emph{generated} on its \emph{source} (a hardware thread or some peripheral)
for a particular \emph{event} (e.g.\ an SGI).
This interrupt is then sent to the interrupt controller,
which is split into
a distributor, the global machinery in charge of routing interrupts to cores,
and the per-thread redistributors,
each of which maintains a thread-local state for each interrupt (which we describe in more detail later).
Interrupts are identified in the GIC by its `interrupt ID number' (INTID).
Each instance of an interrupt sent to the interrupt controller is associated with an INTID,
either by software or a peripheral,
and is provided to the receiving core in a register it can read
(via acknowledgement, described later).

Each hardware thread (PE) has an interrupt status register (the \asm{ISR}),
which has a single pending status bit for each interrupt class (IRQ, FIQ, SError, etc).
For each fetch-decode-execute cycle of the top-level loop
(see \S\ref{subsec:fdx_instances}),
the processor checks these status bits
to determine whether an interrupt is pending;
if an interrupt is pending and is not masked on that PE, the PE takes that interrupt.
It is the interrupt controller's responsibility to set and clear the pending bit in that register,
notifying the thread of a pending interrupt.
To determine when to deliver (set the bit in the interrupt status register) interrupts to the core,
the redistributor maintains three key pieces of state
(this is for an `edge-triggered' interrupt, such as for SGIs;
  we do not discuss `level-sensitive' interrupts):
\begin{itemize}
  \item A priority to assign to each interrupt source,
  and the current `working' priority of the interrupt(s) being handled.
  \item A priority mask, which prevents interrupts with too low a priority from being delivered to the core.
  \item A per-INTID state, which is one of:
  \begin{itemize}
    \item Inactive: there is no current interrupt;
    \item Pending: the GIC has received an interrupt, and maybe delivered it, but the core has not begun handling it; or
    \item Active: the core has signalled it is handling the interrupt, but not yet signalled it is done.
  \end{itemize}
\end{itemize}

\paragraph{Lifecycle of an interrupt}

Interrupts start out Inactive.
When an interrupt is asserted by the source, the GIC sets the state for this interrupt's INTID to Pending.
Within some unspecified, finite amount of time,
the GIC will set the pending bit in the interrupt status register for the core,
enabling the core to take an exception on the next fetch-decode-execute loop.

The core should then \emph{acknowledge} the interrupt,
by reading the appropriate interrupt-acknowledge-register (IAR);
this returns the INTID for use by the core,
and sends a request to the redistributor to mark the INTID as Active.
Transitioning to the active state sets the working priority to the priority of that INTID's source,
preventing lower-priority interrupts from pre-empting the core,
and clears the pending bit in the interrupt-status-register on the core.
If another interrupt with the same INTID is asserted while the interrupt is active,
that instance will be buffered (only a single extra instance may be buffered)
and taken later,
and the INTID is said to be `Active and Pending'.
While the interrupt is active, it will not be re-delivered to the core,
so even if the interrupt service routine performs an \asm{ERET}, it will not re-take the exception.

\begin{figure}
\centering
\begin{tikzpicture}
\node (inactive) at (0,0) {\asm{Inactive}};
\node (pending) at (3.5,0) {\asm{Pending}};
\node [align=center] (activepending) at (6.9,-1) {\asm{Active \&}\\ \asm{pending}};
\node (active) at (6.9,1) {\asm{Active}};

\draw[->] (inactive) to [bend left=12]  node [xshift=5mm, midway, above, align=center, font=\tiny] {source asserts interrupt\\(eg by writing ICC\_SGI1R\_EL1);\\GIC delivers interrupt\\by setting pending bit in ISR} (pending);
\draw[->] (pending) to [bend left=12] node [midway, below, align=center, font=\tiny] {software changes pending state} (inactive);
\draw[->] (activepending) to [bend left=12] node [midway, below left,align=center, font=\tiny] {software deactivates interrupt} (pending);
\draw[->] (active) to [bend left=12] node [midway, right, align=center, font=\tiny] {re-pend \\ INTID} (activepending);
\draw[->] (activepending) to [bend left=12] node [midway, left, align=center, font=\tiny] {software changes \\ pending state} (active);
\draw[->] (pending) to node [pos=0.3, yshift=2mm, above, align=center, font=\tiny] {target acks interrupt\\by reading IAR;\\GIC unsets pending bit in ISR}  (active);
\draw[->] (active) to [out=165, in=55] node [midway, above, align=center, font=\tiny] {target deactivates interrupt\\by writing to EOIR/DIR\\(depending on EIOmode)} (inactive);
\end{tikzpicture}
\caption{GIC automaton, for each PE and each INTID,
based on Figure 4-3 ``Interrupt handling state machine'' from \citet[\S4.1.2]{arm-gic-v3-v4},
specialised to edge-triggered behaviour.
}
\label{fig:GIC-states}
\end{figure}
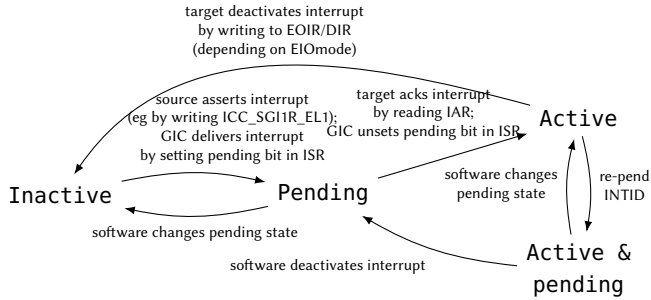

At some later time, the core may finish handling the interrupt and be ready to receive further instances of that INTID.
There are two ways to do this,
depending on whether one wants to separate \emph{priority drop} from \emph{deactivation},
which is controlled by the \asm{EOImode}.
With \asm{EOImode=0},
by writing the INTID to the end-of-interrupt register (EOIR),
the interrupt is deactivated simultaneously with the the priority drop.
With \asm{EOImode=1}, writes to the EOIR only perform priority drop,
requiring separate deactivation through a write to the deactivate-interrupt-register (DIR).
Additionally, the GIC interface provides registers which can manually set the current priority, or mask,
or explicitly set the state of an interrupt.
Figure~\ref{fig:GIC-states} shows the typical transitions between states.

\paragraph{Intended software usage}

Typically, software
use of interrupts
falls into one of two categories:
\begin{itemize}
  \item Nested interrupt servicing,
        where software readily uses priorities and handles the interrupt directly in the interrupt service routine,
        as it typical in real-time OSs.
  \item Deferred interrupt handling,
        where software acknowledges the interrupt directly, but handles it later.
\end{itemize}

Linux falls into the second category,
utilising only a single interrupt priority.
This `split' approach to handling interrupts,
where the interrupt service routine merely acknowledges,
and the actual handling of the interrupt comes later,
leads Linux to adopt \asm{EOImode=1}.
When the interrupt is taken by the core, it is acknowledged,
the INTID is checked against special cases,
priority is quickly dropped,
and interrupts are unmasked.
The actual interrupt may then be handled,
concurrently with new interrupts being signalled to the core,
although duplicates of the incident INTID will still be masked as it is not yet deactivated.
Eventually, the core completes the work for that interrupt and then deactivates it,
advancing the state machine. %

\subsection{Ordering of the propagation of SGIs}

An SGI is generated
by a write to the appropriate register
(e.g.\ \asm{ICC\_SGI1R\_EL1}),
and is received on one or several thread(s).
This gives rise to questions of three kinds:
\begin{itemize}
  \item What is required to order the generation of the SGI after earlier accesses?
  \item Does routing of the SGI imply ordering? e.g. is the interrupt controller an observer wrt. multi-copy-atomicity?
  \item What is required to ensure that the sequence of acknowledgement and deactivation happens correctly?
\end{itemize}

There are few guarantees about the order of propagation of SGIs, or interrupts generally.
Interrupts may be delivered to the core at any time,
and multiple pending interrupts may be delivered in any order
(priorities allowing).
There are no guarantees analogous to the coherence or atomicity of memory,
and generated interrupts may be re-ordered,
or delivered to different cores in different orders.
However, as discussed earlier, interrupts may not be speculated,
and so the interrupt cannot be delivered to the target PE before it is generated.

\paragraph{SGI litmus testing}
We extract the fundamental Message-Pass-via-SGI shape underlying Linux's implementation of RCU on Armv8
as a litmus test,
\testref{MPviaSGIEIOmode1sequence},
in Figure~\ref{fig:MPviaSGIEIOmode1sequence}.
Passing a message through an SGI requires some synchronisation between the
write of the data and the generation of the SGI (here a \asm{DSB ST} on Thread~0), and
requires observation of the data in the exception handler;
the SGI also needs to be is properly acknowledged and deactivated,
with the appropriate barriers.

\begin{figure}
\scalebox{0.8}{\input{tests/MPviaSGIEIOmode1sequence}}
  \caption{MPviaSGIEIOmode1sequence: Synchronisation-via-SGI
  with the full acknowledge-drop-deactivate sequence
  appropriate for \asm{EOImode=1}.}
 \label{test:MPviaSGIEIOmode1sequence}
  \label{fig:MPviaSGIEIOmode1sequence}
\end{figure}

\begin{figure}
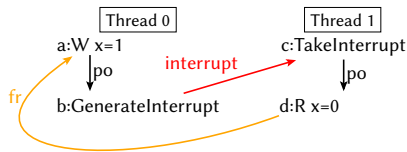
%
  \centering
\scalebox{0.75}{
\begin{tabular}{c}
    \input{tests/MPviaSGI.listing.tikz} \\
    \input{tests/MPviaSGI.diagram.tikz}
  \end{tabular}}
  \label{test:MPviaSGI}

  \caption{\asm{MPviaSGI}: message passing via SGI,
  illustrating two potential phenomena:
  (1)~On the writer side: a po-earlier write
  gets reordered with
  a po-later GenerateInterrupt.
  (2)~On the reader side:
   a po-earlier TakeInterrupt
   gets reordered with
   a po-later read (from the interrupt handler).
  }
  \label{fig:MPviaSGI}
\end{figure}

This test is composed of two interacting parts:
the part that imposes the ordering between the write and the read of the data,
and the part that interacts with the GIC to manage the interrupt.
Figure~\ref{fig:MPviaSGI} asks the most basic question of this shape:
if we try pass a message via an SGI, without any further synchronisation,
can we still read an old value?
The answer is yes,
because the generation and subsequent delivery of the SGI could happen before the propagation of the store.
On the other hand,
the extensive synchronisation on the receiving thread imposed by GIC management is accidental
for the read, which is already strongly ordered after the taking of the exception.

\subsection{Software usage of SGIs}

Synchronisation mechanisms
like those discussed above
rely on this link between memory accesses and interrupts
to achieve low-overhead synchronisation.
More specifically,
they push the cost away from normal memory accesses
and onto a ``system-wide memory barrier'' implemented using interrupts.
This is a fork-join barrier, not a fence.
Interestingly, RCU and the Verona asymmetric lock rely on two different aspects
of this system-wide memory barrier:
RCU relies on masking of interrupts to implement cheap read critical sections,
whereas the Verona asymmetric lock relies on precision of interrupts~(\S\ref{sec:prec}).

\paragraph{System-wide memory barrier}
This system-wide memory barrier
is a two-way barrier:
the issuing PE %
notifies all other PEs,
and waits for a reply from all of them.
The notification is implemented using interrupts,
relying on the ordering described above, which
is guaranteed by Arm-A.
In Kernel RCU
(where this barrier forms the core of \asm{synchronize\_rcu},
exposed to userland as the \asm{sys\_membarrier} syscall),
the wait for a reply is implemented using memory operations,
namely a lock-protected counter that threads increment to acknowledge receipt of the interrupt.
We simplify this (to a write to a flag) in our litmus tests to reduce complexity.

\begin{figure}
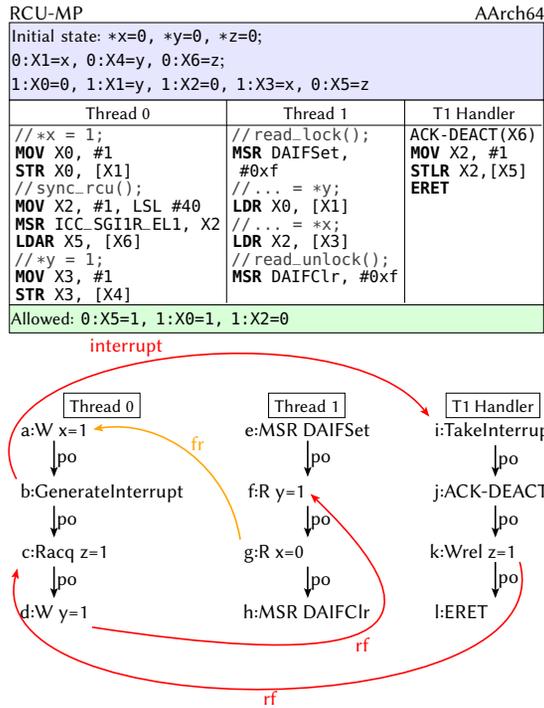

  \centering
\begin{center}
\scalebox{0.75}{\input{tests/RCU-MP.listing.tikz}}

\vspace*{-2.0\baselineskip}

\scalebox{0.75}{\input{tests/RCU-MP.diagram.tikz}}
\end{center}
  \label{test:RCU-MP}
\vspace*{-4.0\baselineskip}

  \caption{\asm{RCU-MP}: the key test of RCU:
  are two writes separated by the generation of an SGI
  ordered with respect to a read critical section
  implemented via interrupts masking?
  \\
  With a \asm{DSB ST} between \asm{a} and \asm{b}, this is forbidden.
  \\
  }
  \label{fig:RCU-MP}
\end{figure}

\paragraph{RCU}
The key concept of RCU
is that of a grace period~\cite{rcutxt}\cite[\S9]{perfbook},
as captured by \citet{AlglaveMMPS18}
in the
\asm{RCU-MP}
litmus test
(Figure~\ref{fig:RCU-MP}).

We focus on the use of interrupts in Kernel RCU.
For performance, RCU also relies on
address dependencies to implement cheap ordering in read sections,
but that is already explained
in the `user' model of Arm-A~\cite{DBLP:conf/popl/FlurGPSSMDS16,PulteFDFSS18} by \asm{MP+dmbst+addr}.

At the level of Arm assembly,
the \asm{synchronize\_rcu} system-wide memory barrier is decomposed into a \asm{DSB ST}
followed by an \asm{MSR} to \asm{SGI1R},
and a wait for the acknowledgement
(in our cut-down tests, a read acquire of the ack flag);
entering the read critical section via \asm{rcu\_read\_lock}
and leaving it via \asm{rcu\_read\_unlock}
decompose to writes to the \asm{DAIF} (pseudo)register
that
mask and unmask interrupts.

The crux of this litmus test is that interrupts are masked between the two reads,
and that the handler is therefore either before both reads, or after both reads, but not in between
(as in, no event of the handler is in between the two reads in program order).
At the Linux C level, this masking ensures that the interrupt generated by
the \asm{synchronize\_rcu} system-wide memory barrier
is taken either before or after the read section,
but not during,
providing the basis for mutual exclusion.
In the litmus tests, this is captured by the fact that if the read of the flag \asm{y} sees the flag,
the read of the data \asm{x} sees the new data.

\paragraph{Verona asymmetric lock}
We capture the key scenario of the asymmetric lock of Verona~\cite{mattp}
(and of `biased locking' and
`asymmetric Dekker synchronisation'%
~\cite{biasedlocking,biasedlockingpatent,deprecatebiasedlocking,Kawachiya02,Russell06,Burrows2004,AsymmetricDekkerSynchronization2001,Kawachiya05} as used in the JVM).
It occurs when an `internal acquire' from the (unique) owner thread
contends with an `external acquire'
from another thread.
The internal acquire is meant to be cheap,
and only involves writing to an `external' flag to express interest,
and then, in program order, reading from an `internal' flag to ensure that
other threads have not expressed interest
(falling onto the slow path if they have).
Crucially, in C++, there is a \asm{Barrier::compiler()} %
that prevents reordering of two instructions
by the compiler, but does not appear in the generated assembly.
The external acquire does the symmetric thing,
writing on the `internal' flag to express interest,
and then reading from the `external' flag to ensure that the owner
has not expressed interest.
To order this, it uses a \asm{Barrier::memory()},
which involves a \asm{FlushProcessWriteBuffers()},
which on Linux
is implemented using a \asm{sys\_membarrier},
which essentially boils down to a \asm{synchronize\_rcu}.

The key guarantee that is relied in the `cheap' thread is that the interrupt
must be taken precisely,
and that it is therefore taken, in program order, either entirely before the read of the internal flag,
entirely between the read of the internal flag and the write to the external flag,
or entirely after the write to the external flag.
In all three cases,
the system-wide memory barrier
ensures that at least one of the two threads
must see that the other thread has expressed interest
(must read the recent write),
and therefore backs off,
ensuring mutual exclusion.

\subsection{Ordering of GIC register writes}

The Arm GIC Architecture Specification text (IHI 0069H.b)
is reasonably clear about the relaxed ordering of GIC events induced by accesses to GIC registers with program-order later events (12.1.6 ``Observability of the effects of accesses to the GIC registers''), though there are still subtle requirements for barriers.
A \asm{DSB.SY} enforces
ordering of GIC events
(generate, acknowledge, drop priority, and deactivate)
induced by accesses to GIC registers
(SGI1R, IAR, EOIR, DIR)
with program-order-later events,
as they are such effects.
DSBs are not needed to merely order the register accesses themselves.

An ISB ensures that any pending interrupts are taken before executing the program-order later instructions.

If there was an interrupt in the Active and Pending state at deactivation,
then it is immediately re-pended on the PE
(and so delivery can immediately happen again).
But, if there is no DSB between the write of the deactivation and the context synchronisation,
it might be that the assertion and delivery did not yet occur,
causing the interrupt to be taken later.

\subsection{A draft axiomatic extension}

We give a draft extension
to the previous axiomatic model to support inter-processor interrupts,
noting the challenges.

\paragraph{GIC candidates}
Unlike with most of the instruction semantics,
there is very little public ASL from Arm which describes %
the priority and INTID state machine system.
While much of the GIC's machinery,
routing, virtualisation and so on,
is not required to discuss the usage of interrupts here,
a large quantity of the base GIC architecture would need to be turned into ASL
and incorporated into the machinery.
The rest of this extension assumes one has such machinery in place.

First, we must extend the thread semantics:
reads and writes of the registers of the CPU interface to the GIC, and interrupt status register,
must be treated differently than other registers,
lifting them to the memory model with a relation constraining the values they could read,
analogous to `reads-from'.
This allows us to tie the thread's events interacting with the GIC, with those events coming from the GIC ASL.

We add the following new events, grouped as \herd{GICEvents}:
\begin{itemize}
  \item \herd{GenerateInterrupt}, for the GIC action from writing the \asm{SGI1R} register,
        which sends an IPI to other cores. It is associated with a \emph{target} set of CPUs.
  \item \herd{Acknowledge}, for the relevant effect in the GIC,
        i.e.~the state machine change and related updates to registers.
        Here, we assume the GIC update is atomic, which ought to be true for simple physical SGIs.
  \item \herd{DropPriority} and \herd{Deactivate}, for the relevant effects on the GIC state machine and priority masking.
\end{itemize}

These new events are placed \herd{iio}-\emph{after} (intra-instruction-ordered) the respective register events.
Such events could instead be inserted into \herd{po}, with suitable modification of the previous relations,
although for simplicity here we do not.

\paragraph{Interrupt witness}

We add a new existentially-quantified relation to the witness: \herd{interrupt}.
This associates the \herd{TakeInterrupt} with the \herd{GenerateInterrupt} which caused it,
constraining any program-order-later \herd{Acknowledge} and corresponding \herd{MRS} event INTID values.
This effectively assigns the INTID at the point the interrupt is taken,
and makes \herd{interrupt} behave like \herd{rf} for INTIDs;
if the INTID is never read, one must consider all possible interrupt sources.

\paragraph{Update to relations and axioms}

The update to the relations is then fairly straightforward:
insert \herd{interrupt} into \herd{ob},
and make \asm{DSB} instructions order GIC events in program-order.
We do not put \asm{GICEvent}s in program order
to express that they may execute out-of-order with respect to other events in the same thread,
including context-synchronisation,
unless explicitly ordered (e.g.\ by DSBs).

\section{Conclusion}

We identify an open problem in giving a definition of precision on relaxed architectures,
and describe the challenge in doing so.
We characterise some basic guarantees of precision,
which should make it possible to apply some of the abstraction techniques
used to reason about nesting of interrupts~\cite{Kroening:2015:EVL, Liang:2017:EVL}.

We extend the Arm-A memory model
to cover exceptions, an important aspect of defining the architectural interface,
clarifying the behaviour at that interface,
and giving an executable-as-a-test-oracle implementation of an axiomatic model
usable as an exploration tool to investigate the effect of synchronisation
on hardware exceptions and interrupts.
We describe the interaction of hardware exceptions with memory errors,
and the consequences %
on the user model.

We begin building %
a model
for software-generated interrupts and the required parts of the interrupt machinery
relied upon by the common computing base,
giving the key shapes and litmus tests, some baseline behaviours of the Arm GIC,
and a draft extension that covers key use cases.

Although there is much work still to do
on exceptions, interrupts, and their interaction with other features,
this work creates a robust foundation that future work can build on.

\ifanon
\else
\section*{Acknowledgements}
We thank Richard Grisenthwaite (Arm EVP, Chief Architect, and Fellow), Martin Weidmann (Director of Product Management, Arm Architecture and Technology Group), and Will Deacon (Google) for detailed discussions about the Arm architecture. We thank Ben Laurie and Sarah de Haas (Google) for their support.

This work was funded in part by Google.
This work was funded in part by Arm.
This work was funded in part by an AUFF starter grant (Pichon-Pharabod).
This work was funded in part by two Amazon Research Awards (Pichon-Pharabod; Sewell and Simner).
This work was funded in part by UK Research and Innovation (UKRI) under the UK
government's Horizon Europe funding guarantee for ERC-AdG-2022, EP/Y035976/1 SAFER.
This project has received funding from the European Research Council
(ERC) under the European Union's Horizon 2020 research and innovation programme (grant agreement No 789108, ERC-AdG-2017 ELVER).
This work is supported by ERC-2024-POC grant ELVER-CHECK, 101189371.
Funded by the European Union. Views and opinions expressed are however those of the author(s) only and do not necessarily reflect those of the European Union or the European Research Council Executive Agency. Neither the European Union nor the granting authority can be held responsible for them.
This work was supported in part by the Innovate UK project Digital Security by Design (DSbD) Technology Platform Prototype, 105694.
The authors would like to thank the Isaac Newton Institute for Mathematical Sciences, Cambridge, for support and hospitality during the programme Big Specification, where work on this paper was undertaken. This work was supported by EPSRC grant EP/Z000580/1.
This work was funded in part by a Royal Society University Research Fellowship.
One of the authors has received funding from the UK Advanced Research and Innovation Agency (ARIA) as part of the project Qbs4Safety: Core Representation Underlying Safeguarded AI.

\goodbreak
\fi %

\ifextended
\clearpage
\appendix
\section{All experimental results}
\FloatBarrier
Figure~\ref{app:hw-results} gives our experimental results.

\begin{figure}[H]
{
\scriptsize
\begin{longtable}{l | l l l l l | l l l l}
\textbf{Name} & \multicolumn{5}{c}{\textbf{Arch Intent}} & \multicolumn{4}{c}{\textbf{Hardware}} \\
               & \tableheader{Base} & \tableheader{ExS} & \tableheader{SEA\_R} & \tableheader{SEA\_W} & \tableheader{SEA\_RW} & \tableheader{RPi 3B+} & \tableheader{RPi 4B} & \tableheader{RPi 5} & \tableheader{ODROID N2+ (big)}  \\
  \tableTestName{LB+svc-dmb-erets}{LB+svc-dmb-erets} & \csname ref_none-LB+svc-dmb-erets ref\endcsname & \csname ref_exs-LB+svc-dmb-erets ref\endcsname & \csname ref_sea_r-LB+svc-dmb-erets ref\endcsname & \csname ref_sea_w-LB+svc-dmb-erets ref\endcsname & \csname ref_sea_rw-LB+svc-dmb-erets ref\endcsname & \csname RPi 3B+-LB+svc-dmb-erets result\endcsname & \csname RPi 4B-LB+svc-dmb-erets result\endcsname & \csname RPi 5-LB+svc-dmb-erets result\endcsname & \csname ODROID N2+ (big)-LB+svc-dmb-erets result\endcsname\\
  \tableTestName{LB+svc-erets}{LB+svc-erets} & \csname ref_none-LB+svc-erets ref\endcsname & \csname ref_exs-LB+svc-erets ref\endcsname & \csname ref_sea_r-LB+svc-erets ref\endcsname & \csname ref_sea_w-LB+svc-erets ref\endcsname & \csname ref_sea_rw-LB+svc-erets ref\endcsname & \csname RPi 3B+-LB+svc-erets result\endcsname & \csname RPi 4B-LB+svc-erets result\endcsname & \csname RPi 5-LB+svc-erets result\endcsname & \csname ODROID N2+ (big)-LB+svc-erets result\endcsname\\
  \tableTestName{LB+svcs}{LB+svcs} & \csname ref_none-LB+svcs ref\endcsname & \csname ref_exs-LB+svcs ref\endcsname & \csname ref_sea_r-LB+svcs ref\endcsname & \csname ref_sea_w-LB+svcs ref\endcsname & \csname ref_sea_rw-LB+svcs ref\endcsname & \csname RPi 3B+-LB+svcs result\endcsname & \csname RPi 4B-LB+svcs result\endcsname & \csname RPi 5-LB+svcs result\endcsname & \csname ODROID N2+ (big)-LB+svcs result\endcsname\\
  \tableTestName{MP+daifset+dmb}{MP+daifset+dmb} & \csname ref_none-MP+daifset+dmb ref\endcsname & \csname ref_exs-MP+daifset+dmb ref\endcsname & \csname ref_sea_r-MP+daifset+dmb ref\endcsname & \csname ref_sea_w-MP+daifset+dmb ref\endcsname & \csname ref_sea_rw-MP+daifset+dmb ref\endcsname & ? & ? & ? & ?\\
  \tableTestName{MP+dmb+ctrl-eret}{MP+dmb+ctrl-eret} & \csname ref_none-MP+dmb+ctrl-eret ref\endcsname & \csname ref_exs-MP+dmb+ctrl-eret ref\endcsname & \csname ref_sea_r-MP+dmb+ctrl-eret ref\endcsname & \csname ref_sea_w-MP+dmb+ctrl-eret ref\endcsname & \csname ref_sea_rw-MP+dmb+ctrl-eret ref\endcsname & \csname RPi 3B+-MP+dmb+ctrl-eret result\endcsname & \csname RPi 4B-MP+dmb+ctrl-eret result\endcsname & \csname RPi 5-MP+dmb+ctrl-eret result\endcsname & \csname ODROID N2+ (big)-MP+dmb+ctrl-eret result\endcsname\\
  \tableTestName{MP+dmb+ctrl-int}{MP+dmb+ctrl-int} & \csname ref_none-MP+dmb+ctrl-int ref\endcsname & \csname ref_exs-MP+dmb+ctrl-int ref\endcsname & \csname ref_sea_r-MP+dmb+ctrl-int ref\endcsname & \csname ref_sea_w-MP+dmb+ctrl-int ref\endcsname & \csname ref_sea_rw-MP+dmb+ctrl-int ref\endcsname & ? & ? & ? & ?\\
  \tableTestName{MP+dmb+ctrl-rfisvceret-addr}{MP+dmb+ctrl-rfisvceret-addr} & \csname ref_none-MP+dmb+ctrl-rfisvceret-addr ref\endcsname & \csname ref_exs-MP+dmb+ctrl-rfisvceret-addr ref\endcsname & \csname ref_sea_r-MP+dmb+ctrl-rfisvceret-addr ref\endcsname & \csname ref_sea_w-MP+dmb+ctrl-rfisvceret-addr ref\endcsname & \csname ref_sea_rw-MP+dmb+ctrl-rfisvceret-addr ref\endcsname & \csname RPi 3B+-MP+dmb+ctrl-rfisvceret-addr result\endcsname & \csname RPi 4B-MP+dmb+ctrl-rfisvceret-addr result\endcsname & \csname RPi 5-MP+dmb+ctrl-rfisvceret-addr result\endcsname & \csname ODROID N2+ (big)-MP+dmb+ctrl-rfisvceret-addr result\endcsname\\
  \tableTestName{MP+dmb+ctrl-svc}{MP+dmb+ctrl-svc} & \csname ref_none-MP+dmb+ctrl-svc ref\endcsname & \csname ref_exs-MP+dmb+ctrl-svc ref\endcsname & \csname ref_sea_r-MP+dmb+ctrl-svc ref\endcsname & \csname ref_sea_w-MP+dmb+ctrl-svc ref\endcsname & \csname ref_sea_rw-MP+dmb+ctrl-svc ref\endcsname & \csname RPi 3B+-MP+dmb+ctrl-svc result\endcsname & \csname RPi 4B-MP+dmb+ctrl-svc result\endcsname & \csname RPi 5-MP+dmb+ctrl-svc result\endcsname & \csname ODROID N2+ (big)-MP+dmb+ctrl-svc result\endcsname\\
  \tableTestName{MP+dmb+ctrlelr}{MP+dmb+ctrlelr} & \csname ref_none-MP+dmb+ctrlelr ref\endcsname & \csname ref_exs-MP+dmb+ctrlelr ref\endcsname & \csname ref_sea_r-MP+dmb+ctrlelr ref\endcsname & \csname ref_sea_w-MP+dmb+ctrlelr ref\endcsname & \csname ref_sea_rw-MP+dmb+ctrlelr ref\endcsname & \csname RPi 3B+-MP+dmb+ctrlelr result\endcsname & \csname RPi 4B-MP+dmb+ctrlelr result\endcsname & \csname RPi 5-MP+dmb+ctrlelr result\endcsname & \csname ODROID N2+ (big)-MP+dmb+ctrlelr result\endcsname\\
  \tableTestName{MP+dmb+daifset}{MP+dmb+daifset} & \csname ref_none-MP+dmb+daifset ref\endcsname & \csname ref_exs-MP+dmb+daifset ref\endcsname & \csname ref_sea_r-MP+dmb+daifset ref\endcsname & \csname ref_sea_w-MP+dmb+daifset ref\endcsname & \csname ref_sea_rw-MP+dmb+daifset ref\endcsname & ? & ? & ? & ?\\
  \tableTestName{MP+dmb+data-svc}{MP+dmb+data-svc} & ? & ? & ? & ? & ? & \csname RPi 3B+-MP+dmb+data-svc result\endcsname & \csname RPi 4B-MP+dmb+data-svc result\endcsname & \csname RPi 5-MP+dmb+data-svc result\endcsname & \csname ODROID N2+ (big)-MP+dmb+data-svc result\endcsname\\
  \tableTestName{MP+dmb+dmb-eret}{MP+dmb+dmb-eret} & \csname ref_none-MP+dmb+dmb-eret ref\endcsname & \csname ref_exs-MP+dmb+dmb-eret ref\endcsname & \csname ref_sea_r-MP+dmb+dmb-eret ref\endcsname & \csname ref_sea_w-MP+dmb+dmb-eret ref\endcsname & \csname ref_sea_rw-MP+dmb+dmb-eret ref\endcsname & \csname RPi 3B+-MP+dmb+dmb-eret result\endcsname & \csname RPi 4B-MP+dmb+dmb-eret result\endcsname & \csname RPi 5-MP+dmb+dmb-eret result\endcsname & \csname ODROID N2+ (big)-MP+dmb+dmb-eret result\endcsname\\
  \tableTestName{MP+dmb+eret}{MP+dmb+eret} & \csname ref_none-MP+dmb+eret ref\endcsname & \csname ref_exs-MP+dmb+eret ref\endcsname & \csname ref_sea_r-MP+dmb+eret ref\endcsname & \csname ref_sea_w-MP+dmb+eret ref\endcsname & \csname ref_sea_rw-MP+dmb+eret ref\endcsname & \csname RPi 3B+-MP+dmb+eret result\endcsname & \csname RPi 4B-MP+dmb+eret result\endcsname & \csname RPi 5-MP+dmb+eret result\endcsname & \csname ODROID N2+ (big)-MP+dmb+eret result\endcsname\\
  \tableTestName{MP+dmb+eret-dmb}{MP+dmb+eret-dmb} & \csname ref_none-MP+dmb+eret-dmb ref\endcsname & \csname ref_exs-MP+dmb+eret-dmb ref\endcsname & \csname ref_sea_r-MP+dmb+eret-dmb ref\endcsname & \csname ref_sea_w-MP+dmb+eret-dmb ref\endcsname & \csname ref_sea_rw-MP+dmb+eret-dmb ref\endcsname & \csname RPi 3B+-MP+dmb+eret-dmb result\endcsname & \csname RPi 4B-MP+dmb+eret-dmb result\endcsname & \csname RPi 5-MP+dmb+eret-dmb result\endcsname & \csname ODROID N2+ (big)-MP+dmb+eret-dmb result\endcsname\\
  \tableTestName{MP+dmb+eret-svc}{MP+dmb+eret-svc} & \csname ref_none-MP+dmb+eret-svc ref\endcsname & \csname ref_exs-MP+dmb+eret-svc ref\endcsname & \csname ref_sea_r-MP+dmb+eret-svc ref\endcsname & \csname ref_sea_w-MP+dmb+eret-svc ref\endcsname & \csname ref_sea_rw-MP+dmb+eret-svc ref\endcsname & \csname RPi 3B+-MP+dmb+eret-svc result\endcsname & \csname RPi 4B-MP+dmb+eret-svc result\endcsname & \csname RPi 5-MP+dmb+eret-svc result\endcsname & \csname ODROID N2+ (big)-MP+dmb+eret-svc result\endcsname\\
  \tableTestName{MP+dmb+eret=addr}{MP+dmb+eret=addr} & \csname ref_none-MP+dmb+eret=addr ref\endcsname & \csname ref_exs-MP+dmb+eret=addr ref\endcsname & \csname ref_sea_r-MP+dmb+eret=addr ref\endcsname & \csname ref_sea_w-MP+dmb+eret=addr ref\endcsname & \csname ref_sea_rw-MP+dmb+eret=addr ref\endcsname & \csname RPi 3B+-MP+dmb+eret=addr result\endcsname & \csname RPi 4B-MP+dmb+eret=addr result\endcsname & \csname RPi 5-MP+dmb+eret=addr result\endcsname & \csname ODROID N2+ (big)-MP+dmb+eret=addr result\endcsname\\
  \tableTestName{MP+dmb+int}{MP+dmb+int} & \csname ref_none-MP+dmb+int ref\endcsname & \csname ref_exs-MP+dmb+int ref\endcsname & \csname ref_sea_r-MP+dmb+int ref\endcsname & \csname ref_sea_w-MP+dmb+int ref\endcsname & \csname ref_sea_rw-MP+dmb+int ref\endcsname & ? & ? & ? & ?\\
  \tableTestName{MP+dmb+svc}{MP+dmb+svc} & \csname ref_none-MP+dmb+svc ref\endcsname & \csname ref_exs-MP+dmb+svc ref\endcsname & \csname ref_sea_r-MP+dmb+svc ref\endcsname & \csname ref_sea_w-MP+dmb+svc ref\endcsname & \csname ref_sea_rw-MP+dmb+svc ref\endcsname & \csname RPi 3B+-MP+dmb+svc result\endcsname & \csname RPi 4B-MP+dmb+svc result\endcsname & \csname RPi 5-MP+dmb+svc result\endcsname & \csname ODROID N2+ (big)-MP+dmb+svc result\endcsname\\
  \tableTestName{MP+dmb+svc-addreret}{MP+dmb+svc-addreret} & \csname ref_none-MP+dmb+svc-addreret ref\endcsname & \csname ref_exs-MP+dmb+svc-addreret ref\endcsname & \csname ref_sea_r-MP+dmb+svc-addreret ref\endcsname & \csname ref_sea_w-MP+dmb+svc-addreret ref\endcsname & \csname ref_sea_rw-MP+dmb+svc-addreret ref\endcsname & \csname RPi 3B+-MP+dmb+svc-addreret result\endcsname & \csname RPi 4B-MP+dmb+svc-addreret result\endcsname & \csname RPi 5-MP+dmb+svc-addreret result\endcsname & \csname ODROID N2+ (big)-MP+dmb+svc-addreret result\endcsname\\
  \tableTestName{MP+dmb+svc-dmb}{MP+dmb+svc-dmb} & \csname ref_none-MP+dmb+svc-dmb ref\endcsname & \csname ref_exs-MP+dmb+svc-dmb ref\endcsname & \csname ref_sea_r-MP+dmb+svc-dmb ref\endcsname & \csname ref_sea_w-MP+dmb+svc-dmb ref\endcsname & \csname ref_sea_rw-MP+dmb+svc-dmb ref\endcsname & \csname RPi 3B+-MP+dmb+svc-dmb result\endcsname & \csname RPi 4B-MP+dmb+svc-dmb result\endcsname & \csname RPi 5-MP+dmb+svc-dmb result\endcsname & \csname ODROID N2+ (big)-MP+dmb+svc-dmb result\endcsname\\
  \tableTestName{MP+dmb+svc-dmb-eret}{MP+dmb+svc-dmb-eret} & \csname ref_none-MP+dmb+svc-dmb-eret ref\endcsname & \csname ref_exs-MP+dmb+svc-dmb-eret ref\endcsname & \csname ref_sea_r-MP+dmb+svc-dmb-eret ref\endcsname & \csname ref_sea_w-MP+dmb+svc-dmb-eret ref\endcsname & \csname ref_sea_rw-MP+dmb+svc-dmb-eret ref\endcsname & \csname RPi 3B+-MP+dmb+svc-dmb-eret result\endcsname & \csname RPi 4B-MP+dmb+svc-dmb-eret result\endcsname & \csname RPi 5-MP+dmb+svc-dmb-eret result\endcsname & \csname ODROID N2+ (big)-MP+dmb+svc-dmb-eret result\endcsname\\
  \tableTestName{MP+dmb+svc-eret}{MP+dmb+svc-eret} & \csname ref_none-MP+dmb+svc-eret ref\endcsname & \csname ref_exs-MP+dmb+svc-eret ref\endcsname & \csname ref_sea_r-MP+dmb+svc-eret ref\endcsname & \csname ref_sea_w-MP+dmb+svc-eret ref\endcsname & \csname ref_sea_rw-MP+dmb+svc-eret ref\endcsname & \csname RPi 3B+-MP+dmb+svc-eret result\endcsname & \csname RPi 4B-MP+dmb+svc-eret result\endcsname & \csname RPi 5-MP+dmb+svc-eret result\endcsname & \csname ODROID N2+ (big)-MP+dmb+svc-eret result\endcsname\\
  \tableTestName{MP+dmb+svcnoeis}{MP+dmb+svcnoeis} & \csname ref_none-MP+dmb+svcnoeis ref\endcsname & \csname ref_exs-MP+dmb+svcnoeis ref\endcsname & \csname ref_sea_r-MP+dmb+svcnoeis ref\endcsname & \csname ref_sea_w-MP+dmb+svcnoeis ref\endcsname & \csname ref_sea_rw-MP+dmb+svcnoeis ref\endcsname & \csname RPi 3B+-MP+dmb+svcnoeis result\endcsname & \csname RPi 4B-MP+dmb+svcnoeis result\endcsname & \csname RPi 5-MP+dmb+svcnoeis result\endcsname & \csname ODROID N2+ (big)-MP+dmb+svcnoeis result\endcsname\\
  \tableTestName{MP+eret+addr}{MP+eret+addr} & \csname ref_none-MP+eret+addr ref\endcsname & \csname ref_exs-MP+eret+addr ref\endcsname & \csname ref_sea_r-MP+eret+addr ref\endcsname & \csname ref_sea_w-MP+eret+addr ref\endcsname & \csname ref_sea_rw-MP+eret+addr ref\endcsname & \csname RPi 3B+-MP+eret+addr result\endcsname & \csname RPi 4B-MP+eret+addr result\endcsname & \csname RPi 5-MP+eret+addr result\endcsname & \csname ODROID N2+ (big)-MP+eret+addr result\endcsname\\
  \tableTestName{MP+eret+dmb}{MP+eret+dmb} & \csname ref_none-MP+eret+dmb ref\endcsname & \csname ref_exs-MP+eret+dmb ref\endcsname & \csname ref_sea_r-MP+eret+dmb ref\endcsname & \csname ref_sea_w-MP+eret+dmb ref\endcsname & \csname ref_sea_rw-MP+eret+dmb ref\endcsname & \csname RPi 3B+-MP+eret+dmb result\endcsname & \csname RPi 4B-MP+eret+dmb result\endcsname & \csname RPi 5-MP+eret+dmb result\endcsname & \csname ODROID N2+ (big)-MP+eret+dmb result\endcsname\\
  \tableTestName{MP+eret+svc}{MP+eret+svc} & \csname ref_none-MP+eret+svc ref\endcsname & \csname ref_exs-MP+eret+svc ref\endcsname & \csname ref_sea_r-MP+eret+svc ref\endcsname & \csname ref_sea_w-MP+eret+svc ref\endcsname & \csname ref_sea_rw-MP+eret+svc ref\endcsname & \csname RPi 3B+-MP+eret+svc result\endcsname & \csname RPi 4B-MP+eret+svc result\endcsname & \csname RPi 5-MP+eret+svc result\endcsname & \csname ODROID N2+ (big)-MP+eret+svc result\endcsname\\
  \tableTestName{MP+erets}{MP+erets} & \csname ref_none-MP+erets ref\endcsname & \csname ref_exs-MP+erets ref\endcsname & \csname ref_sea_r-MP+erets ref\endcsname & \csname ref_sea_w-MP+erets ref\endcsname & \csname ref_sea_rw-MP+erets ref\endcsname & \csname RPi 3B+-MP+erets result\endcsname & \csname RPi 4B-MP+erets result\endcsname & \csname RPi 5-MP+erets result\endcsname & \csname ODROID N2+ (big)-MP+erets result\endcsname\\
  \tableTestName{MP+int+dmb}{MP+int+dmb} & \csname ref_none-MP+int+dmb ref\endcsname & \csname ref_exs-MP+int+dmb ref\endcsname & \csname ref_sea_r-MP+int+dmb ref\endcsname & \csname ref_sea_w-MP+int+dmb ref\endcsname & \csname ref_sea_rw-MP+int+dmb ref\endcsname & ? & ? & ? & ?\\
  \tableTestName{MP+svc+addr}{MP+svc+addr} & \csname ref_none-MP+svc+addr ref\endcsname & \csname ref_exs-MP+svc+addr ref\endcsname & \csname ref_sea_r-MP+svc+addr ref\endcsname & \csname ref_sea_w-MP+svc+addr ref\endcsname & \csname ref_sea_rw-MP+svc+addr ref\endcsname & \csname RPi 3B+-MP+svc+addr result\endcsname & \csname RPi 4B-MP+svc+addr result\endcsname & \csname RPi 5-MP+svc+addr result\endcsname & \csname ODROID N2+ (big)-MP+svc+addr result\endcsname\\
  \tableTestName{MP+svc+dmb}{MP+svc+dmb} & \csname ref_none-MP+svc+dmb ref\endcsname & \csname ref_exs-MP+svc+dmb ref\endcsname & \csname ref_sea_r-MP+svc+dmb ref\endcsname & \csname ref_sea_w-MP+svc+dmb ref\endcsname & \csname ref_sea_rw-MP+svc+dmb ref\endcsname & \csname RPi 3B+-MP+svc+dmb result\endcsname & \csname RPi 4B-MP+svc+dmb result\endcsname & \csname RPi 5-MP+svc+dmb result\endcsname & \csname ODROID N2+ (big)-MP+svc+dmb result\endcsname\\
  \tableTestName{MP+svc+eret}{MP+svc+eret} & \csname ref_none-MP+svc+eret ref\endcsname & \csname ref_exs-MP+svc+eret ref\endcsname & \csname ref_sea_r-MP+svc+eret ref\endcsname & \csname ref_sea_w-MP+svc+eret ref\endcsname & \csname ref_sea_rw-MP+svc+eret ref\endcsname & \csname RPi 3B+-MP+svc+eret result\endcsname & \csname RPi 4B-MP+svc+eret result\endcsname & \csname RPi 5-MP+svc+eret result\endcsname & \csname ODROID N2+ (big)-MP+svc+eret result\endcsname\\
  \tableTestName{MP+svc-dmb+addr}{MP+svc-dmb+addr} & \csname ref_none-MP+svc-dmb+addr ref\endcsname & \csname ref_exs-MP+svc-dmb+addr ref\endcsname & \csname ref_sea_r-MP+svc-dmb+addr ref\endcsname & \csname ref_sea_w-MP+svc-dmb+addr ref\endcsname & \csname ref_sea_rw-MP+svc-dmb+addr ref\endcsname & \csname RPi 3B+-MP+svc-dmb+addr result\endcsname & \csname RPi 4B-MP+svc-dmb+addr result\endcsname & \csname RPi 5-MP+svc-dmb+addr result\endcsname & \csname ODROID N2+ (big)-MP+svc-dmb+addr result\endcsname\\
  \tableTestName{MP+svc-dmb-eret+addr}{MP+svc-dmb-eret+addr} & \csname ref_none-MP+svc-dmb-eret+addr ref\endcsname & \csname ref_exs-MP+svc-dmb-eret+addr ref\endcsname & \csname ref_sea_r-MP+svc-dmb-eret+addr ref\endcsname & \csname ref_sea_w-MP+svc-dmb-eret+addr ref\endcsname & \csname ref_sea_rw-MP+svc-dmb-eret+addr ref\endcsname & \csname RPi 3B+-MP+svc-dmb-eret+addr result\endcsname & \csname RPi 4B-MP+svc-dmb-eret+addr result\endcsname & \csname RPi 5-MP+svc-dmb-eret+addr result\endcsname & \csname ODROID N2+ (big)-MP+svc-dmb-eret+addr result\endcsname\\
  \tableTestName{MP+svc-eret+addr}{MP+svc-eret+addr} & \csname ref_none-MP+svc-eret+addr ref\endcsname & \csname ref_exs-MP+svc-eret+addr ref\endcsname & \csname ref_sea_r-MP+svc-eret+addr ref\endcsname & \csname ref_sea_w-MP+svc-eret+addr ref\endcsname & \csname ref_sea_rw-MP+svc-eret+addr ref\endcsname & \csname RPi 3B+-MP+svc-eret+addr result\endcsname & \csname RPi 4B-MP+svc-eret+addr result\endcsname & \csname RPi 5-MP+svc-eret+addr result\endcsname & \csname ODROID N2+ (big)-MP+svc-eret+addr result\endcsname\\
  \tableTestName{MP+svc-erets}{MP+svc-erets} & \csname ref_none-MP+svc-erets ref\endcsname & \csname ref_exs-MP+svc-erets ref\endcsname & \csname ref_sea_r-MP+svc-erets ref\endcsname & \csname ref_sea_w-MP+svc-erets ref\endcsname & \csname ref_sea_rw-MP+svc-erets ref\endcsname & \csname RPi 3B+-MP+svc-erets result\endcsname & \csname RPi 4B-MP+svc-erets result\endcsname & \csname RPi 5-MP+svc-erets result\endcsname & \csname ODROID N2+ (big)-MP+svc-erets result\endcsname\\
  \tableTestName{MP+svcs}{MP+svcs} & \csname ref_none-MP+svcs ref\endcsname & \csname ref_exs-MP+svcs ref\endcsname & \csname ref_sea_r-MP+svcs ref\endcsname & \csname ref_sea_w-MP+svcs ref\endcsname & \csname ref_sea_rw-MP+svcs ref\endcsname & \csname RPi 3B+-MP+svcs result\endcsname & \csname RPi 4B-MP+svcs result\endcsname & \csname RPi 5-MP+svcs result\endcsname & \csname ODROID N2+ (big)-MP+svcs result\endcsname\\
  \tableTestName{MP.EL1+dmb+ctrlvbarsvc}{MP.EL1+dmb+ctrlvbarsvc} & \csname ref_none-MP.EL1+dmb+ctrlvbarsvc ref\endcsname & \csname ref_exs-MP.EL1+dmb+ctrlvbarsvc ref\endcsname & \csname ref_sea_r-MP.EL1+dmb+ctrlvbarsvc ref\endcsname & \csname ref_sea_w-MP.EL1+dmb+ctrlvbarsvc ref\endcsname & \csname ref_sea_rw-MP.EL1+dmb+ctrlvbarsvc ref\endcsname & \csname RPi 3B+-MP.EL1+dmb+ctrlvbarsvc result\endcsname & \csname RPi 4B-MP.EL1+dmb+ctrlvbarsvc result\endcsname & \csname RPi 5-MP.EL1+dmb+ctrlvbarsvc result\endcsname & \csname ODROID N2+ (big)-MP.EL1+dmb+ctrlvbarsvc result\endcsname\\
  \tableTestName{MP.EL1+dmb+eret}{MP.EL1+dmb+eret} & \csname ref_none-MP.EL1+dmb+eret ref\endcsname & \csname ref_exs-MP.EL1+dmb+eret ref\endcsname & \csname ref_sea_r-MP.EL1+dmb+eret ref\endcsname & \csname ref_sea_w-MP.EL1+dmb+eret ref\endcsname & \csname ref_sea_rw-MP.EL1+dmb+eret ref\endcsname & \csname RPi 3B+-MP.EL1+dmb+eret result\endcsname & \csname RPi 4B-MP.EL1+dmb+eret result\endcsname & \csname RPi 5-MP.EL1+dmb+eret result\endcsname & \csname ODROID N2+ (big)-MP.EL1+dmb+eret result\endcsname\\
  \tableTestName{MP.EL1+dmb+eret-svc}{MP.EL1+dmb+eret-svc} & \csname ref_none-MP.EL1+dmb+eret-svc ref\endcsname & \csname ref_exs-MP.EL1+dmb+eret-svc ref\endcsname & \csname ref_sea_r-MP.EL1+dmb+eret-svc ref\endcsname & \csname ref_sea_w-MP.EL1+dmb+eret-svc ref\endcsname & \csname ref_sea_rw-MP.EL1+dmb+eret-svc ref\endcsname & \csname RPi 3B+-MP.EL1+dmb+eret-svc result\endcsname & \csname RPi 4B-MP.EL1+dmb+eret-svc result\endcsname & \csname RPi 5-MP.EL1+dmb+eret-svc result\endcsname & \csname ODROID N2+ (big)-MP.EL1+dmb+eret-svc result\endcsname\\
  \tableTestName{MP.EL1+dmb+svc}{MP.EL1+dmb+svc} & \csname ref_none-MP.EL1+dmb+svc ref\endcsname & \csname ref_exs-MP.EL1+dmb+svc ref\endcsname & \csname ref_sea_r-MP.EL1+dmb+svc ref\endcsname & \csname ref_sea_w-MP.EL1+dmb+svc ref\endcsname & \csname ref_sea_rw-MP.EL1+dmb+svc ref\endcsname & \csname RPi 3B+-MP.EL1+dmb+svc result\endcsname & \csname RPi 4B-MP.EL1+dmb+svc result\endcsname & \csname RPi 5-MP.EL1+dmb+svc result\endcsname & \csname ODROID N2+ (big)-MP.EL1+dmb+svc result\endcsname\\
  \tableTestName{MP.EL1+dmb+svc-eret}{MP.EL1+dmb+svc-eret} & \csname ref_none-MP.EL1+dmb+svc-eret ref\endcsname & \csname ref_exs-MP.EL1+dmb+svc-eret ref\endcsname & \csname ref_sea_r-MP.EL1+dmb+svc-eret ref\endcsname & \csname ref_sea_w-MP.EL1+dmb+svc-eret ref\endcsname & \csname ref_sea_rw-MP.EL1+dmb+svc-eret ref\endcsname & \csname RPi 3B+-MP.EL1+dmb+svc-eret result\endcsname & \csname RPi 4B-MP.EL1+dmb+svc-eret result\endcsname & \csname RPi 5-MP.EL1+dmb+svc-eret result\endcsname & \csname ODROID N2+ (big)-MP.EL1+dmb+svc-eret result\endcsname\\
  \tableTestName{S+dmb+eret}{S+dmb+eret} & \csname ref_none-S+dmb+eret ref\endcsname & \csname ref_exs-S+dmb+eret ref\endcsname & \csname ref_sea_r-S+dmb+eret ref\endcsname & \csname ref_sea_w-S+dmb+eret ref\endcsname & \csname ref_sea_rw-S+dmb+eret ref\endcsname & \csname RPi 3B+-S+dmb+eret result\endcsname & \csname RPi 4B-S+dmb+eret result\endcsname & \csname RPi 5-S+dmb+eret result\endcsname & \csname ODROID N2+ (big)-S+dmb+eret result\endcsname\\
  \tableTestName{S+dmb+svc}{S+dmb+svc} & \csname ref_none-S+dmb+svc ref\endcsname & \csname ref_exs-S+dmb+svc ref\endcsname & \csname ref_sea_r-S+dmb+svc ref\endcsname & \csname ref_sea_w-S+dmb+svc ref\endcsname & \csname ref_sea_rw-S+dmb+svc ref\endcsname & \csname RPi 3B+-S+dmb+svc result\endcsname & \csname RPi 4B-S+dmb+svc result\endcsname & \csname RPi 5-S+dmb+svc result\endcsname & \csname ODROID N2+ (big)-S+dmb+svc result\endcsname\\
  \tableTestName{S+erets}{S+erets} & \csname ref_none-S+erets ref\endcsname & \csname ref_exs-S+erets ref\endcsname & \csname ref_sea_r-S+erets ref\endcsname & \csname ref_sea_w-S+erets ref\endcsname & \csname ref_sea_rw-S+erets ref\endcsname & \csname RPi 3B+-S+erets result\endcsname & \csname RPi 4B-S+erets result\endcsname & \csname RPi 5-S+erets result\endcsname & \csname ODROID N2+ (big)-S+erets result\endcsname\\
  \tableTestName{S+svc-dmb-erets}{S+svc-dmb-erets} & \csname ref_none-S+svc-dmb-erets ref\endcsname & \csname ref_exs-S+svc-dmb-erets ref\endcsname & \csname ref_sea_r-S+svc-dmb-erets ref\endcsname & \csname ref_sea_w-S+svc-dmb-erets ref\endcsname & \csname ref_sea_rw-S+svc-dmb-erets ref\endcsname & \csname RPi 3B+-S+svc-dmb-erets result\endcsname & \csname RPi 4B-S+svc-dmb-erets result\endcsname & \csname RPi 5-S+svc-dmb-erets result\endcsname & \csname ODROID N2+ (big)-S+svc-dmb-erets result\endcsname\\
  \tableTestName{S+svc-erets}{S+svc-erets} & \csname ref_none-S+svc-erets ref\endcsname & \csname ref_exs-S+svc-erets ref\endcsname & \csname ref_sea_r-S+svc-erets ref\endcsname & \csname ref_sea_w-S+svc-erets ref\endcsname & \csname ref_sea_rw-S+svc-erets ref\endcsname & \csname RPi 3B+-S+svc-erets result\endcsname & \csname RPi 4B-S+svc-erets result\endcsname & \csname RPi 5-S+svc-erets result\endcsname & \csname ODROID N2+ (big)-S+svc-erets result\endcsname\\
  \tableTestName{S+svcs}{S+svcs} & \csname ref_none-S+svcs ref\endcsname & \csname ref_exs-S+svcs ref\endcsname & \csname ref_sea_r-S+svcs ref\endcsname & \csname ref_sea_w-S+svcs ref\endcsname & \csname ref_sea_rw-S+svcs ref\endcsname & \csname RPi 3B+-S+svcs result\endcsname & \csname RPi 4B-S+svcs result\endcsname & \csname RPi 5-S+svcs result\endcsname & \csname ODROID N2+ (big)-S+svcs result\endcsname\\
  \tableTestName{SB+daifsets}{SB+daifsets} & \csname ref_none-SB+daifsets ref\endcsname & \csname ref_exs-SB+daifsets ref\endcsname & \csname ref_sea_r-SB+daifsets ref\endcsname & \csname ref_sea_w-SB+daifsets ref\endcsname & \csname ref_sea_rw-SB+daifsets ref\endcsname & ? & ? & ? & ?\\
  \tableTestName{SB+dmb+eret}{SB+dmb+eret} & \csname ref_none-SB+dmb+eret ref\endcsname & \csname ref_exs-SB+dmb+eret ref\endcsname & \csname ref_sea_r-SB+dmb+eret ref\endcsname & \csname ref_sea_w-SB+dmb+eret ref\endcsname & \csname ref_sea_rw-SB+dmb+eret ref\endcsname & \csname RPi 3B+-SB+dmb+eret result\endcsname & \csname RPi 4B-SB+dmb+eret result\endcsname & \csname RPi 5-SB+dmb+eret result\endcsname & \csname ODROID N2+ (big)-SB+dmb+eret result\endcsname\\
  \tableTestName{SB+dmb+rfi-ctrl-eret}{SB+dmb+rfi-ctrl-eret} & \csname ref_none-SB+dmb+rfi-ctrl-eret ref\endcsname & \csname ref_exs-SB+dmb+rfi-ctrl-eret ref\endcsname & \csname ref_sea_r-SB+dmb+rfi-ctrl-eret ref\endcsname & \csname ref_sea_w-SB+dmb+rfi-ctrl-eret ref\endcsname & \csname ref_sea_rw-SB+dmb+rfi-ctrl-eret ref\endcsname & \csname RPi 3B+-SB+dmb+rfi-ctrl-eret result\endcsname & \csname RPi 4B-SB+dmb+rfi-ctrl-eret result\endcsname & \csname RPi 5-SB+dmb+rfi-ctrl-eret result\endcsname & \csname ODROID N2+ (big)-SB+dmb+rfi-ctrl-eret result\endcsname\\
  \tableTestName{SB+dmb+rfi-ctrl-svc}{SB+dmb+rfi-ctrl-svc} & \csname ref_none-SB+dmb+rfi-ctrl-svc ref\endcsname & \csname ref_exs-SB+dmb+rfi-ctrl-svc ref\endcsname & \csname ref_sea_r-SB+dmb+rfi-ctrl-svc ref\endcsname & \csname ref_sea_w-SB+dmb+rfi-ctrl-svc ref\endcsname & \csname ref_sea_rw-SB+dmb+rfi-ctrl-svc ref\endcsname & \csname RPi 3B+-SB+dmb+rfi-ctrl-svc result\endcsname & \csname RPi 4B-SB+dmb+rfi-ctrl-svc result\endcsname & \csname RPi 5-SB+dmb+rfi-ctrl-svc result\endcsname & \csname ODROID N2+ (big)-SB+dmb+rfi-ctrl-svc result\endcsname\\
  \tableTestName{SB+dmb+rfieret-addr}{SB+dmb+rfieret-addr} & \csname ref_none-SB+dmb+rfieret-addr ref\endcsname & \csname ref_exs-SB+dmb+rfieret-addr ref\endcsname & \csname ref_sea_r-SB+dmb+rfieret-addr ref\endcsname & \csname ref_sea_w-SB+dmb+rfieret-addr ref\endcsname & \csname ref_sea_rw-SB+dmb+rfieret-addr ref\endcsname & \csname RPi 3B+-SB+dmb+rfieret-addr result\endcsname & \csname RPi 4B-SB+dmb+rfieret-addr result\endcsname & \csname RPi 5-SB+dmb+rfieret-addr result\endcsname & \csname ODROID N2+ (big)-SB+dmb+rfieret-addr result\endcsname\\
  \tableTestName{SB+dmb+rfisvc-addr}{SB+dmb+rfisvc-addr} & \csname ref_none-SB+dmb+rfisvc-addr ref\endcsname & \csname ref_exs-SB+dmb+rfisvc-addr ref\endcsname & \csname ref_sea_r-SB+dmb+rfisvc-addr ref\endcsname & \csname ref_sea_w-SB+dmb+rfisvc-addr ref\endcsname & \csname ref_sea_rw-SB+dmb+rfisvc-addr ref\endcsname & \csname RPi 3B+-SB+dmb+rfisvc-addr result\endcsname & \csname RPi 4B-SB+dmb+rfisvc-addr result\endcsname & \csname RPi 5-SB+dmb+rfisvc-addr result\endcsname & \csname ODROID N2+ (big)-SB+dmb+rfisvc-addr result\endcsname\\
  \tableTestName{SB+dmb+svc}{SB+dmb+svc} & \csname ref_none-SB+dmb+svc ref\endcsname & \csname ref_exs-SB+dmb+svc ref\endcsname & \csname ref_sea_r-SB+dmb+svc ref\endcsname & \csname ref_sea_w-SB+dmb+svc ref\endcsname & \csname ref_sea_rw-SB+dmb+svc ref\endcsname & \csname RPi 3B+-SB+dmb+svc result\endcsname & \csname RPi 4B-SB+dmb+svc result\endcsname & \csname RPi 5-SB+dmb+svc result\endcsname & \csname ODROID N2+ (big)-SB+dmb+svc result\endcsname\\
  \tableTestName{SB+svc+dmb-erets}{SB+svc+dmb-erets} & \csname ref_none-SB+svc+dmb-erets ref\endcsname & \csname ref_exs-SB+svc+dmb-erets ref\endcsname & \csname ref_sea_r-SB+svc+dmb-erets ref\endcsname & \csname ref_sea_w-SB+svc+dmb-erets ref\endcsname & \csname ref_sea_rw-SB+svc+dmb-erets ref\endcsname & ? & ? & ? & ?\\
  \tableTestName{SB+svc-dmb-erets}{SB+svc-dmb-erets} & ? & ? & ? & ? & ? & \csname RPi 3B+-SB+svc-dmb-erets result\endcsname & \csname RPi 4B-SB+svc-dmb-erets result\endcsname & \csname RPi 5-SB+svc-dmb-erets result\endcsname & \csname ODROID N2+ (big)-SB+svc-dmb-erets result\endcsname\\
  \tableTestName{SB+svc-erets}{SB+svc-erets} & ? & ? & ? & ? & ? & \csname RPi 3B+-SB+svc-erets result\endcsname & \csname RPi 4B-SB+svc-erets result\endcsname & \csname RPi 5-SB+svc-erets result\endcsname & \csname ODROID N2+ (big)-SB+svc-erets result\endcsname\\
  \tableTestName{SB+svcs}{SB+svcs} & \csname ref_none-SB+svcs ref\endcsname & \csname ref_exs-SB+svcs ref\endcsname & \csname ref_sea_r-SB+svcs ref\endcsname & \csname ref_sea_w-SB+svcs ref\endcsname & \csname ref_sea_rw-SB+svcs ref\endcsname & \csname RPi 3B+-SB+svcs result\endcsname & \csname RPi 4B-SB+svcs result\endcsname & \csname RPi 5-SB+svcs result\endcsname & \csname ODROID N2+ (big)-SB+svcs result\endcsname\\
  \tableTestName{SB.EL1+erets}{SB.EL1+erets} & \csname ref_none-SB.EL1+erets ref\endcsname & \csname ref_exs-SB.EL1+erets ref\endcsname & \csname ref_sea_r-SB.EL1+erets ref\endcsname & \csname ref_sea_w-SB.EL1+erets ref\endcsname & \csname ref_sea_rw-SB.EL1+erets ref\endcsname & \csname RPi 3B+-SB.EL1+erets result\endcsname & \csname RPi 4B-SB.EL1+erets result\endcsname & \csname RPi 5-SB.EL1+erets result\endcsname & \csname ODROID N2+ (big)-SB.EL1+erets result\endcsname\\
  \tableTestName{SB.EL1+svc-erets}{SB.EL1+svc-erets} & \csname ref_none-SB.EL1+svc-erets ref\endcsname & \csname ref_exs-SB.EL1+svc-erets ref\endcsname & \csname ref_sea_r-SB.EL1+svc-erets ref\endcsname & \csname ref_sea_w-SB.EL1+svc-erets ref\endcsname & \csname ref_sea_rw-SB.EL1+svc-erets ref\endcsname & \csname RPi 3B+-SB.EL1+svc-erets result\endcsname & \csname RPi 4B-SB.EL1+svc-erets result\endcsname & \csname RPi 5-SB.EL1+svc-erets result\endcsname & \csname ODROID N2+ (big)-SB.EL1+svc-erets result\endcsname\\
  \tableTestName{SEA_R_detect}{SEA\_R\_detect} & ? & ? & ? & ? & ? & \csname RPi 3B+-SEA_R_detect result\endcsname & \csname RPi 4B-SEA_R_detect result\endcsname & \csname RPi 5-SEA_R_detect result\endcsname & \csname ODROID N2+ (big)-SEA_R_detect result\endcsname\\
  \tableTestName{SEA_W_detect}{SEA\_W\_detect} & ? & ? & ? & ? & ? & \csname RPi 3B+-SEA_W_detect result\endcsname & \csname RPi 4B-SEA_W_detect result\endcsname & \csname RPi 5-SEA_W_detect result\endcsname & \csname ODROID N2+ (big)-SEA_W_detect result\endcsname\\
\end{longtable}

}
\caption{Isla model results on the listed tests,
with combinations of variants
\\
(`default' is context-synchronising entry and exit,
and no synchronous external aborts).}
\label{app:model-results}
\label{app:hw-results}
\end{figure}

\newpage

\section{Litmus Test Glossary for Exceptions}
\label{app:litmus-glossary}

\etocsettocdepth{3}
\localtableofcontents{}

\medskip

For ease of reference, we repeat most of the tests for exceptions we refer to
from the main document.

For each litmus test we give:
\begin{itemize}
  \item The hand-transcribed code listing and execution diagram of the relaxed outcome.
  \item The results from our preliminary running on hardware (where it exists).
        We have ran these on a small collection of devices:
      \begin{itemize}
        \item A Raspberry Pi 3B+, on Arm Cortex-A53 r0p4 cores.
        \item A Raspberry Pi 4B, on Arm Cortex-A72 r0p3 cores.
        \item A Raspberry Pi 5, with Arm Cortex-A76 r1p4 cores.
        \item An ODROID N2+, with an Amlogic S922X SoC.
              This SoC has a big.LITTLE architecture;
               our results are from the `big' cores running Arm Cortex-A73 r0p2 cores.
      \end{itemize}
  \item The understood architectural intent derived by the model (where it exists).
        The `Allow/Forbid' in the listing is the default (base) architectural intent,
        with the variants listed beneath the test:
      \begin{itemize}
        \item ExS: Without context synchronisation on entry or exit.
        \item \SEAR: With the assumption that any load could generate a synchronous external abort.
        \item \SEAW: With the assumption that any store could generate a synchronous external abort.
        \item \SEARW: With the assumption that any load or store could generate a synchronous external abort.
      \end{itemize}
\end{itemize}

\newpage
\subsection{Sychronous Exceptions}

\subsubsection{Out-of-order execution across boundaries}

Reads and writes may be executed out-of-order across any exception boundary:

\begin{longtable}{l c c}
  & \textbf{Entry} & \textbf{Exit/Both} \\
  \tableheader{Read-Read}   & %
\scalebox{0.75}{
\begin{tabular}{c}
    \input{tests/MP+dmb+svc.listing.tikz} \\
    \input{tests/MP+dmb+svc.diagram.tikz}
  \end{tabular}}
  \label{test:MP+dmb+svc}
   & %
\scalebox{0.75}{
\begin{tabular}{c}
    \input{tests/MP+dmb+eret.listing.tikz} \\
    \input{tests/MP+dmb+eret.diagram.tikz}
  \end{tabular}}
  \label{test:MP+dmb+eret}
  \\
                            &                       & %
\scalebox{0.75}{
\begin{tabular}{c}
    \input{tests/MP+dmb+svceret.listing.tikz} \\
    \input{tests/MP+dmb+svceret.diagram.tikz}
  \end{tabular}}
  \label{test:MP+dmb+svceret}
 \\
  \tableheader{Read-Write}  & %
\scalebox{0.75}{
\begin{tabular}{c}
    \input{tests/S+dmb+svc.listing.tikz} \\
    \input{tests/S+dmb+svc.diagram.tikz}
  \end{tabular}}
  \label{test:S+dmb+svc}
    & %
\scalebox{0.75}{
\begin{tabular}{c}
    \input{tests/S+dmb+eret.listing.tikz} \\
    \input{tests/S+dmb+eret.diagram.tikz}
  \end{tabular}}
  \label{test:S+dmb+eret}
 \\
                            &                       & %
\scalebox{0.75}{
\begin{tabular}{c}
    \input{tests/LB+svcerets.listing.tikz} \\
    \input{tests/LB+svcerets.diagram.tikz}
  \end{tabular}}
  \label{test:LB+svcerets}
 \\
  \tableheader{Write-Read}  & %
\scalebox{0.75}{
\begin{tabular}{c}
    \input{tests/SB+dmb+svc.listing.tikz} \\
    \input{tests/SB+dmb+svc.diagram.tikz}
  \end{tabular}}
  \label{test:SB+dmb+svc}
   & %
\scalebox{0.75}{
\begin{tabular}{c}
    \input{tests/SB+dmb+eret.listing.tikz} \\
    \input{tests/SB+dmb+eret.diagram.tikz}
  \end{tabular}}
  \label{test:SB+dmb+eret}
 \\
                            &                       & %
\scalebox{0.75}{
\begin{tabular}{c}
    \input{tests/SB+svcerets.listing.tikz} \\
    \input{tests/SB+svcerets.diagram.tikz}
  \end{tabular}}
  \label{test:SB+svcerets}
 \\
  \tableheader{Write-Write} & %
\scalebox{0.75}{
\begin{tabular}{c}
    \input{tests/MP+svc+dmb.listing.tikz} \\
    \begin{tikzpicture}[%
  /litmus/append to tikz path search,
  kind=events,
]
\node[simple instructions] {
  a:W x=1 \\
  b:W y=1 \\
};

\node[simple instructions] {
  c:R y=1 \\
  d:R x=0 \\
};

\draw [vertical align]
  (a) edge[po=svc] (b);
\draw [vertical align]
  (c) edge[dmb] (d);

\draw (b) edge[rf] (c);
\draw (d) edge[fr] (a);
\end{tikzpicture}
  \end{tabular}}
  \label{test:MP+svc+dmb}
   & %
\scalebox{0.75}{
\begin{tabular}{c}
    \input{tests/MP+eret+dmb.listing.tikz} \\
    \begin{tikzpicture}[%
  /litmus/append to tikz path search,
  kind=events,
]
\node[simple instructions] {
  a:W x=1 \\
  b:W y=1 \\
};

\node[simple instructions] {
  c:R y=1 \\
  d:R x=0 \\
};

\draw [vertical align]
  (a) edge[po=eret] (b);
\draw [vertical align]
  (c) edge[dmb] (d);

\draw (b) edge[rf] (c);
\draw (d) edge[fr] (a);
\end{tikzpicture}
  \end{tabular}}
  \label{test:MP+eret+dmb}
 \\
                            &                       & %
\scalebox{0.75}{
\begin{tabular}{c}
    \input{tests/MP+svceret+addr.listing.tikz} \\
    \input{tests/MP+svceret+addr.diagram.tikz}
  \end{tabular}}
  \label{test:MP+svceret+addr}
 \\
\end{longtable}

\newpage
\paragraph{Other combinations}

Any composition of exception boundaries does not induce ordering,
e.g. exception exit-plus-entry is not synchronising:
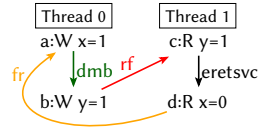
\begin{center}
\scalebox{0.75}{
\begin{tabular}{l@{}l}
    \input{tests/MP+dmb+eretsvc.listing.tikz}
    &
    \input{tests/MP+dmb+eretsvc.diagram.tikz}
  \end{tabular}}
  \label{test:MP+dmb+eretsvc}

\end{center}

\paragraph{Barriers}

Barriers do not interact with the interrupt machinery,
and the barriers in the tests could be replaced with any other thread-local ordering:
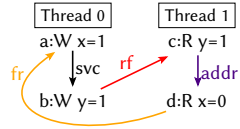
\begin{center}
\scalebox{0.75}{
\begin{tabular}{l@{}l}
    \input{tests/MP+svc+addr.listing.tikz}
    &
    \begin{tikzpicture}[%
  /litmus/append to tikz path search,
  kind=events,
]
\node[simple instructions] {
  a:W x=1 \\
  b:W y=1 \\
};

\node[simple instructions] {
  c:R y=1 \\
  d:R x=0 \\
};

\draw [vertical align]
  (a) edge[po=svc] (b);
\draw [vertical align]
  (c) edge[addr] (d);

\draw (b) edge[rf] (c);
\draw (d) edge[fr] (a);
\end{tikzpicture}
  \end{tabular}}
  \label{test:MP+svc+addr}

\end{center}

\paragraph{Exception level}

The privilege level does not effect what out-of-order execution is permissible
in the presence of exceptions.
In particular,
same-exception-level tests do not give weaker results,
e.g.:
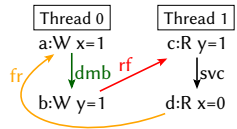
\begin{center}
\scalebox{0.75}{
\begin{tabular}{l@{}l}
    \input{tests/MP.EL1+dmb+svc.listing.tikz}
    &
    \input{tests/MP.EL1+dmb+svc.diagram.tikz}
  \end{tabular}}
  \label{test:MP.EL1+dmb+svc}

\end{center}

\newpage
\subsubsection{Speculation}

\noindent
Context synchronising exception entry is not executed speculatively:
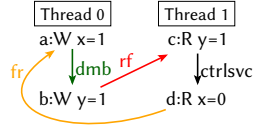
\begin{center}
\scalebox{0.75}{
\begin{tabular}{l@{}l}
    \input{tests/MP+dmb+ctrlsvc.listing.tikz}
    &
    \input{tests/MP+dmb+ctrlsvc.diagram.tikz}
  \end{tabular}}
  \label{test:MP+dmb+ctrlsvc}

\end{center}

\noindent
Context synchronising exception exit is not executed speculatively:
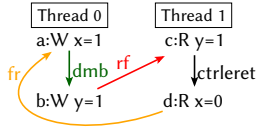
\begin{center}
\scalebox{0.75}{
\begin{tabular}{l@{}l}
    \input{tests/MP+dmb+ctrleret.listing.tikz}
    &
    \input{tests/MP+dmb+ctrleret.diagram.tikz}
  \end{tabular}}
  \label{test:MP+dmb+ctrleret}

\end{center}

\noindent
Still-speculative writes may not be forwarded into exception handlers
(This a variant of PPOCA with the rfi across an SVC):
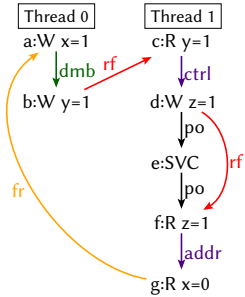
\begin{center}
\scalebox{0.75}{
\begin{tabular}{l@{}l}
    \input{tests/MP+dmb+ctrl-rfisvc-addr.listing.tikz}
    &
    \input{tests/MP+dmb+ctrl-rfisvc-addr.diagram.tikz}
  \end{tabular}}
  \label{test:MP+dmb+ctrl-rfisvc-addr}

\end{center}

\newpage
\noindent
Code after an exception return may not be executed
while the exception is still speculative
(in particular, not even permitting thread-local accesses to forward):
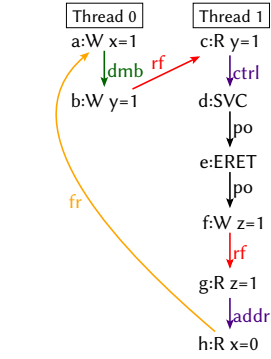
\begin{center}
\scalebox{0.75}{
\begin{tabular}{l@{}l}
    \input{tests/MP+dmb+ctrl-rfisvceret-addr.listing.tikz}
    &
    \input{tests/MP+dmb+ctrl-rfisvceret-addr.diagram.tikz}
  \end{tabular}}
  \label{test:MP+dmb+ctrl-rfisvceret-addr}

\end{center}

\noindent
However, speculative writes inside the handler can be satisfied early:
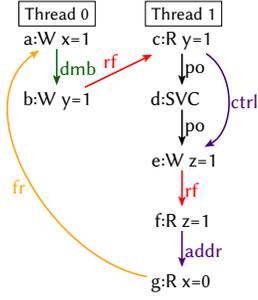
\begin{center}
\scalebox{0.75}{
\begin{tabular}{l@{}l}
    \input{tests/MP+dmb+svc-ctrl-rfi-addr.listing.tikz}
    &
    \input{tests/MP+dmb+svc-ctrl-rfi-addr.diagram.tikz}
  \end{tabular}}
  \label{test:MP+dmb+svc-ctrl-rfi-addr}

\end{center}

\noindent
Non-speculative writes can be visible to the same thread early,
even to an exception handler at a higher privilege level:
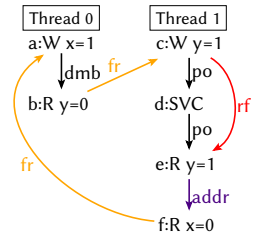
\begin{center}
\scalebox{0.75}{
\begin{tabular}{l@{}l}
    \input{tests/SB+dmb+rfisvc-addr.listing.tikz}
    &
    \input{tests/SB+dmb+rfisvc-addr.diagram.tikz}
  \end{tabular}}
  \label{test:SB+dmb+rfisvc-addr}

\end{center}

\newpage
\subsubsection{Dependencies through registers}

Dependencies through system registers compose with CSEs
(see \S\ref{subsec:exception-register-dependencies}):
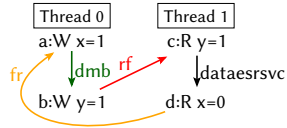
\begin{center}
\scalebox{0.75}{
\begin{tabular}{l@{}l}
    \input{tests/MP.EL1+dmb+dataesrsvc.listing.tikz}
    &
    \input{tests/MP.EL1+dmb+dataesrsvc.diagram.tikz}
  \end{tabular}}
  \label{test:MP.EL1+dmb+dataesrsvc}

\end{center}

\noindent
Similarly with control dependencies induced by writes to the VBAR:
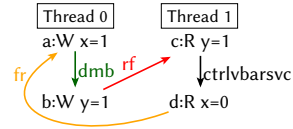
\begin{center}
\scalebox{0.75}{
\begin{tabular}{l@{}l}
    \input{tests/MP.EL1+dmb+ctrlvbarsvc.listing.tikz}
    &
    \input{tests/MP.EL1+dmb+ctrlvbarsvc.diagram.tikz}
  \end{tabular}}
  \label{test:MP.EL1+dmb+ctrlvbarsvc}

\end{center}

\noindent
The thread id register (TPIDR) is technically a system register,
but is never written to indirectly by instructions,
so it may be that hardware could optimise it.
This is an open question:
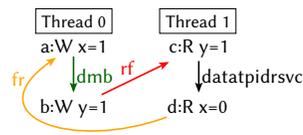
\begin{center}
\scalebox{0.75}{
\begin{tabular}{l@{}l}
    \input{tests/MP.EL1+dmb+datatpidrsvc.listing.tikz}
    &
    \input{tests/MP.EL1+dmb+datatpidrsvc.diagram.tikz}
  \end{tabular}}
  \label{test:MP.EL1+dmb+datatpidrsvc}

\end{center}

\noindent
Special-purpose registers are self-synchronising:
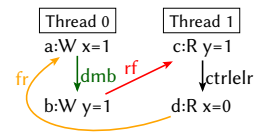
\begin{center}
\scalebox{0.75}{
\begin{tabular}{l@{}l}
    \input{tests/MP+dmb+ctrlelr.listing.tikz}
    &
    \input{tests/MP+dmb+ctrlelr.diagram.tikz}
  \end{tabular}}
  \label{test:MP+dmb+ctrlelr}

\end{center}

\newpage
\noindent
Or implicitly via dataflow to the ELR:
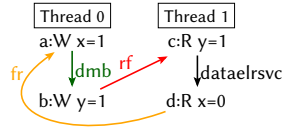
\begin{center}
\scalebox{0.75}{
\begin{tabular}{l@{}l}
    \input{tests/MP.EL1+dmb+dataelrsvc.listing.tikz}
    &
    \input{tests/MP.EL1+dmb+dataelrsvc.diagram.tikz}
  \end{tabular}}
  \label{test:MP.EL1+dmb+dataelrsvc}

\end{center}

\subsubsection{Other exceptions and ETS}

Some types of exceptions have additional synchronisation effects,
for example taking an interrupt, or a page fault:
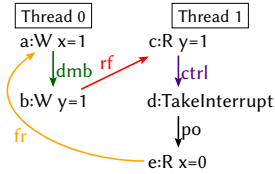
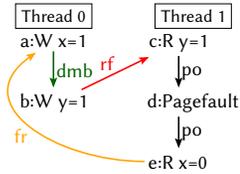
\begin{center}
\scalebox{0.75}{
\begin{tabular}{l@{}l}
    \input{tests/MP+dmb.sy+br-int.listing.tikz}
    &
    \input{tests/MP+dmb.sy+br-int.diagram.tikz}
  \end{tabular}}
  \label{test:MP+dmb.sy+br-int}

\scalebox{0.75}{
\begin{tabular}{l@{}l}
    \input{tests/MP+dmb.sy+fault.listing.tikz}
    &
    \input{tests/MP+dmb.sy+fault.diagram.tikz}
  \end{tabular}}
  \label{test:MP+dmb.sy+fault}

\end{center}

\newpage
\subsubsection{Detection of synchronous external aborts}
\label{sec:seaxdetect}

We can detect synchronous external aborts directly:
by issuing a load or store,
and catching any external abort reported as a DataAbort.
It is, we believe, impossible
to force such a situation to occur from software.

The following tests issue a load or store
to an out-of-range address
(which may or may not cause the SoC to generate an external abort)
and sets \asm{X0} to 1 if it detects it.

There is,
as far as we understand,
no \emph{architectural} way to detect if these tests are permitted to return 1.

\begin{center}
  \input{tests/SEA-R-detect}
\end{center}

\begin{center}
  \input{tests/SEA-W-detect}
\end{center}

\newpage
\subsection{Software Generated Interrupts}

\subsubsection{Basic shapes for SGIs}
\label{sec:basicshapesforsgis}

\label{sec:LBviaSGI}

Basic shapes with one SGIs
are formed by replacing a \asm{[W] ; rf ; [R]} sequence
by a \asm{[GenerateInterrupt] ; interrupt ; [TakeInterrupt]} sequence
(and so there is no equivalent of 2+2W).

All the following outcomes are allowed because of the sender:
events program-order-before a \asm{GenerateInterrupt}
are not ordered-before the \asm{GenerateInterrupt}.

To forbid these outcomes, an appropriate \asm{DSB}
(\asm{DSB ST} for MP, \asm{DSB LD} for LB) is sufficient
(and necessary: \asm{DMB} is not enough).

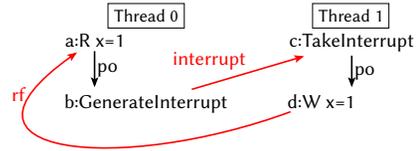
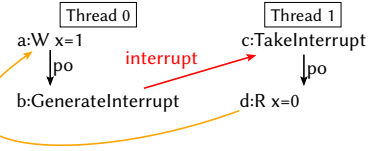
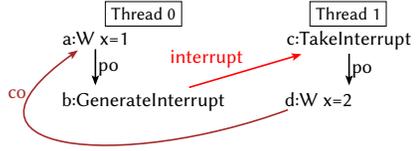
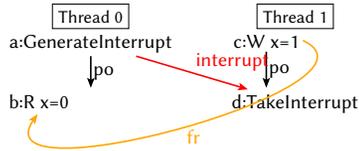
\begin{center}
\scalebox{0.75}{
\begin{tabular}{l@{}l}
    \input{tests/LBviaSGI.listing.tikz}
    &
    \input{tests/LBviaSGI.diagram.tikz}
  \end{tabular}}
  \label{test:LBviaSGI}

\\
\scalebox{0.75}{
\begin{tabular}{l@{}l}
    \input{tests/MPviaSGI.listing.tikz}
    &
    \input{tests/MPviaSGI.diagram.tikz}
  \end{tabular}}
  \label{test:MPviaSGI}

\\
\scalebox{0.75}{
\begin{tabular}{l@{}l}
    \input{tests/SviaSGI.listing.tikz}
    &
    \input{tests/SviaSGI.diagram.tikz}
  \end{tabular}}
  \label{test:SviaSGI}

\\
\scalebox{0.75}{
\begin{tabular}{l@{}l}
    \input{tests/SBviaSGI.listing.tikz}
    &
    \input{tests/SBviaSGI.diagram.tikz}
  \end{tabular}}
  \label{test:SBviaSGI}

\end{center}

\newpage
\subsubsection{Order of propagation of SGIs}
\label{sec:orderofpropagation}

The order of propagation of SGIs
depends on the details of the GIC,
and are out of scope for the main body of this document, but
we nonetheless highlight some key questions.

\medskip

Coherence-like test for SGIs:
the `sender' thread 0 sends two SGIs with INTIDs 1 and 2.
The `receiver' thread 1 can observe them in the opposite order
(even when INTIDs 1 and 2 have the same priority for the receiver),
because of propagation.
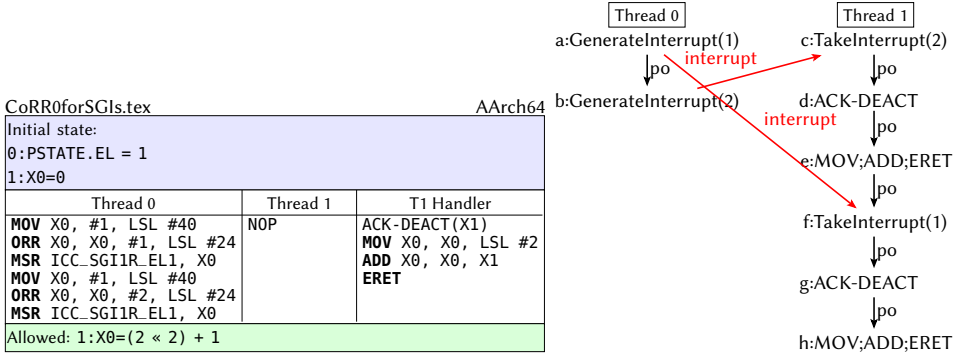
\begin{center}
\scalebox{0.75}{
\begin{tabular}{l@{}l}
    \input{tests/CoRR0forSGIs.listing.tikz}
    &
    \input{tests/CoRR0forSGIs.diagram.tikz}
  \end{tabular}}
  \label{test:CoRR0forSGIs}

\end{center}

\medskip

Independent-reads-of-independent-write-like test for SGIs:
two `sender' threads, 0 and 2, send a SGI each,
with INTIDs 1 and 2, respectively.
The two `receiver' threads, 1 and 3,
can observe them in different orders
(even when, for both receivers, INTIDs 1 and 2 have the same priority).
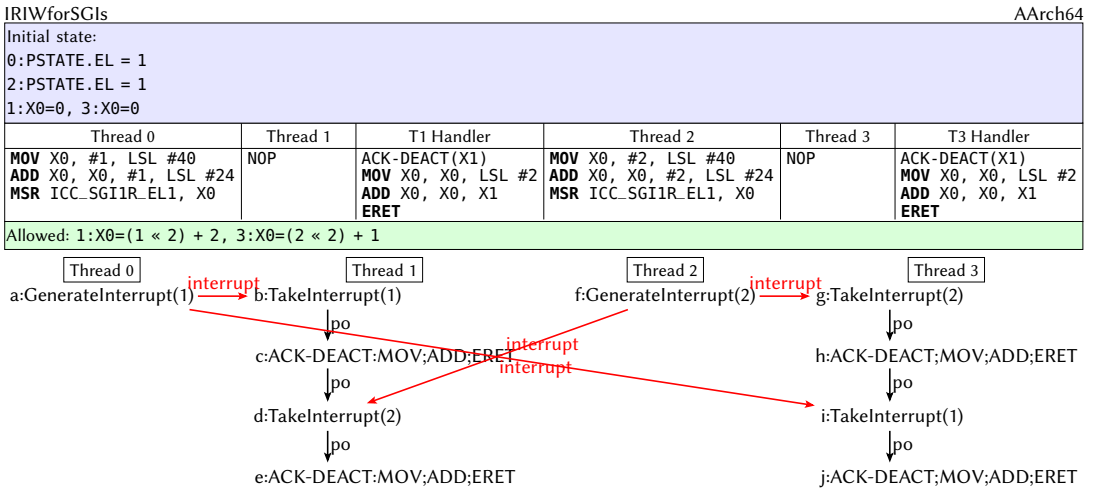
\begin{center}
\scalebox{0.75}{
\begin{tabular}{c}
    \input{tests/IRIWforSGIs.listing.tikz} \\
    \input{tests/IRIWforSGIs.diagram.tikz}
  \end{tabular}}
  \label{test:IRIWforSGIs}

\end{center}

\newpage
Write-to-read-coherence-like test for SGIs:
a `sender' thread sends an SGI with INTID 1.
A `forwarding' thread receives that SGI, and its interrupt handler sends an SGI with INTID 2.
A `receiver' thread can observe the two SGIs in the opposite order:
2 then 1
(even when INTIDs 1 and 2 have the same priority).
(We assume that Thread 0 has a handler that acknowledges-deactivates-erets,
or is somehow not targeted by Thread 1's SGI.)
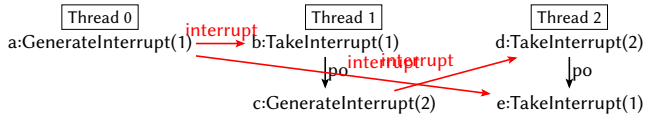
\begin{center}
\scalebox{0.75}{
\begin{tabular}{c}
    \input{tests/WRCforSGIs.listing.tikz} \\
    \input{tests/WRCforSGIs.diagram.tikz}
  \end{tabular}}
  \label{test:WRCforSGIs}

\end{center}

\newpage
\subsubsection{Propagation of SGIs}

This is a variant of MP (although interpreted backwards).
It illustrates that interrupts do not propagate instantly with respect to writes
as observed by reads.
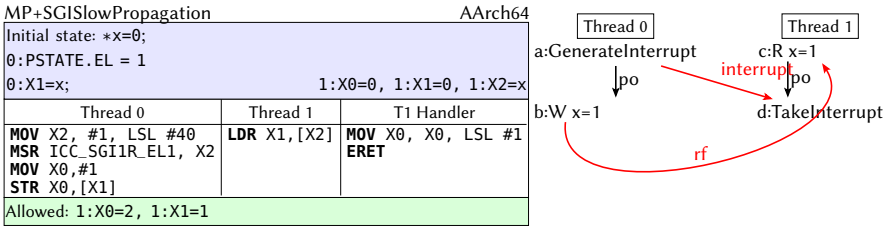
\begin{center}
\scalebox{0.75}{
\begin{tabular}{l@{}l}
    \input{tests/MP+SGISlowPropagation.listing.tikz}
    &
    \input{tests/MP+SGISlowPropagation.diagram.tikz}
  \end{tabular}}
  \label{test:MP+SGISlowPropagation}

\end{center}

\medskip

\noindent
This is a variant of R (although interpreted backwards).
It illustrates that interrupts do not propagate instantly with respect to writes
in terms of coherence.
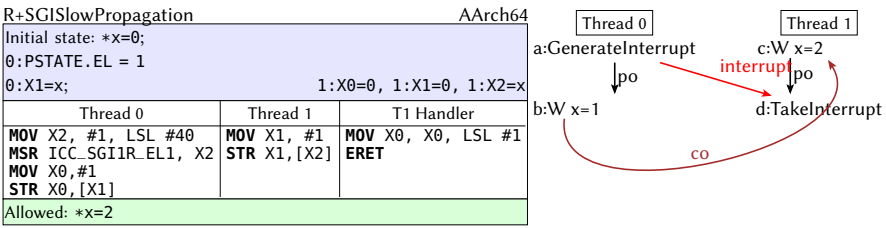
\begin{center}
\scalebox{0.75}{
\begin{tabular}{l@{}l}
    \input{tests/R+SGISlowPropagation.listing.tikz}
    &
    \input{tests/R+SGISlowPropagation.diagram.tikz}
  \end{tabular}}
  \label{test:R+SGISlowPropagation}

\end{center}

\medskip

\noindent
This is a variant of SB with only interrupts.
It illustrates that interrupts do not propagate instantly with respect to each other.
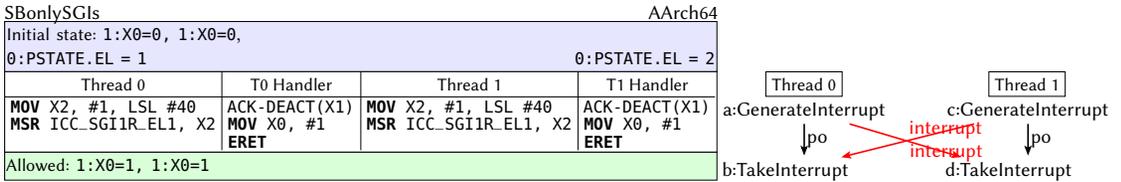
\begin{center}
\scalebox{0.75}{
\begin{tabular}{l@{}l}
    \input{tests/SBonlySGIs.listing.tikz}
    &
    \input{tests/SBonlySGIs.diagram.tikz}
  \end{tabular}}
  \label{test:SBonlySGIs}

\end{center}

\newpage
\subsubsection{Load buffering of SGIs}
\label{sec:lbonlysgis}

Load buffering involving only SGIs: the code of both threads is \asm{NOP}
--- only the interrupt handlers have actual code.
(To contrast with LBviaSGI in \S\ref{sec:LBviaSGI}).
Interrupts are not taken speculatively.
Practically, it would be extremely disruptive to programming,
allowing spontaneous interrupts which cause themselves.
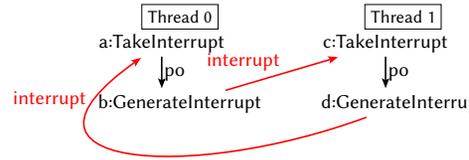
\begin{center}
\scalebox{0.75}{
\begin{tabular}{l@{}l}
    \input{tests/LBonlySGIs.listing.tikz}
    &
    \input{tests/LBonlySGIs.diagram.tikz}
  \end{tabular}}
  \label{test:LBonlySGIs}

\end{center}

\newpage
\subsubsection{Acknowledgement, priority drop, and deactivation}

There are two tests, depending on the EOIMode,
which determines whether priority drop and deactivation are separate.

For both, missing any of the steps (except the trailing \asm{ISB}) makes the relaxed outcome allowed.

\begin{center}
\scriptsize
  \input{tests/MPviaSGIEIOmode0sequence}
\end{center}

\begin{center}
\scriptsize
  \input{tests/MPviaSGIEIOmode1sequence}
\end{center}

\newpage
\subsubsection{RCU and Verona}

\asm{RCU-MP} characterises the message-passing aspect of RCU.
\\
With a \asm{DMB ST} between the write to \asm{x} and the \asm{MSR SGI1R},
this becomes forbidden.
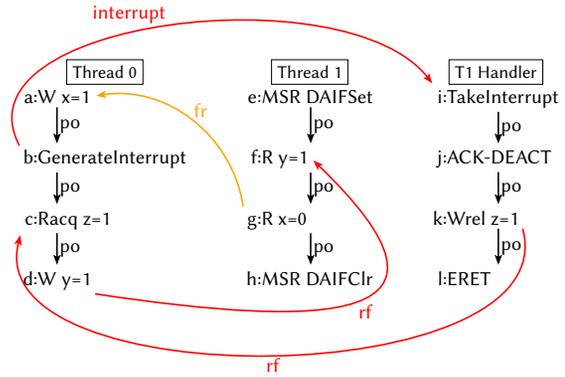
\begin{center}
\scalebox{0.75}{
\begin{tabular}{l@{}l}
    \input{tests/RCU-MP.listing.tikz}
    &
    \input{tests/RCU-MP.diagram.tikz}
  \end{tabular}}
  \label{test:RCU-MP}

\end{center}

\medskip

\noindent
Although \asm{RCU-deferred-free} characterises the reclamation part of RCU,
the ordering of the reads does not affect the outcome,
which means that it is subsumed by RCU-MP~\cite{AlglaveMMPS18}.
\\
Again,
with a \asm{DMB ST} between the write to \asm{x} and the \asm{MSR SGI1R},
this becomes forbidden.
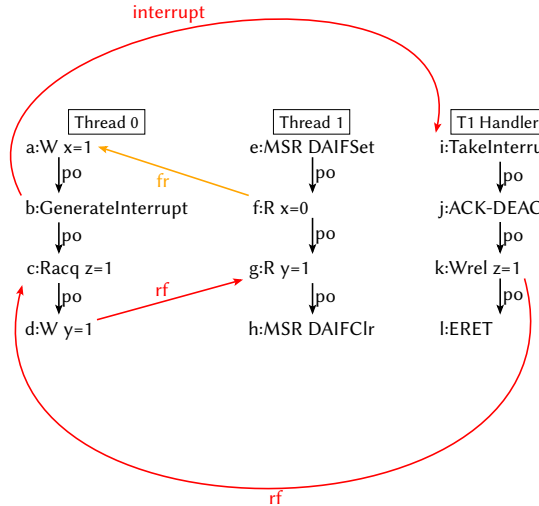
\begin{center}
\scalebox{0.75}{
\begin{tabular}{l@{}l}
    \input{tests/RCU-deferred-free.listing.tikz}
    &
    \input{tests/RCU-deferred-free.diagram.tikz}
  \end{tabular}}
  \label{test:RCU-deferred-free}

\end{center}

\newpage

\noindent
SBVerona captures the key synchronisation pattern of the asymmetric lock of the Project Verona runtime.
Again,
with a \asm{DMB ST} between the write to \asm{x} and the \asm{MSR SGI1R},
this becomes forbidden.
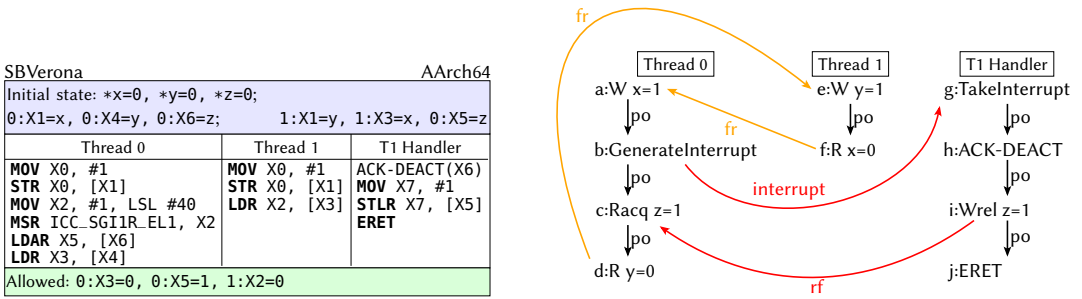
\begin{center}
\scalebox{0.75}{
\begin{tabular}{l@{}l}
    \input{tests/SBVerona.listing.tikz}
    &
    \input{tests/SBVerona.diagram.tikz}
  \end{tabular}}
  \label{test:SBVerona}

\end{center}

\subsubsection{Conflation of SGIs}
\FloatBarrier

Two SGIs with the same INTID and the same source PE
can be taken together (associated to the same \asm{TakeException}).
This is allowed even with ordering between the two MSRs,
because they can be distributed-redistributed before the first is acknowledged.
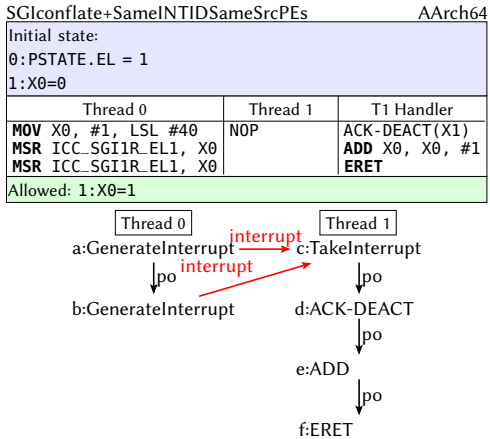
\begin{center}
\scalebox{0.75}{
\begin{tabular}{c}
    \input{tests/SGIconflate+SameINTIDSameSrcPEs.listing.tikz} \\
    \input{tests/SGIconflate+SameINTIDSameSrcPEs.diagram.tikz}
  \end{tabular}}
  \label{test:SGIconflate+SameINTIDSameSrcPEs}

\end{center}

\newpage
\noindent
Two SGIs with the same INTID but with different source PEs
can be taken together (associated to the same \asm{TakeException}).
With GICv3,
this is treated the same as the litmus test with the same source PE
(SGIconflate+SameINTIDSameSrcPEs), but with GICv2, this was forbidden
(SGIs were banked not just by INTID, but also by source PE).
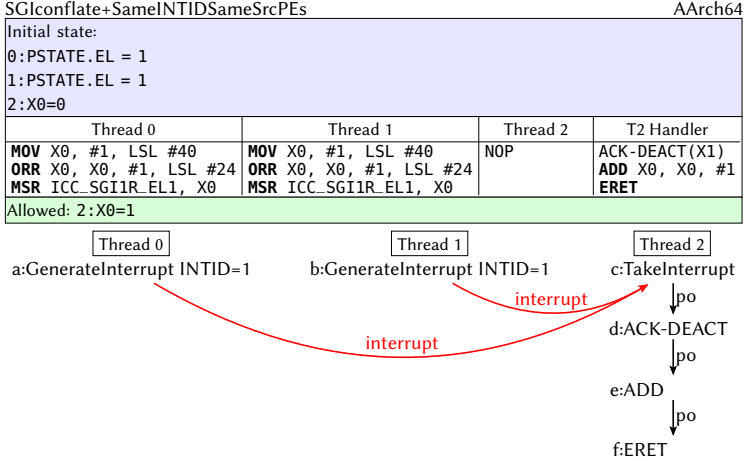
\begin{center}
\scalebox{0.75}{
\begin{tabular}{c}
    \input{tests/SGIconflate+SameINTIDDifferentSrcPEs.listing.tikz} \\
    \input{tests/SGIconflate+SameINTIDDifferentSrcPEs.diagram.tikz}
  \end{tabular}}
  \label{test:SGIconflate+SameINTIDDifferentSrcPEs}

\end{center}

\noindent
Two SGIs with different INTIDs and the same source PE
cannot be taken together (associated to the same \asm{TakeException}).
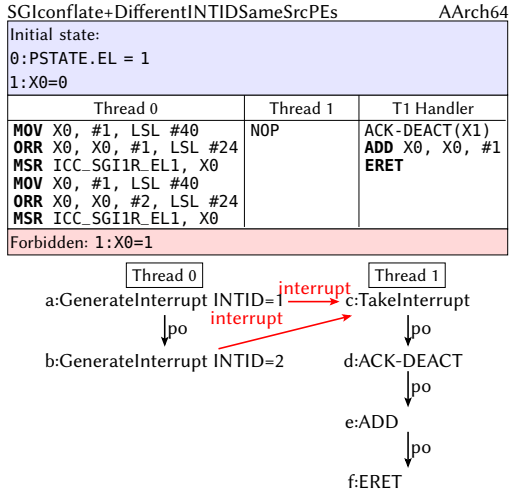
\begin{center}
\scalebox{0.75}{
\begin{tabular}{c}
    \input{tests/SGIconflate+DifferentINTIDSameSrcPEs.listing.tikz} \\
    \input{tests/SGIconflate+DifferentINTIDSameSrcPEs.diagram.tikz}
  \end{tabular}}
  \label{test:SGIconflate+DifferentINTIDSameSrcPEs}

\end{center}

\newpage
\subsubsection{Preemption of SGIs}
We do not cover priority in this document,
but priority and (un)masking
interact.

\

\begin{center}
  \input{tests/SGIpreempt+Priority}
\end{center}
\begin{center}
  \input{tests/SGIpreempt}
\end{center}

\begin{center}
  \input{tests/SGIpreempt+Unmask}
\end{center}

\newpage
\subsubsection{Acknowledge-deactivate sequence}

The same interrupt (as in, a single \asm{GenerateInterrupt}) can be taken twice (or more)
when it is not deactivated (and, to be in a position to do that, first acknowledged).

The basic test is below. 
The interrupt is not deactivated, and so it gets taken again
when it is not masked anymore (at the \asm{ERET},
which restores processor state),
forever.
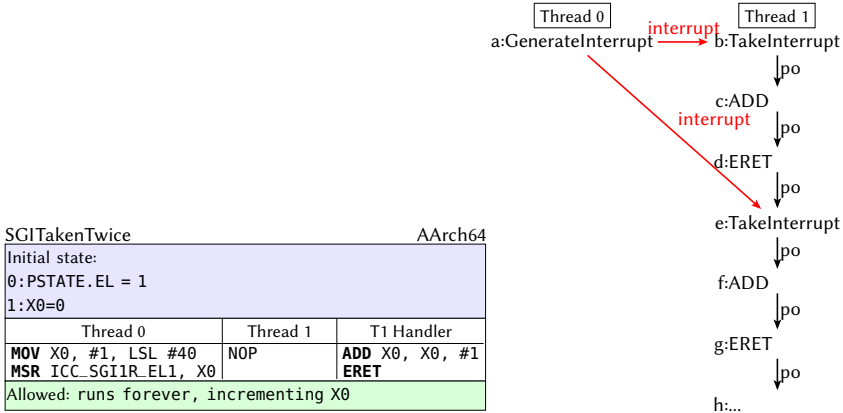
\begin{center}
\scalebox{0.75}{
\begin{tabular}{l@{}l}
    \input{tests/SGITakenTwice.listing.tikz}
    &
    \input{tests/SGITakenTwice.diagram.tikz}
  \end{tabular}}
  \label{test:SGITakenTwice}

\end{center}

\noindent
The way one deactivates an interrupt depends on the \asm{EOImode}.
With \asm{EOImode}=0, this is the standard acknowledge-deactivate sequence.
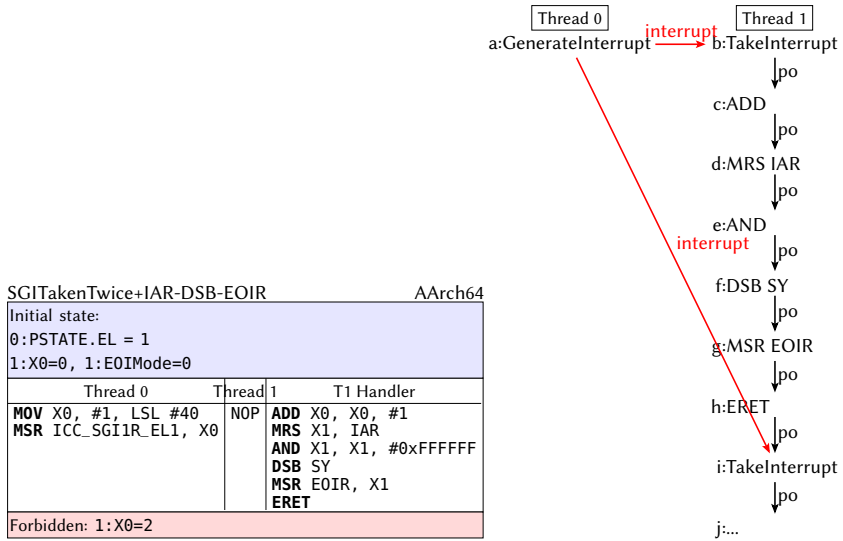
\begin{center}
\scalebox{0.75}{
\begin{tabular}{l@{}l}
    \input{tests/SGITakenTwice+IAR-DSB-EOIR.listing.tikz}
    &
    \input{tests/SGITakenTwice+IAR-DSB-EOIR.diagram.tikz}
  \end{tabular}}
  \label{test:SGITakenTwice+IAR-DSB-EOIR}

\end{center}

\newpage

\noindent
With \asm{EOImode}=1, this is the standard acknowledge-priority drop-deactivate sequence
(although the priority drop is not required, see \asm{SGITakenTwice+IAR-DSB-DIR}).
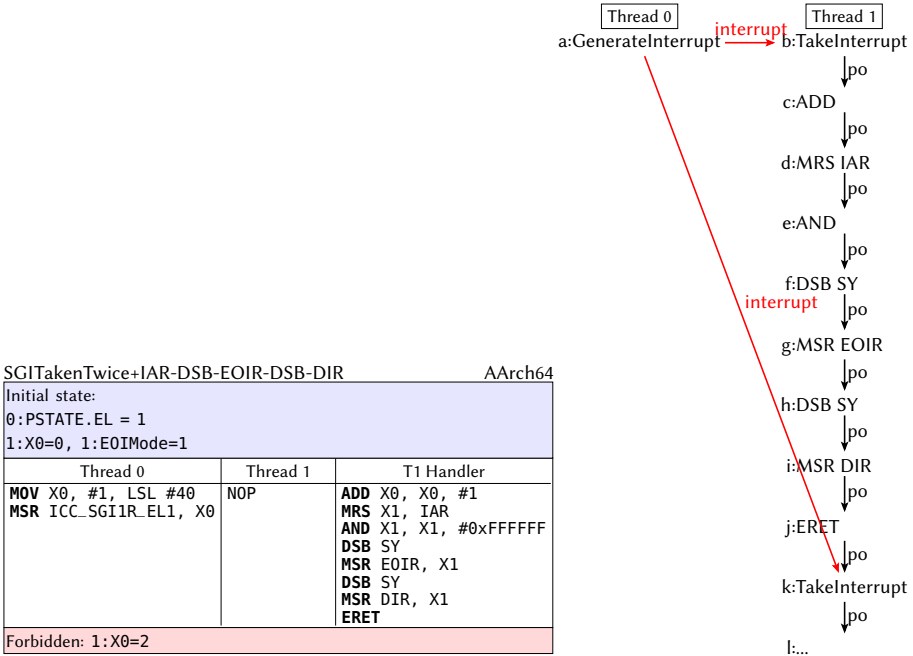
\begin{center}
\scalebox{0.75}{
\begin{tabular}{l@{}l}
    \input{tests/SGITakenTwice+IAR-DSB-EOIR-DSB-DIR.listing.tikz}
    &
    \input{tests/SGITakenTwice+IAR-DSB-EOIR-DSB-DIR.diagram.tikz}
  \end{tabular}}
  \label{test:SGITakenTwice+IAR-DSB-EOIR-DSB-DIR}

\end{center}

With \asm{EOImode}=1, dropping the priority is not required.
(With \asm{EOImode}=0, this is allowed!)
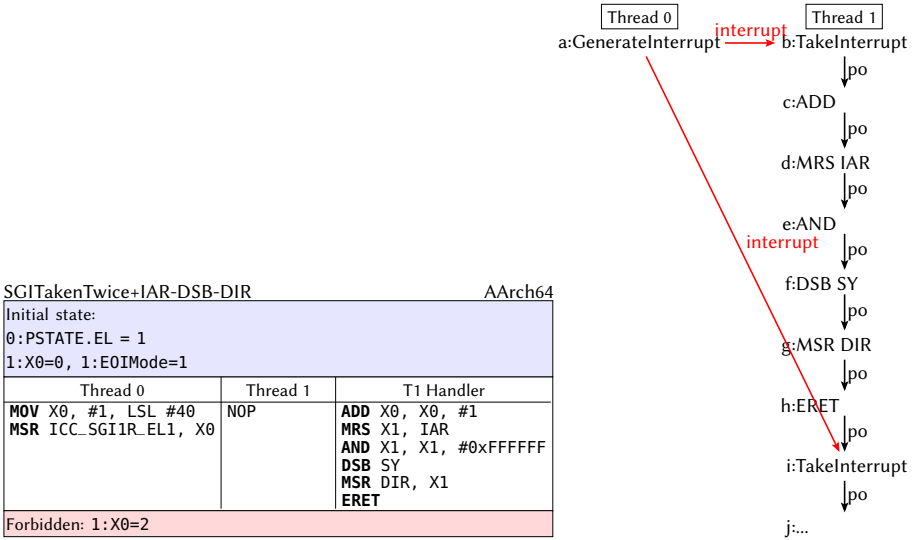
\begin{center}
\scalebox{0.75}{
\begin{tabular}{l@{}l}
    \input{tests/SGITakenTwice+IAR-DSB-DIR.listing.tikz}
    &
    \input{tests/SGITakenTwice+IAR-DSB-DIR.diagram.tikz}
  \end{tabular}}
  \label{test:SGITakenTwice+IAR-DSB-DIR}

\end{center}

\newpage
On the other hand, partial or out-of-order acknowledgement-deactivation
does not prevent the interrupt from being taken again --- examples:

\noindent
Merely acknowledging does not prevent the interrupt from being taken again.
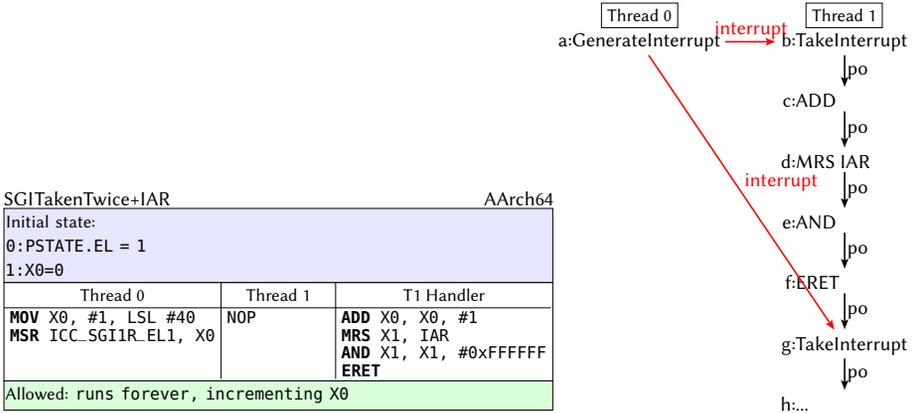
\begin{center}
\scalebox{0.75}{
\begin{tabular}{l@{}l}
    \input{tests/SGITakenTwice+IAR.listing.tikz}
    &
    \input{tests/SGITakenTwice+IAR.diagram.tikz}
  \end{tabular}}
  \label{test:SGITakenTwice+IAR}

\end{center}

\medskip

\noindent
Deactivating (\asm{EOImode}=0) without acknowledging does not prevent the interrupt from being taken again.
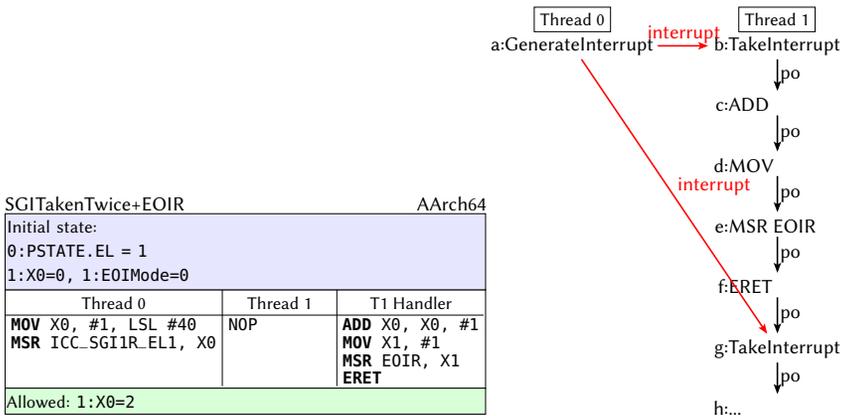
\begin{center}
\scalebox{0.75}{
\begin{tabular}{l@{}l}
    \input{tests/SGITakenTwice+EOIR.listing.tikz}
    &
    \input{tests/SGITakenTwice+EOIR.diagram.tikz}
  \end{tabular}}
  \label{test:SGITakenTwice+EOIR}

\end{center}

\medskip

\noindent
Deactivating (\asm{EOImode}=1) without acknowledging does not prevent the interrupt from being taken again.
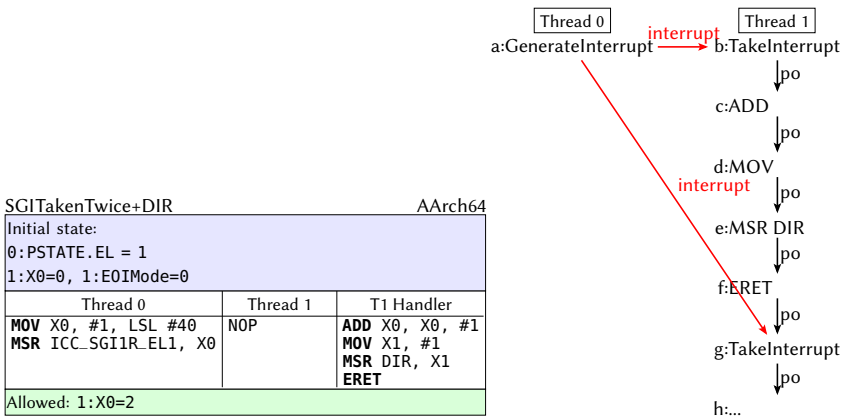
\begin{center}
\scalebox{0.75}{
\begin{tabular}{l@{}l}
    \input{tests/SGITakenTwice+DIR.listing.tikz}
    &
    \input{tests/SGITakenTwice+DIR.diagram.tikz}
  \end{tabular}}
  \label{test:SGITakenTwice+DIR}

\end{center}

\medskip

\noindent
Deactivating then acknowledging does not prevent the interrupt from being taken again.
The incorrect order of actions means that the automaton state does not move.
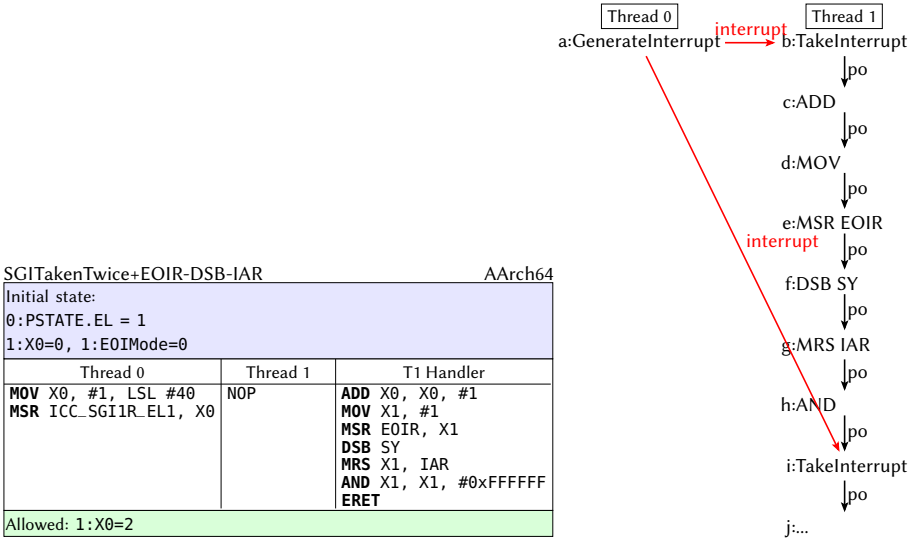
\begin{center}
\scalebox{0.75}{
\begin{tabular}{l@{}l}
    \input{tests/SGITakenTwice+EOIR-DSB-IAR.listing.tikz}
    &
    \input{tests/SGITakenTwice+EOIR-DSB-IAR.diagram.tikz}
  \end{tabular}}
  \label{test:SGITakenTwice+EOIR-DSB-IAR}

\end{center}

\medskip

\noindent
Acknowledging followed in program order by deactivating
does not prevent the interrupt from being taken again.
The GIC might see the two effects in different orders.
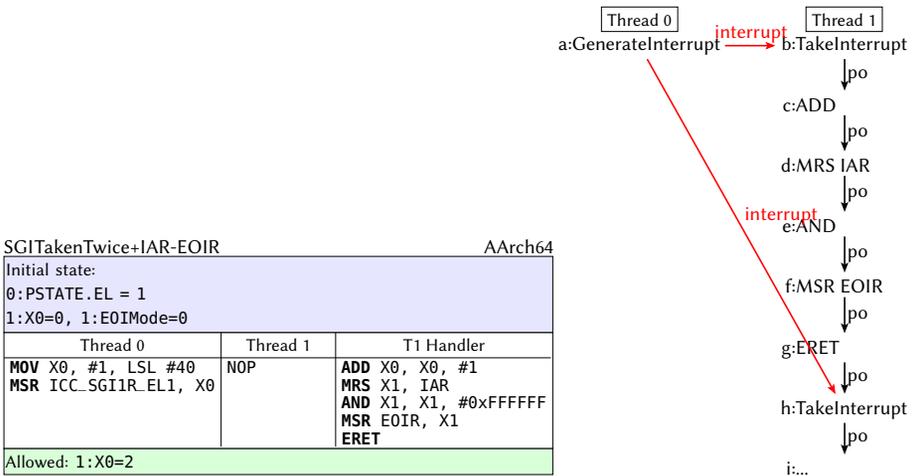
\begin{center}
\scalebox{0.75}{
\begin{tabular}{l@{}l}
    \input{tests/SGITakenTwice+IAR-EOIR.listing.tikz}
    &
    \input{tests/SGITakenTwice+IAR-EOIR.diagram.tikz}
  \end{tabular}}
  \label{test:SGITakenTwice+IAR-EOIR}

\end{center}

\fi %

\clearpage

\ifarxiv
  \ifarxivbuild

  \else
    \bibliography{refs}
  \fi
\else
  \bibliography{refs}

\begin{thebibliography}{58}


\ifx \showCODEN    \undefined \def \showCODEN     #1{\unskip}     \fi
\ifx \showDOI      \undefined \def \showDOI       #1{#1}\fi
\ifx \showISBNx    \undefined \def \showISBNx     #1{\unskip}     \fi
\ifx \showISBNxiii \undefined \def \showISBNxiii  #1{\unskip}     \fi
\ifx \showISSN     \undefined \def \showISSN      #1{\unskip}     \fi
\ifx \showLCCN     \undefined \def \showLCCN      #1{\unskip}     \fi
\ifx \shownote     \undefined \def \shownote      #1{#1}          \fi
\ifx \showarticletitle \undefined \def \showarticletitle #1{#1}   \fi
\ifx \showURL      \undefined \def \showURL       {\relax}        \fi
\providecommand\bibfield[2]{#2}
\providecommand\bibinfo[2]{#2}
\providecommand\natexlab[1]{#1}
\providecommand\showeprint[2][]{arXiv:#2}

\bibitem[Adir et~al\mbox{.}(2003)]%
        {Adir:2003}
\bibfield{author}{\bibinfo{person}{A. Adir}, \bibinfo{person}{H. Attiya}, {and}
  \bibinfo{person}{G. Shurek}.} \bibinfo{year}{2003}\natexlab{}.
\newblock \showarticletitle{Information-Flow Models for Shared Memory with an
  Application to the {PowerPC} Architecture}.
\newblock \bibinfo{journal}{\emph{IEEE Trans. Parallel Distrib. Syst.}}
  \bibinfo{volume}{14}, \bibinfo{number}{5} (\bibinfo{year}{2003}),
  \bibinfo{pages}{502--515}.
\newblock
\showISSN{1045-9219}
\urldef\tempurl%
\url{https://doi.org/10.1109/TPDS.2003.1199067}
\showDOI{\tempurl}


\bibitem[Alglave(2010)]%
        {JadeThesis}
\bibfield{author}{\bibinfo{person}{Jade Alglave}.}
  \bibinfo{year}{2010}\natexlab{}.
\newblock \emph{\bibinfo{title}{A Shared Memory Poetics}}.
\newblock \bibinfo{thesistype}{Ph.\,D. Dissertation}.
  \bibinfo{school}{Universit{\'{e}} Paris 7 -- Denis Diderot}.
\newblock


\bibitem[Alglave et~al\mbox{.}(2021)]%
        {AlglaveDGHM21}
\bibfield{author}{\bibinfo{person}{Jade Alglave}, \bibinfo{person}{Will
  Deacon}, \bibinfo{person}{Richard Grisenthwaite}, \bibinfo{person}{Antoine
  Hacquard}, {and} \bibinfo{person}{Luc Maranget}.}
  \bibinfo{year}{2021}\natexlab{}.
\newblock \showarticletitle{Armed Cats: Formal Concurrency Modelling at Arm}.
\newblock \bibinfo{journal}{\emph{{ACM} Trans. Program. Lang. Syst.}}
  \bibinfo{volume}{43}, \bibinfo{number}{2} (\bibinfo{year}{2021}),
  \bibinfo{pages}{8:1--8:54}.
\newblock
\urldef\tempurl%
\url{https://doi.org/10.1145/3458926}
\showDOI{\tempurl}


\bibitem[Alglave et~al\mbox{.}(2024)]%
        {alglave:hal-04567296}
\bibfield{author}{\bibinfo{person}{Jade Alglave}, \bibinfo{person}{Richard
  Grisenthwaite}, \bibinfo{person}{Artem Khyzha}, \bibinfo{person}{Luc
  Maranget}, {and} \bibinfo{person}{Nikos Nikoleris}.}
  \bibinfo{year}{2024}\natexlab{}.
\newblock \bibinfo{title}{{Puss In Boots: on formalising {Arm's Virtual Memory
  System Architecture} (extended version)}}.  (\bibinfo{date}{May}
  \bibinfo{year}{2024}).
\newblock
\urldef\tempurl%
\url{https://inria.hal.science/hal-04567296}
\showURL{%
\tempurl}
\newblock
\shownote{working paper or preprint}.


\bibitem[Alglave et~al\mbox{.}(2018)]%
        {AlglaveMMPS18}
\bibfield{author}{\bibinfo{person}{Jade Alglave}, \bibinfo{person}{Luc
  Maranget}, \bibinfo{person}{Paul~E. McKenney}, \bibinfo{person}{Andrea
  Parri}, {and} \bibinfo{person}{Alan~S. Stern}.}
  \bibinfo{year}{2018}\natexlab{}.
\newblock \showarticletitle{Frightening Small Children and Disconcerting
  Grown-ups: Concurrency in the Linux Kernel}. In
  \bibinfo{booktitle}{\emph{Proceedings of the Twenty-Third International
  Conference on Architectural Support for Programming Languages and Operating
  Systems, {ASPLOS} 2018, Williamsburg, VA, USA, March 24-28, 2018}},
  \bibfield{editor}{\bibinfo{person}{Xipeng Shen}, \bibinfo{person}{James
  Tuck}, \bibinfo{person}{Ricardo Bianchini}, {and} \bibinfo{person}{Vivek
  Sarkar}} (Eds.). \bibinfo{publisher}{{ACM}}, \bibinfo{pages}{405--418}.
\newblock
\urldef\tempurl%
\url{https://doi.org/10.1145/3173162.3177156}
\showDOI{\tempurl}


\bibitem[Alglave et~al\mbox{.}(2010)]%
        {cav2010}
\bibfield{author}{\bibinfo{person}{Jade Alglave}, \bibinfo{person}{Luc
  Maranget}, \bibinfo{person}{Susmit Sarkar}, {and} \bibinfo{person}{Peter
  Sewell}.} \bibinfo{year}{2010}\natexlab{}.
\newblock \showarticletitle{Fences in Weak Memory Models}. In
  \bibinfo{booktitle}{\emph{Computer Aided Verification, 22nd International
  Conference, {CAV} 2010, Edinburgh, UK, July 15-19, 2010. Proceedings}}
  \emph{(\bibinfo{series}{Lecture Notes in Computer Science},
  Vol.~\bibinfo{volume}{6174})}, \bibfield{editor}{\bibinfo{person}{Tayssir
  Touili}, \bibinfo{person}{Byron Cook}, {and} \bibinfo{person}{Paul~B.
  Jackson}} (Eds.). \bibinfo{publisher}{Springer}, \bibinfo{pages}{258--272}.
\newblock
\urldef\tempurl%
\url{https://doi.org/10.1007/978-3-642-14295-6\_25}
\showDOI{\tempurl}


\bibitem[Alglave et~al\mbox{.}(2011)]%
        {AMSStacas2011}
\bibfield{author}{\bibinfo{person}{J. Alglave}, \bibinfo{person}{L. Maranget},
  \bibinfo{person}{S. Sarkar}, {and} \bibinfo{person}{P. Sewell}.}
  \bibinfo{year}{2011}\natexlab{}.
\newblock \showarticletitle{Litmus: Running Tests Against Hardware}. In
  \bibinfo{booktitle}{\emph{Proc.~TACAS}}.
\newblock
\urldef\tempurl%
\url{https://doi.org/10.1007/978-3-642-19835-9_5}
\showDOI{\tempurl}


\bibitem[Alglave et~al\mbox{.}(2014)]%
        {DBLP:journals/toplas/AlglaveMT14}
\bibfield{author}{\bibinfo{person}{Jade Alglave}, \bibinfo{person}{Luc
  Maranget}, {and} \bibinfo{person}{Michael Tautschnig}.}
  \bibinfo{year}{2014}\natexlab{}.
\newblock \showarticletitle{Herding Cats: Modelling, Simulation, Testing, and
  Data Mining for Weak Memory}.
\newblock \bibinfo{journal}{\emph{{ACM} Trans. Program. Lang. Syst.}}
  \bibinfo{volume}{36}, \bibinfo{number}{2} (\bibinfo{year}{2014}),
  \bibinfo{pages}{7:1--7:74}.
\newblock
\urldef\tempurl%
\url{https://doi.org/10.1145/2627752}
\showDOI{\tempurl}


\bibitem[Arm(2024)]%
        {Arm-K.a}
\bibfield{author}{\bibinfo{person}{Arm}.} \bibinfo{year}{2024}\natexlab{}.
\newblock \bibinfo{title}{{Arm Architecture Reference Manual: for A-profile
  architecture}}.
\newblock
  \bibinfo{howpublished}{\url{https://developer.arm.com/documentation/ddi0487/latest}}.
\newblock
\newblock
\shownote{Accessed 2024-05-11. Issue K.a. 14777 pages.}.


\bibitem[{Arm}(2024)]%
        {arm-gic-v3-v4}
\bibfield{author}{\bibinfo{person}{{Arm}}.} \bibinfo{year}{2024}\natexlab{}.
\newblock \bibinfo{booktitle}{\emph{Arm Generic Interrupt Controller
  Architecture Specification, {GIC} architecture version 3 and version 4}}.
\newblock \bibinfo{type}{{T}echnical {R}eport}. \bibinfo{institution}{{Arm}}.
\newblock
\newblock
\shownote{IHI 0069H.b (ID041224)}.


\bibitem[Armstrong et~al\mbox{.}(2019)]%
        {sail-popl2019}
\bibfield{author}{\bibinfo{person}{Alasdair Armstrong}, \bibinfo{person}{Thomas
  Bauereiss}, \bibinfo{person}{Brian Campbell}, \bibinfo{person}{Alastair
  Reid}, \bibinfo{person}{Kathryn~E. Gray}, \bibinfo{person}{Robert~M. Norton},
  \bibinfo{person}{Prashanth Mundkur}, \bibinfo{person}{Mark Wassell},
  \bibinfo{person}{Jon French}, \bibinfo{person}{Christopher Pulte},
  \bibinfo{person}{Shaked Flur}, \bibinfo{person}{Ian Stark},
  \bibinfo{person}{Neel Krishnaswami}, {and} \bibinfo{person}{Peter Sewell}.}
  \bibinfo{year}{2019}\natexlab{}.
\newblock \showarticletitle{{ISA} Semantics for {ARMv8-A, RISC-V, and
  CHERI-MIPS}}. In \bibinfo{booktitle}{\emph{Proceedings of the 46th ACM
  SIGPLAN Symposium on Principles of Programming Languages}}.
\newblock
\urldef\tempurl%
\url{https://doi.org/10.1145/3290384}
\showDOI{\tempurl}
\newblock
\shownote{Proc. ACM Program. Lang. 3, POPL, Article 71}.


\bibitem[Armstrong et~al\mbox{.}(2021)]%
        {isla-cav}
\bibfield{author}{\bibinfo{person}{Alasdair Armstrong}, \bibinfo{person}{Brian
  Campbell}, \bibinfo{person}{Ben Simner}, \bibinfo{person}{Christopher Pulte},
  {and} \bibinfo{person}{Peter Sewell}.} \bibinfo{year}{2021}\natexlab{}.
\newblock \showarticletitle{Isla: Integrating full-scale {ISA} semantics and
  axiomatic concurrency models}. In \bibinfo{booktitle}{\emph{Proc. 33rd
  International Conference on Computer-Aided Verification}}
  \emph{(\bibinfo{series}{Lecture Notes in Computer Science},
  Vol.~\bibinfo{volume}{12759})}. \bibinfo{publisher}{Springer},
  \bibinfo{pages}{303--316}.
\newblock
\urldef\tempurl%
\url{https://doi.org/10.1007/978-3-030-81685-8\_14}
\showDOI{\tempurl}


\bibitem[Batty et~al\mbox{.}(2015)]%
        {BattyMNPS15}
\bibfield{author}{\bibinfo{person}{Mark Batty}, \bibinfo{person}{Kayvan
  Memarian}, \bibinfo{person}{Kyndylan Nienhuis}, \bibinfo{person}{Jean
  Pichon{-}Pharabod}, {and} \bibinfo{person}{Peter Sewell}.}
  \bibinfo{year}{2015}\natexlab{}.
\newblock \showarticletitle{The Problem of Programming Language Concurrency
  Semantics}. In \bibinfo{booktitle}{\emph{Programming Languages and Systems -
  24th European Symposium on Programming, {ESOP} 2015, Held as Part of the
  European Joint Conferences on Theory and Practice of Software, {ETAPS} 2015,
  London, UK, April 11-18, 2015. Proceedings}} \emph{(\bibinfo{series}{Lecture
  Notes in Computer Science}, Vol.~\bibinfo{volume}{9032})},
  \bibfield{editor}{\bibinfo{person}{Jan Vitek}} (Ed.).
  \bibinfo{publisher}{Springer}, \bibinfo{pages}{283--307}.
\newblock
\urldef\tempurl%
\url{https://doi.org/10.1007/978-3-662-46669-8\_12}
\showDOI{\tempurl}


\bibitem[Batty(2015)]%
        {Batty15}
\bibfield{author}{\bibinfo{person}{Mark~John Batty}.}
  \bibinfo{year}{2015}\natexlab{}.
\newblock \emph{\bibinfo{title}{The {C11} and {C++11} concurrency model}}.
\newblock \bibinfo{thesistype}{Ph.\,D. Dissertation}.
  \bibinfo{school}{University of Cambridge, {UK}}.
\newblock
\urldef\tempurl%
\url{https://ethos.bl.uk/OrderDetails.do?uin=uk.bl.ethos.708458}
\showURL{%
\tempurl}


\bibitem[Bauereiss et~al\mbox{.}(2024)]%
        {sail-arm-9.4}
\bibfield{author}{\bibinfo{person}{Thomas Bauereiss}, \bibinfo{person}{Brian
  Campbell}, \bibinfo{person}{Alasdair Armstrong}, \bibinfo{person}{Alastair
  Reid}, \bibinfo{person}{Kathryn~E. Gray}, \bibinfo{person}{Anthony Fox},
  \bibinfo{person}{Peter Sewell}, {and} \bibinfo{person}{Arm Limited}.}
  \bibinfo{year}{2024}\natexlab{}.
\newblock \bibinfo{title}{Sail {Armv9.4-A} instruction-set architecture ({ISA})
  model}.
\newblock
\newblock
\newblock
\shownote{\url{https://github.com/rems-project/sail-arm}. Accessed
  2024-05-11.}.


\bibitem[Burrows(2004)]%
        {Burrows2004}
\bibfield{author}{\bibinfo{person}{Mike Burrows}.}
  \bibinfo{year}{2004}\natexlab{}.
\newblock \bibinfo{booktitle}{\emph{How to Implement Unnecessary Mutexes}}.
\newblock \bibinfo{publisher}{Springer New York}, \bibinfo{address}{New York,
  NY}, \bibinfo{pages}{51--57}.
\newblock
\showISBNx{978-0-387-21821-2}
\urldef\tempurl%
\url{https://doi.org/10.1007/0-387-21821-1_7}
\showDOI{\tempurl}


\bibitem[Chakraborty(2019)]%
        {Chakraborty19}
\bibfield{author}{\bibinfo{person}{Soham Chakraborty}.}
  \bibinfo{year}{2019}\natexlab{}.
\newblock \emph{\bibinfo{title}{Correct Compilation of Relaxed Memory
  Concurrency}}.
\newblock \bibinfo{thesistype}{Ph.\,D. Dissertation}.
  \bibinfo{school}{Kaiserslautern University of Technology, Germany}.
\newblock
\urldef\tempurl%
\url{https://kluedo.ub.rptu.de/frontdoor/index/index/docId/5697}
\showURL{%
\tempurl}


\bibitem[Cheeseman et~al\mbox{.}(2023)]%
        {CheesemanEtAl2023}
\bibfield{author}{\bibinfo{person}{Luke Cheeseman}, \bibinfo{person}{Matthew~J.
  Parkinson}, \bibinfo{person}{Sylvan Clebsch}, \bibinfo{person}{Marios
  Kogias}, \bibinfo{person}{Sophia Drossopoulou}, \bibinfo{person}{David
  Chisnall}, \bibinfo{person}{Tobias Wrigstad}, {and} \bibinfo{person}{Paul
  Liétar}.} \bibinfo{year}{2023}\natexlab{}.
\newblock \showarticletitle{When Concurrency Matters: Behaviour-Oriented
  Concurrency}.
\newblock \bibinfo{journal}{\emph{Proc. ACM Program. Lang.}}
  \bibinfo{volume}{7}, \bibinfo{number}{OOPSLA2} (\bibinfo{date}{October}
  \bibinfo{year}{2023}).
\newblock
\urldef\tempurl%
\url{https://www.microsoft.com/en-us/research/publication/when-concurrency-matters-behaviour-oriented-concurrency/}
\showURL{%
\tempurl}


\bibitem[Collier(1992)]%
        {DBLP:books/daglib/0073498}
\bibfield{author}{\bibinfo{person}{William~W. Collier}.}
  \bibinfo{year}{1992}\natexlab{}.
\newblock \bibinfo{booktitle}{\emph{Reasoning about parallel architectures}}.
\newblock \bibinfo{publisher}{Prentice Hall}.
\newblock
\showISBNx{978-0-13-766098-8}


\bibitem[Dave~Dice(2001)]%
        {AsymmetricDekkerSynchronization2001}
\bibfield{author}{\bibinfo{person}{Mingyao~Yang Dave~Dice, Hui~Huang}.}
  \bibinfo{year}{2001}\natexlab{}.
\newblock \bibinfo{title}{Asymmetric Dekker Synchronization}.
\newblock
\newblock
\urldef\tempurl%
\url{http://web.archive.org/web/20070214114205/http://blogs.sun.com/dave/resource/Asymmetric-Dekker-Synchronization.txt}
\showURL{%
\tempurl}


\bibitem[Deacon et~al\mbox{.}(2023)]%
        {arm-cat}
\bibfield{author}{\bibinfo{person}{Will Deacon}, \bibinfo{person}{Jade
  Alglave}, \bibinfo{person}{Nikos Nikoleris}, {and} \bibinfo{person}{Artem
  Khyzha}.} \bibinfo{year}{2023}\natexlab{}.
\newblock \bibinfo{title}{The {ARMv8} Application Level Memory Model}.
\newblock
  \bibinfo{howpublished}{\url{https://github.com/herd/herdtools7/blob/master/herd/libdir/aarch64.cat}
  (accessed 2019-07-01)}.
\newblock
\newblock
\shownote{Accessed 2024-11-19}.


\bibitem[Dice(2006)]%
        {biasedlocking}
\bibfield{author}{\bibinfo{person}{Dave Dice}.}
  \bibinfo{year}{2006}\natexlab{}.
\newblock \bibinfo{title}{Biased Locking in Hotspot}.
\newblock \bibinfo{howpublished}{Oracle Blog, Wayback Machine}.
\newblock
\urldef\tempurl%
\url{http://web.archive.org/web/20150320095550/https://blogs.oracle.com/dave/entry/biased_locking_in_hotspot}
\showURL{%
\tempurl}


\bibitem[Dice et~al\mbox{.}(2010)]%
        {biasedlockingpatent}
\bibfield{author}{\bibinfo{person}{David Dice}, \bibinfo{person}{Mark~S. Moir},
  {and} \bibinfo{person}{William N.~Scherer III}.}
  \bibinfo{year}{2010}\natexlab{}.
\newblock \bibinfo{title}{United States Patent US 7814488B1 Quickly
  Reacquirable Locks}.
\newblock \bibinfo{howpublished}{United Statess Patent Office}.
\newblock


\bibitem[Flur et~al\mbox{.}(2016)]%
        {DBLP:conf/popl/FlurGPSSMDS16}
\bibfield{author}{\bibinfo{person}{Shaked Flur}, \bibinfo{person}{Kathryn~E.
  Gray}, \bibinfo{person}{Christopher Pulte}, \bibinfo{person}{Susmit Sarkar},
  \bibinfo{person}{Ali Sezgin}, \bibinfo{person}{Luc Maranget},
  \bibinfo{person}{Will Deacon}, {and} \bibinfo{person}{Peter Sewell}.}
  \bibinfo{year}{2016}\natexlab{}.
\newblock \showarticletitle{Modelling the {ARMv8} architecture, operationally:
  concurrency and {ISA}}. In \bibinfo{booktitle}{\emph{Proceedings of the 43rd
  {ACM} {SIGPLAN-SIGACT} Symposium on Principles of Programming Languages (St.
  Petersburg, FL, USA)}}. \bibinfo{pages}{608--621}.
\newblock
\urldef\tempurl%
\url{https://doi.org/10.1145/2837614.2837615}
\showDOI{\tempurl}


\bibitem[Flur et~al\mbox{.}(2017)]%
        {mixed17}
\bibfield{author}{\bibinfo{person}{Shaked Flur}, \bibinfo{person}{Susmit
  Sarkar}, \bibinfo{person}{Christopher Pulte}, \bibinfo{person}{Kyndylan
  Nienhuis}, \bibinfo{person}{Luc Maranget}, \bibinfo{person}{Kathryn~E. Gray},
  \bibinfo{person}{Ali Sezgin}, \bibinfo{person}{Mark Batty}, {and}
  \bibinfo{person}{Peter Sewell}.} \bibinfo{year}{2017}\natexlab{}.
\newblock \showarticletitle{Mixed-size concurrency: ARM, POWER, C/C++11, and
  {SC}}. In \bibinfo{booktitle}{\emph{Proceedings of the 44th {ACM} {SIGPLAN}
  Symposium on Principles of Programming Languages, {POPL} 2017, Paris, France,
  January 18-20, 2017}}, \bibfield{editor}{\bibinfo{person}{Giuseppe Castagna}
  {and} \bibinfo{person}{Andrew~D. Gordon}} (Eds.). \bibinfo{publisher}{{ACM}},
  \bibinfo{pages}{429--442}.
\newblock
\urldef\tempurl%
\url{https://doi.org/10.1145/3009837.3009839}
\showDOI{\tempurl}


\bibitem[Gharachorloo(1995)]%
        {GharachorlooPhD}
\bibfield{author}{\bibinfo{person}{Kourosh Gharachorloo}.}
  \bibinfo{year}{1995}\natexlab{}.
\newblock \emph{\bibinfo{title}{Memory Consistency Models for Shared-Memory
  Multiprocessors}}.
\newblock \bibinfo{thesistype}{Ph.\,D. Dissertation}. \bibinfo{school}{Stanford
  University}.
\newblock


\bibitem[Gharachorloo et~al\mbox{.}(1990)]%
        {DBLP:conf/isca/GharachorlooLLGGH90}
\bibfield{author}{\bibinfo{person}{Kourosh Gharachorloo},
  \bibinfo{person}{Daniel Lenoski}, \bibinfo{person}{James Laudon},
  \bibinfo{person}{Phillip~B. Gibbons}, \bibinfo{person}{Anoop Gupta}, {and}
  \bibinfo{person}{John~L. Hennessy}.} \bibinfo{year}{1990}\natexlab{}.
\newblock \showarticletitle{Memory Consistency and Event Ordering in Scalable
  Shared-Memory Multiprocessors}. In \bibinfo{booktitle}{\emph{Proceedings of
  the 17th Annual International Symposium on Computer Architecture, Seattle,
  WA, USA, June 1990}}, \bibfield{editor}{\bibinfo{person}{Jean{-}Loup Baer},
  \bibinfo{person}{Larry Snyder}, {and} \bibinfo{person}{James~R. Goodman}}
  (Eds.). \bibinfo{publisher}{{ACM}}, \bibinfo{pages}{15--26}.
\newblock
\urldef\tempurl%
\url{https://doi.org/10.1145/325164.325102}
\showDOI{\tempurl}


\bibitem[Gray et~al\mbox{.}(2015)]%
        {DBLP:conf/micro/GrayKMPSS15}
\bibfield{author}{\bibinfo{person}{Kathryn~E. Gray}, \bibinfo{person}{Gabriel
  Kerneis}, \bibinfo{person}{Dominic~P. Mulligan}, \bibinfo{person}{Christopher
  Pulte}, \bibinfo{person}{Susmit Sarkar}, {and} \bibinfo{person}{Peter
  Sewell}.} \bibinfo{year}{2015}\natexlab{}.
\newblock \showarticletitle{An integrated concurrency and core-{ISA}
  architectural envelope definition, and test oracle, for {IBM} {POWER}
  multiprocessors}. In \bibinfo{booktitle}{\emph{Proceedings of the 48th
  International Symposium on Microarchitecture (Waikiki)}}.
  \bibinfo{pages}{635--646}.
\newblock
\urldef\tempurl%
\url{https://doi.org/10.1145/2830772.2830775}
\showDOI{\tempurl}


\bibitem[Hennessy and Patterson(2012)]%
        {HennessyPatterson12}
\bibfield{author}{\bibinfo{person}{John~L. Hennessy} {and}
  \bibinfo{person}{David~A. Patterson}.} \bibinfo{year}{2012}\natexlab{}.
\newblock \bibinfo{booktitle}{\emph{Computer Architecture: A Quantitative
  Approach} (\bibinfo{edition}{5} ed.)}.
\newblock \bibinfo{publisher}{Morgan Kaufmann}, \bibinfo{address}{Amsterdam}.
\newblock
\showISBNx{978-0-12-383872-8}


\bibitem[\ifanon Anonymous~\else Luc~Maranget\fi(2024)]%
        {lucpc}
\bibfield{author}{\bibinfo{person}{\ifanon Anonymous~\else Luc~Maranget\fi}.}
  \bibinfo{year}{2024}\natexlab{}.
\newblock \bibinfo{title}{Personal communication}.
\newblock
\newblock


\bibitem[Intel(2002)]%
        {ItaniumFormal}
\bibfield{author}{\bibinfo{person}{Intel}.} \bibinfo{year}{2002}\natexlab{}.
\newblock \bibinfo{title}{A Formal Specification of {I}ntel {I}tanium Processor
  Family Memory Ordering}.
\newblock
\newblock
\newblock
\shownote{\url{developer.intel.com/design/itanium/downloads/251429.htm}}.


\bibitem[Jagadeesan et~al\mbox{.}(2020)]%
        {JagadeesanJR20}
\bibfield{author}{\bibinfo{person}{Radha Jagadeesan}, \bibinfo{person}{Alan
  Jeffrey}, {and} \bibinfo{person}{James Riely}.}
  \bibinfo{year}{2020}\natexlab{}.
\newblock \showarticletitle{Pomsets with preconditions: a simple model of
  relaxed memory}.
\newblock \bibinfo{journal}{\emph{Proc. {ACM} Program. Lang.}}
  \bibinfo{volume}{4}, \bibinfo{number}{{OOPSLA}} (\bibinfo{year}{2020}),
  \bibinfo{pages}{194:1--194:30}.
\newblock
\urldef\tempurl%
\url{https://doi.org/10.1145/3428262}
\showDOI{\tempurl}


\bibitem[Kang et~al\mbox{.}(2017)]%
        {kang-et-al:promising-semantics}
\bibfield{author}{\bibinfo{person}{Jeehoon Kang}, \bibinfo{person}{Chung-Kil
  Hur}, \bibinfo{person}{Ori Lahav}, \bibinfo{person}{Viktor Vafeiadis}, {and}
  \bibinfo{person}{Derek Dreyer}.} \bibinfo{year}{2017}\natexlab{}.
\newblock \showarticletitle{A promising semantics for relaxed-memory
  concurrency}. In \bibinfo{booktitle}{\emph{Proceedings of the 44th ACM
  SIGPLAN Symposium on Principles of Programming Languages}} (Paris, France)
  \emph{(\bibinfo{series}{POPL '17})}. \bibinfo{publisher}{Association for
  Computing Machinery}, \bibinfo{address}{New York, NY, USA},
  \bibinfo{pages}{175–189}.
\newblock
\showISBNx{9781450346603}
\urldef\tempurl%
\url{https://doi.org/10.1145/3009837.3009850}
\showDOI{\tempurl}


\bibitem[Kawachiya(2005)]%
        {Kawachiya05}
\bibfield{author}{\bibinfo{person}{Kiyokuni Kawachiya}.}
  \bibinfo{year}{2005}\natexlab{}.
\newblock \emph{\bibinfo{title}{Java Locks: Analysis and Acceleration}}.
\newblock \bibinfo{thesistype}{Ph.\,D. Dissertation}. \bibinfo{school}{Keio
  University}.
\newblock


\bibitem[Kawachiya et~al\mbox{.}(2002)]%
        {Kawachiya02}
\bibfield{author}{\bibinfo{person}{Kiyokuni Kawachiya}, \bibinfo{person}{Akira
  Koseki}, {and} \bibinfo{person}{Tamiya Onodera}.}
  \bibinfo{year}{2002}\natexlab{}.
\newblock \showarticletitle{Lock reservation: Java locks can mostly do without
  atomic operations}. In \bibinfo{booktitle}{\emph{Proceedings of the 17th ACM
  SIGPLAN Conference on Object-Oriented Programming, Systems, Languages, and
  Applications}} (Seattle, Washington, USA) \emph{(\bibinfo{series}{OOPSLA
  '02})}. \bibinfo{publisher}{Association for Computing Machinery},
  \bibinfo{address}{New York, NY, USA}, \bibinfo{pages}{130–141}.
\newblock
\showISBNx{1581134711}
\urldef\tempurl%
\url{https://doi.org/10.1145/582419.582433}
\showDOI{\tempurl}


\bibitem[Kokologiannakis et~al\mbox{.}(2017)]%
  {kokologiannakis-et-al:effective-stateless-model-checking-for-c/c++-concurrency}
\bibfield{author}{\bibinfo{person}{Michalis Kokologiannakis},
  \bibinfo{person}{Ori Lahav}, \bibinfo{person}{Konstantinos Sagonas}, {and}
  \bibinfo{person}{Viktor Vafeiadis}.} \bibinfo{year}{2017}\natexlab{}.
\newblock \showarticletitle{Effective stateless model checking for C/C++
  concurrency}.
\newblock \bibinfo{journal}{\emph{Proc. ACM Program. Lang.}}
  \bibinfo{volume}{2}, \bibinfo{number}{POPL}, Article \bibinfo{articleno}{17}
  (\bibinfo{date}{dec} \bibinfo{year}{2017}), \bibinfo{numpages}{32}~pages.
\newblock
\urldef\tempurl%
\url{https://doi.org/10.1145/3158105}
\showDOI{\tempurl}


\bibitem[Kokologiannakis et~al\mbox{.}(2019)]%
        {kokologiannakis-et-al:model-checking-for-weakly-consistent-library}
\bibfield{author}{\bibinfo{person}{Michalis Kokologiannakis},
  \bibinfo{person}{Azalea Raad}, {and} \bibinfo{person}{Viktor Vafeiadis}.}
  \bibinfo{year}{2019}\natexlab{}.
\newblock \showarticletitle{Model checking for weakly consistent libraries}. In
  \bibinfo{booktitle}{\emph{Proceedings of the 40th ACM SIGPLAN Conference on
  Programming Language Design and Implementation}} (Phoenix, AZ, USA)
  \emph{(\bibinfo{series}{PLDI 2019})}. \bibinfo{publisher}{Association for
  Computing Machinery}, \bibinfo{address}{New York, NY, USA},
  \bibinfo{pages}{96–110}.
\newblock
\showISBNx{9781450367127}
\urldef\tempurl%
\url{https://doi.org/10.1145/3314221.3314609}
\showDOI{\tempurl}


\bibitem[Kokologiannakis and Vafeiadis(2021)]%
        {kokologiannakis-vafeiadis:genmc}
\bibfield{author}{\bibinfo{person}{Michalis Kokologiannakis} {and}
  \bibinfo{person}{Viktor Vafeiadis}.} \bibinfo{year}{2021}\natexlab{}.
\newblock \showarticletitle{GenMC: A Model Checker for Weak Memory Models}. In
  \bibinfo{booktitle}{\emph{Computer Aided Verification}},
  \bibfield{editor}{\bibinfo{person}{Alexandra Silva} {and}
  \bibinfo{person}{K.~Rustan~M. Leino}} (Eds.). \bibinfo{publisher}{Springer
  International Publishing}, \bibinfo{address}{Cham},
  \bibinfo{pages}{427--440}.
\newblock
\showISBNx{978-3-030-81685-8}
\urldef\tempurl%
\url{https://doi.org/10.1007/978-3-030-81685-8_20}
\showDOI{\tempurl}


\bibitem[Kroening et~al\mbox{.}(2015)]%
        {Kroening:2015:EVL}
\bibfield{author}{\bibinfo{person}{Daniel Kroening}, \bibinfo{person}{Lihao
  Liang}, \bibinfo{person}{Tom Melham}, \bibinfo{person}{Peter Schrammel},
  {and} \bibinfo{person}{Michael Tautschnig}.} \bibinfo{year}{2015}\natexlab{}.
\newblock \showarticletitle{Effective Verification of Low-Level Software with
  Nested Interrupts}. In \bibinfo{booktitle}{\emph{Proceedings of the 2015
  Design, Automation {\&} Test in Europe Conference {\&} Exhibition, {DATE}
  2015, Grenoble, France, March 9-13, 2015}},
  \bibfield{editor}{\bibinfo{person}{Wolfgang Nebel} {and}
  \bibinfo{person}{David Atienza}} (Eds.). \bibinfo{publisher}{{EDA
  Consortium}}, \bibinfo{pages}{229--234}.
\newblock
\showISBNx{978-3-9815370-4-8}
\urldef\tempurl%
\url{http://www.cs.ox.ac.uk/tom.melham/pub/Kroening-2015-EVL.pdf}
\showURL{%
\tempurl}


\bibitem[Lahav et~al\mbox{.}(2017)]%
        {lahav-et-al:repairing-c11}
\bibfield{author}{\bibinfo{person}{Ori Lahav}, \bibinfo{person}{Viktor
  Vafeiadis}, \bibinfo{person}{Jeehoon Kang}, \bibinfo{person}{Chung-Kil Hur},
  {and} \bibinfo{person}{Derek Dreyer}.} \bibinfo{year}{2017}\natexlab{}.
\newblock \showarticletitle{Repairing sequential consistency in C/C++11}. In
  \bibinfo{booktitle}{\emph{Proceedings of the 38th ACM SIGPLAN Conference on
  Programming Language Design and Implementation}} (Barcelona, Spain)
  \emph{(\bibinfo{series}{PLDI 2017})}. \bibinfo{publisher}{Association for
  Computing Machinery}, \bibinfo{address}{New York, NY, USA},
  \bibinfo{pages}{618–632}.
\newblock
\showISBNx{9781450349888}
\urldef\tempurl%
\url{https://doi.org/10.1145/3062341.3062352}
\showDOI{\tempurl}


\bibitem[Liang et~al\mbox{.}(2017)]%
        {Liang:2017:EVL}
\bibfield{author}{\bibinfo{person}{Lihao Liang}, \bibinfo{person}{Tom Melham},
  \bibinfo{person}{Daniel Kroening}, \bibinfo{person}{Peter Schrammel}, {and}
  \bibinfo{person}{Michael Tautschnig}.} \bibinfo{year}{2017}\natexlab{}.
\newblock \showarticletitle{Effective Verification for Low-Level Software with
  Competing Interrupts}.
\newblock \bibinfo{journal}{\emph{ACM Transactions on Embedded Computing
  Systems}} \bibinfo{volume}{17}, \bibinfo{number}{2} (\bibinfo{date}{December}
  \bibinfo{year}{2017}), \bibinfo{pages}{36:1--36:26}.
\newblock
\showISSN{1539-9087}
\urldef\tempurl%
\url{https://doi.org/10.1145/3147432}
\showDOI{\tempurl}


\bibitem[Mateo(2021)]%
        {deprecatebiasedlocking}
\bibfield{author}{\bibinfo{person}{Patricio~Chilano Mateo}.}
  \bibinfo{year}{2021}\natexlab{}.
\newblock \bibinfo{title}{JEP 374: Deprecate and Disable Biased Locking}.
\newblock \bibinfo{howpublished}{JDK Enhancement Proposal}.
\newblock
\urldef\tempurl%
\url{https://openjdk.org/jeps/374}
\showURL{%
\tempurl}


\bibitem[McKenney(2023)]%
        {perfbook}
\bibfield{author}{\bibinfo{person}{Paul~E. McKenney}.}
  \bibinfo{year}{2023}\natexlab{}.
\newblock \bibinfo{booktitle}{\emph{Is Parallel Programming Hard, And, If So,
  What Can You Do About It?}}
\newblock
\urldef\tempurl%
\url{https://mirrors.edge.kernel.org/pub/linux/kernel/people/paulmck/perfbook/perfbook.html}
\showURL{%
\tempurl}


\bibitem[McKenney(2024)]%
        {rcutxt}
\bibfield{author}{\bibinfo{person}{Paul~E. McKenney}.}
  \bibinfo{year}{2024}\natexlab{}.
\newblock \bibinfo{title}{RCU Concepts}.
\newblock
\newblock
\urldef\tempurl%
\url{https://www.kernel.org/doc/Documentation/RCU/rcu.txt}
\showURL{%
\tempurl}
\newblock
\shownote{Accessed 2024-11-19}.


\bibitem[Parkinson(2024)]%
        {mattp}
\bibfield{author}{\bibinfo{person}{Matthew~J. Parkinson}.}
  \bibinfo{year}{2024}\natexlab{}.
\newblock \bibinfo{title}{Some things I wish I hadn’t seen}.
\newblock \bibinfo{howpublished}{presented at The Future of Weak Memory 2024}.
\newblock


\bibitem[Pichon{-}Pharabod and Sewell(2016)]%
        {Pichon-PharabodS16}
\bibfield{author}{\bibinfo{person}{Jean Pichon{-}Pharabod} {and}
  \bibinfo{person}{Peter Sewell}.} \bibinfo{year}{2016}\natexlab{}.
\newblock \showarticletitle{A concurrency semantics for relaxed atomics that
  permits optimisation and avoids thin-air executions}. In
  \bibinfo{booktitle}{\emph{Proceedings of the 43rd Annual {ACM}
  {SIGPLAN-SIGACT} Symposium on Principles of Programming Languages, {POPL}
  2016, St. Petersburg, FL, USA, January 20 - 22, 2016}},
  \bibfield{editor}{\bibinfo{person}{Rastislav Bod{\'{\i}}k} {and}
  \bibinfo{person}{Rupak Majumdar}} (Eds.). \bibinfo{publisher}{{ACM}},
  \bibinfo{pages}{622--633}.
\newblock
\urldef\tempurl%
\url{https://doi.org/10.1145/2837614.2837616}
\showDOI{\tempurl}


\bibitem[Pugh(1999)]%
        {Pugh99}
\bibfield{author}{\bibinfo{person}{William~W. Pugh}.}
  \bibinfo{year}{1999}\natexlab{}.
\newblock \showarticletitle{Fixing the Java Memory Model}. In
  \bibinfo{booktitle}{\emph{Proceedings of the {ACM} 1999 Conference on Java
  Grande, {JAVA} '99, San Francisco, CA, USA, June 12-14, 1999}},
  \bibfield{editor}{\bibinfo{person}{Geoffrey~C. Fox},
  \bibinfo{person}{Klaus~E. Schauser}, {and} \bibinfo{person}{Marc Snir}}
  (Eds.). \bibinfo{publisher}{{ACM}}, \bibinfo{pages}{89--98}.
\newblock
\urldef\tempurl%
\url{https://doi.org/10.1145/304065.304106}
\showDOI{\tempurl}


\bibitem[Pulte(2018)]%
        {Pulte-phd}
\bibfield{author}{\bibinfo{person}{Christopher Pulte}.}
  \bibinfo{year}{2018}\natexlab{}.
\newblock \emph{\bibinfo{title}{{The Semantics of Multicopy Atomic ARMv8 and
  RISC-V}}}.
\newblock \bibinfo{thesistype}{Ph.\,D. Dissertation}.
  \bibinfo{school}{University of Cambridge}.
\newblock
\newblock
\shownote{\url{https://www.repository.cam.ac.uk/handle/1810/292229}}.


\bibitem[Pulte et~al\mbox{.}(2018)]%
        {PulteFDFSS18}
\bibfield{author}{\bibinfo{person}{Christopher Pulte}, \bibinfo{person}{Shaked
  Flur}, \bibinfo{person}{Will Deacon}, \bibinfo{person}{Jon French},
  \bibinfo{person}{Susmit Sarkar}, {and} \bibinfo{person}{Peter Sewell}.}
  \bibinfo{year}{2018}\natexlab{}.
\newblock \showarticletitle{Simplifying {ARM} concurrency: multicopy-atomic
  axiomatic and operational models for ARMv8}.
\newblock \bibinfo{journal}{\emph{Proc. {ACM} Program. Lang.}}
  \bibinfo{volume}{2}, \bibinfo{number}{{POPL}} (\bibinfo{year}{2018}),
  \bibinfo{pages}{19:1--19:29}.
\newblock
\urldef\tempurl%
\url{https://doi.org/10.1145/3158107}
\showDOI{\tempurl}


\bibitem[Russell and Detlefs(2006)]%
        {Russell06}
\bibfield{author}{\bibinfo{person}{Kenneth Russell} {and}
  \bibinfo{person}{David Detlefs}.} \bibinfo{year}{2006}\natexlab{}.
\newblock \showarticletitle{Eliminating synchronization-related atomic
  operations with biased locking and bulk rebiasing}. In
  \bibinfo{booktitle}{\emph{Proceedings of the 21st Annual ACM SIGPLAN
  Conference on Object-Oriented Programming Systems, Languages, and
  Applications}} (Portland, Oregon, USA) \emph{(\bibinfo{series}{OOPSLA '06})}.
  \bibinfo{publisher}{Association for Computing Machinery},
  \bibinfo{address}{New York, NY, USA}, \bibinfo{pages}{263–272}.
\newblock
\showISBNx{1595933484}
\urldef\tempurl%
\url{https://doi.org/10.1145/1167473.1167496}
\showDOI{\tempurl}


\bibitem[Sarkar et~al\mbox{.}(2012)]%
        {pldi2012}
\bibfield{author}{\bibinfo{person}{Susmit Sarkar}, \bibinfo{person}{Kayvan
  Memarian}, \bibinfo{person}{Scott Owens}, \bibinfo{person}{Mark Batty},
  \bibinfo{person}{Peter Sewell}, \bibinfo{person}{Luc Maranget},
  \bibinfo{person}{Jade Alglave}, {and} \bibinfo{person}{Derek Williams}.}
  \bibinfo{year}{2012}\natexlab{}.
\newblock \showarticletitle{Synchronising {C/C++} and {POWER}}. In
  \bibinfo{booktitle}{\emph{{ACM} {SIGPLAN} Conference on Programming Language
  Design and Implementation, {PLDI} '12, Beijing, China - June 11 - 16, 2012}},
  \bibfield{editor}{\bibinfo{person}{Jan Vitek}, \bibinfo{person}{Haibo Lin},
  {and} \bibinfo{person}{Frank Tip}} (Eds.). \bibinfo{publisher}{{ACM}},
  \bibinfo{pages}{311--322}.
\newblock
\urldef\tempurl%
\url{https://doi.org/10.1145/2254064.2254102}
\showDOI{\tempurl}


\bibitem[Sarkar et~al\mbox{.}(2011)]%
        {pldi105}
\bibfield{author}{\bibinfo{person}{Susmit Sarkar}, \bibinfo{person}{Peter
  Sewell}, \bibinfo{person}{Jade Alglave}, \bibinfo{person}{Luc Maranget},
  {and} \bibinfo{person}{Derek Williams}.} \bibinfo{year}{2011}\natexlab{}.
\newblock \showarticletitle{Understanding {POWER} multiprocessors}. In
  \bibinfo{booktitle}{\emph{Proceedings of the 32nd {ACM} {SIGPLAN} Conference
  on Programming Language Design and Implementation, {PLDI} 2011, San Jose, CA,
  USA, June 4-8, 2011}}, \bibfield{editor}{\bibinfo{person}{Mary~W. Hall} {and}
  \bibinfo{person}{David~A. Padua}} (Eds.). \bibinfo{publisher}{{ACM}},
  \bibinfo{pages}{175--186}.
\newblock
\urldef\tempurl%
\url{https://doi.org/10.1145/1993498.1993520}
\showDOI{\tempurl}


\bibitem[Sarkar et~al\mbox{.}(2009)]%
        {x86popl}
\bibfield{author}{\bibinfo{person}{Susmit Sarkar}, \bibinfo{person}{Peter
  Sewell}, \bibinfo{person}{Francesco~Zappa Nardelli}, \bibinfo{person}{Scott
  Owens}, \bibinfo{person}{Tom Ridge}, \bibinfo{person}{Thomas Braibant},
  \bibinfo{person}{Magnus~O. Myreen}, {and} \bibinfo{person}{Jade Alglave}.}
  \bibinfo{year}{2009}\natexlab{}.
\newblock \showarticletitle{The semantics of x86-CC multiprocessor machine
  code}. In \bibinfo{booktitle}{\emph{Proceedings of the 36th {ACM}
  {SIGPLAN-SIGACT} Symposium on Principles of Programming Languages, {POPL}
  2009, Savannah, GA, USA, January 21-23, 2009}},
  \bibfield{editor}{\bibinfo{person}{Zhong Shao} {and}
  \bibinfo{person}{Benjamin~C. Pierce}} (Eds.). \bibinfo{publisher}{{ACM}},
  \bibinfo{pages}{379--391}.
\newblock
\urldef\tempurl%
\url{https://doi.org/10.1145/1480881.1480929}
\showDOI{\tempurl}


\bibitem[Sewell et~al\mbox{.}(2022)]%
        {acs-2022}
\bibfield{author}{\bibinfo{person}{Peter Sewell}, \bibinfo{person}{Christopher
  Pulte}, \bibinfo{person}{Shaked Flur}, \bibinfo{person}{Mark Batty},
  \bibinfo{person}{Luc Maranget}, {and} \bibinfo{person}{Alasdair Armstrong}.}
  \bibinfo{year}{2022}\natexlab{}.
\newblock \bibinfo{title}{Multicore Semantics: Making Sense of Relaxed Memory
  (MPhil slides)}.
\newblock
\newblock
\urldef\tempurl%
\url{https://www.cl.cam.ac.uk/~pes20/slides-acs-2022.pdf}
\showURL{%
\tempurl}


\bibitem[Sewell et~al\mbox{.}(2010)]%
        {cacm}
\bibfield{author}{\bibinfo{person}{P. Sewell}, \bibinfo{person}{S. Sarkar},
  \bibinfo{person}{S. Owens}, \bibinfo{person}{F. Zappa~Nardelli}, {and}
  \bibinfo{person}{M.~O. Myreen}.} \bibinfo{year}{2010}\natexlab{}.
\newblock \showarticletitle{{x86-TSO}: A Rigorous and Usable Programmer's Model
  for x86 Multiprocessors}.
\newblock \bibinfo{journal}{\emph{Commun. ACM}} \bibinfo{volume}{53},
  \bibinfo{number}{7} (\bibinfo{date}{July} \bibinfo{year}{2010}),
  \bibinfo{pages}{89--97}.
\newblock
\urldef\tempurl%
\url{https://doi.org/10.1145/1785414.1785443}
\showDOI{\tempurl}


\bibitem[Simner et~al\mbox{.}(2022)]%
        {relaxedVM-esop2022}
\bibfield{author}{\bibinfo{person}{Ben Simner}, \bibinfo{person}{Alasdair
  Armstrong}, \bibinfo{person}{Jean Pichon-Pharabod},
  \bibinfo{person}{Christopher Pulte}, \bibinfo{person}{Richard Grisenthwaite},
  {and} \bibinfo{person}{Peter Sewell}.} \bibinfo{year}{2022}\natexlab{}.
\newblock \showarticletitle{Relaxed virtual memory in {Armv8-A}}. In
  \bibinfo{booktitle}{\emph{Proceedings of the 31st European Symposium on
  Programming}} \emph{(\bibinfo{series}{Lecture Notes in Computer Science},
  Vol.~\bibinfo{volume}{13240})}. \bibinfo{publisher}{Springer},
  \bibinfo{pages}{143--173}.
\newblock
\urldef\tempurl%
\url{https://doi.org/10.1007/978-3-030-99336-8\_6}
\showDOI{\tempurl}


\bibitem[Simner et~al\mbox{.}(2020)]%
        {SimnerFPAPMS20}
\bibfield{author}{\bibinfo{person}{Ben Simner}, \bibinfo{person}{Shaked Flur},
  \bibinfo{person}{Christopher Pulte}, \bibinfo{person}{Alasdair Armstrong},
  \bibinfo{person}{Jean Pichon{-}Pharabod}, \bibinfo{person}{Luc Maranget},
  {and} \bibinfo{person}{Peter Sewell}.} \bibinfo{year}{2020}\natexlab{}.
\newblock \showarticletitle{ARMv8-A System Semantics: Instruction Fetch in
  Relaxed Architectures}. In \bibinfo{booktitle}{\emph{Programming Languages
  and Systems - 29th European Symposium on Programming, {ESOP} 2020, Held as
  Part of the European Joint Conferences on Theory and Practice of Software,
  {ETAPS} 2020, Dublin, Ireland, April 25-30, 2020, Proceedings}}
  \emph{(\bibinfo{series}{Lecture Notes in Computer Science},
  Vol.~\bibinfo{volume}{12075})}, \bibfield{editor}{\bibinfo{person}{Peter
  M{\"{u}}ller}} (Ed.). \bibinfo{publisher}{Springer},
  \bibinfo{pages}{626--655}.
\newblock
\urldef\tempurl%
\url{https://doi.org/10.1007/978-3-030-44914-8\_23}
\showDOI{\tempurl}


\bibitem[Sindhu et~al\mbox{.}(1991)]%
        {SFC91}
\bibfield{author}{\bibinfo{person}{P.~S. Sindhu}, \bibinfo{person}{J.-M.
  Frailong}, {and} \bibinfo{person}{M. Cekleov}.}
  \bibinfo{year}{1991}\natexlab{}.
\newblock \showarticletitle{Formal Specification of Memory Models}.
\newblock In \bibinfo{booktitle}{\emph{Scalable Shared Memory
  Multiprocessors}}. \bibinfo{publisher}{Kluwer}, \bibinfo{pages}{25--42}.
\newblock
\urldef\tempurl%
\url{https://doi.org/10.1007/978-1-4615-3604-8_2}
\showDOI{\tempurl}


\end{thebibliography}
\fi

\ifoldbits
\newpage
\appendix
\include{old-bits.tex}
\fi %

\end{document}